\documentclass{article}
\RequirePackage[colorlinks,citecolor=blue,urlcolor=blue]{hyperref}
\usepackage{amsmath}
\usepackage{amsfonts}
\usepackage{url}
\usepackage{bbm}
\usepackage{accents}
\usepackage{subcaption}
\usepackage{float}
\usepackage[authoryear]{natbib}
\usepackage{tabularx}
\usepackage{array}
\usepackage{framed}
\usepackage[]{mdframed}
\usepackage{enumitem}
\usepackage{diagbox}
\usepackage{wrapfig,lipsum,booktabs}
\usepackage{csquotes}
\usepackage{amsthm}
\usepackage{nccmath}
\usepackage{mathtools}
\usepackage{mdframed}
\usepackage{lipsum}
\usepackage{pdfpages}
\usepackage{tcolorbox}
\usepackage{graphicx}
\usepackage{xfrac}
\usepackage{xcolor}
\usepackage{appendix}
\usepackage{lingmacros}
\usepackage{tree-dvips}
\usepackage{multicol}
\usepackage{multirow}
\usepackage[linesnumbered,ruled,vlined]{algorithm2e}
\usepackage{algpseudocode}

\usepackage{soul}

\date{}

\theoremstyle{plain}

\newtheorem{theorem}{Theorem}[section]
\newtheorem{lemma}[theorem]{Lemma}

\theoremstyle{definition}
\newtheorem{definition}[theorem]{Definition}

\newcommand\subsetsim{\mathrel{\substack{\textstyle\subset\\[0ex]\textstyle\sim}}}
\newcommand\supsetsim{\mathrel{\substack{\textstyle\supset\\[0ex]\textstyle\sim}}}

\newtheorem{remark}{Remark}

\newtheorem{assumption}{Assumption}

\usepackage[left = 2 cm, right = 2 cm]{geometry}
\let\Abstract\abstract
\long\def\abstract{\mdframed[backgroundcolor=white!20,hidealllines=true]
  \vspace*{-0.1\baselineskip}\Abstract}
\let\endAbstract\endabstract
\def\endabstract{\endAbstract\endmdframed\par\bigskip}
		\title{\textbf{\Large Consistent detection and estimation of multiple structural changes in functional data: unsupervised and supervised approaches}}
\begin{document}
	\maketitle

		
\begin{center}
	\null\vskip-1.5cm
	\author{  \textbf{\large Sourav Chakrabarty$^{1}$, Anirvan Chakraborty$^{2}$, 
			 Shyamal\ K.\ De}$^{1}$
		\\
		$^{1}$Applied Statistics Unit \\
		Indian Statistical Institute, Kolkata, India  \\
		$^{2}$Department of Mathematics and Statistics	\\
		Indian Institute of Science Education and Research Kolkata, India
		Emails:$^{1}$chakrabartysourav024@gmail.com, $^{2}$anirvan.c@iiserkol.ac.in, $^{1}$shyamalkd@isical.ac.in
	}
\end{center}

\begin{abstract}
We develop algorithms for detecting multiple changepoints in  functional data when the number of changepoints is unknown (unsupervised case), when it is specified apriori (supervised case), and when certain bounds are available (semi-supervised case). These algorithms utilize the maximum mean discrepancy (MMD) measure between distributions on Hilbert spaces. We develop an oracle analysis of the changepoint detection problem which reveals an interesting relationship between the true changepoint locations and the local maxima of the oracle MMD curve. The proposed algorithms are shown to detect general distributional changes by exploiting this connection. In the unsupervised case, we test the significance of a potential changepoint and establish its consistency under the single changepoint setting. We investigate the strong consistency of the changepoint estimators in both single and multiple changepoint settings. In both supervised and semi-supervised scenarios, we include a step to merge consecutive groups that are similar to appropriately utilize the prior information about the number of changepoints. In the supervised scenario, the algorithm satisfies an order-preserving property: the estimated changepoints are contained in the true set of changepoints in the underspecified case, while they contain the true set under overspecification. We evaluate the performance of the algorithms on a variety of datasets demonstrating  the superiority of the proposed algorithms compared to some of the existing methods.
\end{abstract}




\section{Introduction}\label{Introduction}
Changepoint detection focuses on identifying abrupt changes in the statistical properties of data over time. These changepoints determine transitions in the underlying processes generating the data. Detection of these shifts is essential across various domains including finance, healthcare, climate science, and engineering, where timely identification of changes can lead to better decision-making. The importance of changepoint detection has grown with the increasing availability of data streams from diverse sources such as sensors, financial markets and social media platforms. Consequently, researchers have developed a variety of techniques for both offline and online changepoint detection. Offline methods analyze complete datasets retrospectively while online methods focus on real-time monitoring to identify changes as they occur. We refer to \citet{truong2020selective} for an extensive review of multivariate changepoint detection methods while \citet{xu2025change} reviews both online and offline changepoint detection methods for a wide variety of practical applications using modern deep learning models. 

In this paper, we consider the offline changepoint problem with an emphasis towards functional data. For functional data, \citet{berkes2009detecting} consider a Cumulative Sum (CUSUM) test for independently observed functional data under the At Most One Changepoint (AMOC) situation when there is change in the mean function of the data. This test was subsequently extended and studied by \citet{aue2009estimation}, \citet{hormann2010weakly} and \citet{aston2012detecting}. Among the other literature on detecting one or multiple changes in the mean function include \citet{44e75106888d4236b3935cb41db7bc34} (self-normalization technique), \citet{sharipov2016sequential} (block-bootstrap procedure), \citet{gromenko2017detection} (spatially correlated functional data), \citet{aue2018detecting} (fully-functional approach), \citet{li2021bayesian} (Bayesian approach) and \citet{bastian2024multiple} (sup-norm approach). \citet{chiou2019identifying} presented a two-stage dynamic segmentation and backward elimination technique for detecting multiple changepoints in the mean function while \citet{chen2023greedy} proposed a greedy segmentation algorithm as an alternative approach. \citet{padilla2022} studied the problem of changepoint detection for sparsely or densely sampled functional curves through an algorithm called functional seeded binary segmentation. Recently, changes in the covariance structure has been studied for functional data, see e.g., \citet{aue2020structural}, \citet{dette2021detecting} and \citet{horvath2022change}. \citet{harris2022scalable} proposed a multiple changepoint detection procedure for functional data which handles changes in both mean function and covariance operator.  \citet{BHT2025} proposed a novel family of test statistics to detect the change in distribution in a sequence of dependent functional observations based on a generalization of the empirical energy distance, while \citet{horvath2025change} developed a method based on the empirical characteristic function. \citet{ramsay2025robust} developed a robust changepoint detection procedure for changes in the covariance structure based on a functional Kruskal-Wallis test. 

There is a growing literature on changepoint detection methods for non-Eucliean object data (for example, data in arbitrary metric spaces or manifolds) using distance-based or graph-based approaches including the works by \citet{dubey2020frechet}, \citet{jiang2024two}, \citet{zhang2025change}, \citet{chu2019asymptotic}, \citet{dai2019discovering}, \citet{wang2023online, wang2024non}, \citet{kostic2025change} and \citet{chen2023graph}, which provides a detailed review of graph-based approaches.

Recently, a number of kernel-based changepoint detection methods have been developed for identifying distributional shifts in complex, high-dimensional, or non-Euclidean data where the idea is to embed observations into a reproducing kernel Hilbert space (RKHS) and then detect changes in the mean embedding of the underlying distributions. \citet{harchaoui2007retrospective} and \citet{harchaoui2008kernel} pioneered this approach by formulating a kernelized least-squares segmentation method for offline changepoint analysis. Building on these works, \cite{arlot2019kernel} proposed the Kernel Change-Point (KCP) algorithm, which integrates the kernel least-squares criterion with a data-driven penalization scheme to estimate the number and locations of changepoints simultaneously. \cite{garreau2018consistent} further advanced the theory by proving asymptotic consistency and localization rates for the KCP estimator, thereby ensuring correct identification of the number and positions of changepoints. More recently, \citet{song2024practical} proposed a kernel-based changepoint detection procedure that leverages kernel embeddings for improved computational efficiency and sensitivity in high-dimensional and non-Euclidean settings such as networks or manifold-valued data. 
\par

In this paper, we aim at detecting multiple changes in the data without making any assumption about the nature of the changes. In the context of functional data, the change could either be in the mean function or in the covariance operator (eigenvalues or eigenfunctions) or a general change in the distribution. To this end, we use the Maximum Mean Discrepancy (MMD) metric between two probability distributions on a separable Hilbert space which allows us to capture any form of distributional change. We develop a unified MMD-based framework by introducing unsupervised, supervised, and semi-supervised algorithms that automatically estimate both the number and positions of changepoints. First, we consider the situation where the number of changepoints $K_0$ is unknown (unsupervised scenario). An iterative binary splitting strategy is employed in a way to ensure maximal separation of the segments of the data in terms of the MMD metric. For each such split, we use a permutation test to assess whether the resulting changepoint (arising out of the split) is significant or not. We stop the recursion when all of the changepoints obtained from all of the new splits are found to be insignificant by the permutation test procedure. The algorithm thus yields an estimate of the number of changepoints as well as their locations. We prove the consistency of the permutation test under the single changepoint setup. The strong consistency of the estimation procedure is investigated for both single and multiple changepoints settings. We also consider two other situations, namely, when the number of changepoints is specified in advance (supervised scenario) as well as when lower or upper bounds are provided for it (semi-supervised scenario). For both of these scenarios, we modify the previous algorithm to include a merging step which merges two adjacent segments that are similar in terms of the MMD metric. This allows us to properly use the prior information on the number of changepoints. Indeed, for the supervised scenario, we no longer need the testing step. We prove that in this scenario, the Hausdorff distance between the set of estimated changepoints and the set of true changepoints converges to zero almost surely if the number of changepoints is correctly specified. We also establish an order-preserving property of the procedure under misspecification of the number of changepoints. We show that when the number of changepoints is underspecified (respectively, overspecified), the set of estimated changepoints will be asymptotically contained in (respectively, contains) the set of true changepoints almost surely (cf. Theorem \ref{pop}).

\indent The main contributions of the paper are summarized below: \\
1. We consider general structural changes in the context of time-ordered infinite-dimensional functional observations. \\
2. In the context of changepoint detection, we develop a novel oracle framework that provides a rigorous analytical understanding of the changepoint problem. This framework substantially extends the theoretical scope of kernel-based changepoint methodologies beyond that of the existing approaches. The oracle analysis establishes an exact correspondence between the true changepoint locations and the local maxima of the oracle MMD curve (cf. \eqref{rho curve in oracle-single} and \eqref{rho_curve_in_oracle-two} in Section~\ref{motivation}, which address the single and two changepoint settings, respectively). In particular, the true changepoints coincide precisely with these local maxima. The proposed algorithms capitalize on this fundamental characterization to achieve efficient and statistically principled detection of changepoints.\\
3. We propose an exact test of size $\alpha$ for assessing the significance of an estimated changepoint and provide a novel mathematical proof of  the consistency of the test as well as the estimation procedure in the single changepoint setting.  
The consistency of the estimates of the changepoints in the multiple changepoint setting can also be inferred from this analysis.\\
4. We propose algorithms (DESC-S and DESC-SS) that successfully utilize prior information about the number of changepoints, whether the specification be the exact number of changepoints or upper/lower bounds. Such a general treatment is novel to the best of our knowledge. \\
5.  We theoretically establish a novel order-preserving property of the algorithms in the supervised scenario which is desirable. \\
It should be noted that although we choose to work with functional data for our simulation and the real data analyses, the techniques developed in the paper can be applied to more general data. Indeed, the theoretical results hold seamlessly across multivariate data as well as data in general metric spaces. The only requirement is to have a kernel (in the definition of MMD) which can identify distributions uniquely. This generalizability of our method distinguishes it from classical CUSUM-type algorithms which are only applicable in linear spaces (cf. the discussion in Section \ref{conclusion}). It is worth mentioning that a permutation test based changepoint algorithm using the energy distance has been studied in \citet{matteson2014nonparametric} in the context of multivariate data. However, we would like to point out that our paper significantly generalizes their work in terms of (a) broader scope -- uses the MMD measure and is applicable to data taking values in general metric spaces, (b) improved theoretical understanding -- the novel oracle analysis as well as the consistency of the permutation test, and (c) additional novelty -- algorithms that allow the use of prior information on the number of changepoints and enjoy a desirable order-preserving property. 

The remaining part of this paper is organized as follows. Section \ref{motivation} provides a brief introduction to the MMD measure and an oracle analysis of changepoint problem which guides us to develop such algorithms. Changepoint estimate and a testing procedure to determine its significance are the fundamental components of the changepoint detection methods described in Section \ref{methodology}. The changepoint detection algorithm under the AMOC situation is discussed as well as the consistency of the estimated breakfraction and the testing procedure under single changepoint setup are discussed in this section. 
In Section \ref{Method for unknown number of changepoints}, multiple changepoint detection algorithm is proposed when the number of changepoints is not specified. 
In Section \ref{methodology for prefixed number of changepoints}, a changepoint detection algorithm is developed for specified number of changepoints and we establish that this algorithm enjoys an order-preserving property for large sample size. In section \ref{Methodology for specified bounds of number of changepoints}, an algorithm has been developed to identify changepoints in situations where some bounds on the number of changepoints are provided instead of a predetermined number. A detailed simulation study and some real data  analyses are provided in Section \ref{simulation}, and we compare the proposed algorithms with some of the existing changepoint detection procedures for functional data. We conclude the paper with some additional remarks and future scope of research in Section \ref{conclusion}.

\section{An Oracle Analysis Using MMD} \label{motivation}
For two probability distributions $P$ and $Q$ defined on a separable Hilbert space ${\cal H}$ the Maximum Mean Discrepancy (MMD) measure (see \citet{gretton2006kernel}) is defined by 
\begin{center}
	$\mbox{MMD}(P,Q)$ = $\underset{f \in {\cal F}} {\sup}$$|\int f\,dP - \int f\,dQ|$,
\end{center}
where ${\cal F}$ is a suitably class of functions $f : \mathcal{H} \xrightarrow[]{} \mathbb{R}$ which is rich enough to discriminate between any two distinct probability measures. If we take ${\cal F}$ to be the RKHS generated by a positive semi-definite kernel $k : {\cal H} \times {\cal H} \rightarrow \mathbb{R}$, then it is known that the square of the MMD measure, which we denote by $d(\cdot,\cdot)$ can also be expressed as
\begin{align*}
	d(P,Q) =& \iint k(x,x')\,dP(x)dP(x') + \iint k(y,y')\,dQ(y)dQ(y') - 2\iint k(x,y)\,dP(x)dQ(y),
\end{align*}
provided that $\int \sqrt{k(x,x)}dP(x) < \infty$  (see for example \citet{sriperumbudur2010hilbert}). The choices of $k$ for which the metric property of $d(\cdot,\cdot)$ holds are called characteristic kernels. If ${\cal H} = \mathbb{R}^d$ for some $d \geq 1$, several characteristic kernels have been well-studied (see for example \citet{NIPS2008_d07e70ef}), while for infinite dimensional Hilbert spaces, Gaussian kernels and Laplace kernels have been shown to be characteristic (see Theorems 3.1 and 4.2 of \citet{ziegel2024characteristic}). Throughout this paper, we will take $k$ to be a bounded characteristic kernel. Note that the Gaussian and the Laplace kernels are bounded characteristic kernels.

Given i.i.d. samples $X_1, X_2,\ldots,X_m$ from $P$ and $Y_1,Y_2,\ldots,Y_n$ from $Q$, a simple consistent and unbiased estimator of $d(P,Q)$ is given by
\begin{align*}
d(\widehat{P},\widehat{Q}) =  \frac{1}{m^2}\sum_{i,i'=1}^{m} k(X_i,X_{i'}) + \frac{1}{n^2}\sum\limits_{j,j'=1}^{n}   k(Y_j,Y_{j'}) - \frac{2}{mn}\sum\limits_{i=1}^{m}\sum\limits_{j=1}^{n}  k(X_i,Y_j),
\end{align*}
\noindent
where $\widehat{P} = m^{-1}\sum_{i=1}^{m} \delta_{X_{i}}$ and $\widehat{Q} = n^{-1}\sum_{j=1}^{n}   \delta_{Y_{j}}$ are the empirical measures with $\delta_z$ denoting the Dirac measure of $z$. In the next section, we will develop a novel methodology using the MMD, which will allow us to identify changepoints in a set of observations. 

To motivate the usefulness of MMD in the context of changepoint detection, let us consider a simple scenario with exactly one changepoint. We will consider an oracle setup, where the location of the changepoint in the dataset $\{Z_1,Z_2,\dots,Z_n\}$ is known. More specifically, $Z_1,Z_2,\ldots,Z_{n_1} \sim P_1$ and \\$Z_{n_1+1},Z_{n_1+2},\ldots,Z_n \sim P_2$. Now consider any partition of the data into subsets $S_1$ and $S_2$, where $S_1 = \{Z_1,Z_2,\ldots,Z_r\}$ and $S_2 = \{Z_{r+1},Z_{r+2},\ldots,Z_n\}$ with $r$ being a potential changepoint location. The value of $r$ relative to $n_1$ will then determine whether $S_1$ contains observations only from $P_1$ (pure set) or if it contains observations from both $P_1$ and $P_2$ (impure set). If $S_1$ is a pure set, which happens if $r \leq n_1$, it may be viewed as a sample from $\widehat{P}_1$. On the other hand, if $S_1$ is an impure set, which happens if $r > n_1$, it can be viewed as a sample from a mixture of $\widehat{P}_1$ and $\widehat{P}_2$, say $\alpha\widehat{P}_1 + (1-\alpha)\widehat{P}_2$. The most canonical choice of $\alpha$ in this case will be the sample proportion of observations in $S_1$ coming from $\widehat{P}_1$, namely, $n_1/r$. We define the notion of representative distribution of $S_1$ to precisely describe this configuration, and it is denoted by $\widetilde{P}_{S_1}$ which is defined to be $\widehat{P}_!$ if $r \leq n_1$ and equal to $(n_1/r) \widehat{P}_1 + \{(r-n_1)/r\}\widehat{P}_2$ if $r > n_1$. In a similar way, the representative distribution of $S_2$ is denoted by $\widetilde{P}_{S_2}$ which is defined to be $\{(n_1-r)/(n_1+n_2-r)\}\widehat{P}_1 + \{n_2/(n_1+n_2-r)\}\widehat{P}_2$ if $r \leq n_1$ and equal to $\widehat{P}_2$ if $r > n_1$. 

Any oracle changepoint algorithm should detect the changepoint location perfectly, that is, choose $r = n_1$. To achieve this end, our approach is to compute a suitable measure of discrepancy (in terms of MMD) between $S_1$ and $S_2$, more specifically between $\widetilde{P}_{S_1}$ and $\widetilde{P}_{S_2}$, and maximize it with respect to $r$. This idea is motivated from the following notable property of MMD: if $F$ and $G$ are two probability distributions and $\alpha,\beta \in [0,1]$, then
\begin{eqnarray}
	d(\alpha F + (1-\alpha)G, \beta F + (1-\beta) G) \ = \ (\alpha - \beta)^2d(F,G) \ \leq \ d(F,G), \label{MMDproperty}
\end{eqnarray}
where the equality holds if and only if $(\alpha,\beta) = (0,1)$ or $(1,0)$. The above property implies that $d(\widetilde{P}_{S_1},\widetilde{P}_{S_2})$ will be maximum if and only if $r = n_1$. Our oracle algorithm will use a modified distance between $S_1$ and $S_2$ given by 
\begin{align*}
	\rho^{*}(S_1,S_2) = \frac{r(n-r)}{n^2}d\left(\widetilde{P}_{S_1},\widetilde{P}_{S_2}\right),
\end{align*}
where the multiplicative factor acts as a balancing term in case one of $S_1$ or $S_2$ is small in size compared to the other.  It can actually be shown that 
   \begin{eqnarray}
	\rho^*(S_{1},S_{2}) = \left \{
	\begin{array}{l@{\quad} l }
		\frac{rn_2^2}{n^2(n-r)}d(\widehat{P}_1,\widehat{P}_2) ; & \text{if $r = 1, 2, \dots, n_1$},\\\\
		\frac{n_1^2(n-r)}{rn^2}d(\widehat{P}_1,\widehat{P}_2) ; & \text{if $r = n_1 + 1, n_1 + 2, \dots, n$}.
	\end{array}
	\right . \label{rho curve in oracle-single}
\end{eqnarray}
The output of the algorithm will be the maximizer of $\rho^{*}(S_1,S_2)$ with respect to $r$. It follows from \eqref{rho curve in oracle-single} that the $\rho^*$ curve increases till $r = n_1$ and decreases thereafter, and thus the maxima is attained at $r = n_1$.

Next, consider the situation with two changepoints at the locations $n_1$ and $n_1 + n_2 (< n)$ in the data, where we denote the three populations by $P_1$, $P_2$ and $P_3$. Let us also denote the corresponding empirical measures by $\widehat{P}_j$ for $j=1,2,3$. A simple algebra shows that $\rho^*(S_1,S_2)$ takes the following functional form
\begin{align}
\label{rho_curve_in_oracle-two}
&\rho^*(S_{1},S_{2}) \nonumber\\&= 
\left\{
\begin{aligned}
&\frac{r}{n^2(n-r)}\big\{n_2(n_2+n_3)d(\widehat{P}_1,\widehat{P}_2) + n_3(n_2+n_3)d(\widehat{P}_1,\widehat{P}_3) - n_2n_3d(\widehat{P}_2,\widehat{P}_3) \big\},\\
&\hspace{8.5cm}  \text{if } r = 1,2,\dots,n_1, \\
\\
&\frac{nn_1 - r(n_1 + n_3)}{n^2} \left\{ \frac{n_1}{r}d(\widehat{P}_1,\widehat{P}_2) - \frac{n_3}{n - r}d(\widehat{P}_2,\widehat{P}_3) \right\} + \frac{n_1n_3}{n^2}d(\widehat{P}_1,\widehat{P}_3),\\
&\hspace{7.5cm} \text{if } r = n_1+1, \dots, n_1+n_2, \\
\\
&\frac{n - r}{rn^2} \big\{ n_2(n_1+n_2)d(\widehat{P}_2,\widehat{P}_3) + n_1(n_1+n_2)d(\widehat{P}_1,\widehat{P}_3) - n_1n_2d(\widehat{P}_1,\widehat{P}_2) \big\},\\
&\hspace{7.5cm}   \text{if } r = n_1 + n_2 + 1, \dots, n.
\end{aligned}
\right.
\end{align}
Equation \eqref{rho_curve_in_oracle-two} reveals that the $\rho^*$ curve increases till the $n_1$-th iteration and decreases after the $(n_1+n_2)$-th iteration. Between the $n_1$-th and the $(n_1+n_2)$-th iteration, the $\rho^*$ curve is convex. Hence, the $\rho^*$ curve attains its maximum value either at $r=n_1$ or $r=n_1+n_2$. Due to the convexity of the $\rho^*$ curve  between the $n_1$-th and the $(n_1+n_2)$-th iteration, it may happen that the $\rho^*$ curve has only one local maxima (which is also the global maxima) in this scenario. Thus, to detect both of the changepoints, we adopt a recursive binary splitting strategy (to be discussed later) and subsequent local maxima hunting instead of finding the local maxima of the entire $\rho^*$ curve.

Finally, consider the $K \geq 3$ changepoint setting with the changepoint locations at $n_1, n_1+n_2, \ldots, n_1+n_2+\cdots+n_K$. The behavior of the $\rho^*$ curve in this setup is similar to its behavior when $K = 2$ in the sense that it increases till the first changepoint, decreases after the last changepoint, and is convex between any two consecutive changepoints. Thus, the $\rho^*$ curve attains its maximum value at one of the true changepoint locations which motivates us to use a similar recursive binary splitting strategy to obtain all the local maxima. 

\begin{figure}
	\centering	%
\scalebox{0.5}{	\includegraphics[]{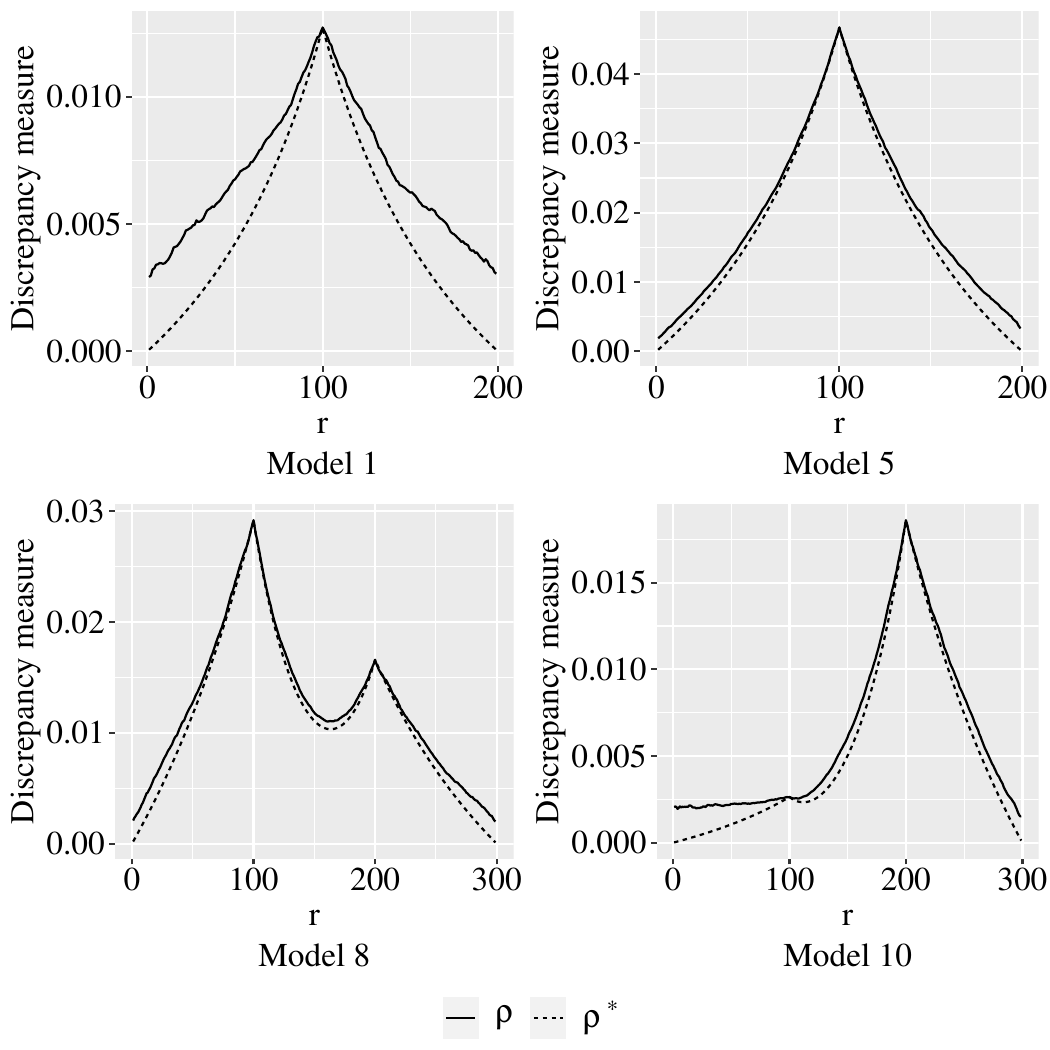}}
	\caption{Curves of $\rho$ and $\rho^*$ distance in two changepoints setup}
	\label{fig:rho_curves_single_cp}
\end{figure}

In practice, we do not have knowledge of the true changepoints and hence the representative distributions. So, the data-driven changepoint algorithm will replace the $\rho^*$ distance with the $\rho$ distance defined by 
\begin{equation} \label{rho_def}
	\rho(S_1,S_2) = \frac{r(n-r)}{n^2} d(\widehat{P}_{S_1},\widehat{P}_{S_2}),
\end{equation}
where $\widehat{P}_{S_1}$ and $\widehat{P}_{S_2}$ denote the empirical distributions corresponding to the sets $S_1$ and $S_2$. We have plotted the $\rho^*$ and the $\rho$ curves for some of the models described in Section \ref{simulation} in Figure \ref{fig:rho_curves_single_cp}. It is observed that for both Models 1 and 5 ($K = 1$), the two curves are similar in shape and both of them attain their maxima at the true changepoint $r = n_1 = 100$. For $K = 2$ (Models 8 and 10), the two curves have similar shape. Although for Model 8, both the curves have local maxima at the true changepoints $n_1 = 100$ and $n_1+n_2 = 200$, but for Model 10, the only prominent local maxima is at $n_1+n_2 = 200$. The latter observation demonstrates the necessity of a recursive binary splitting procedure to detect the other true changepoint at $n_1 = 100$.  

\section{Changepoint Detection: AMOC Setup}
\label{methodology}
In this section, we will propose a method for changepoint detection in the At Most One Changepoint (AMOC) setup. Let ${\cal D}_n = \{Z_{1},Z_{2},\dots,Z_{n}\}$ be a sequence of independent random variables defined on a probability space $(\Omega,\mathcal{F},\mathbb{P})$ and taking values on a separable Hilbert space ${\cal H}$. The problem of changepoint detection in this setup is to first test the presence of a changepoint and to estimate it consequently. To this end, we consider the following test of hypothesis
\begin{align}\label{amoc-hypothesis}
&H_0: Z_1,Z_2,\ldots,Z_n \stackrel{i.i.d.}{\sim} F_1 \  \text{versus} \ H_1 := \bigcup_{\gamma \in (0,1)} H_{1\gamma}, \ \text{where}\\
&H_{1\gamma} : Z_{1},\dots,Z_{\lfloor n\gamma \rfloor} \stackrel{i.i.d.}{\sim} F_{1} \  \text{and} \ Z_{\lfloor n\gamma \rfloor+1},\dots,Z_{n}  \stackrel{i.i.d.}{\sim} F_2 \  \text{with} \ F_1 \neq F_2
\end{align}
To carry out the test, we divide the set of observations ${\cal D}_n$ into two sets 
\begin{equation*}
G_{n}(\widetilde{\gamma})
  = \left\{Z_1, Z_2, \dots, Z_{\lfloor n\widetilde{\gamma} \rfloor}\right\}
  \;\text{and}\;
H_{n}(\widetilde{\gamma})
  = \left\{Z_{\lfloor n\widetilde{\gamma}\rfloor+1},
            Z_{\lfloor n\widetilde{\gamma}\rfloor+2},
            \dots, Z_n\right\}
\end{equation*}
for some $\widetilde{\gamma} \in (0,1)$. We should reject the null hypothesis if the maximum value of the $\rho$ distance between $G_n(t/n)$ and $H_n(t/n)$ over $t  \in \{1,2,\ldots,n\}$ exceeds a threshold. However, detection of a changepoint which is located near the two endpoints of the time-ordered data is difficult due to the lack of observations in one of $G_n(t/n)$ and $H_n(t/n)$. It has thus become standard practice in the literature of changepoint detection to assume that the true changepoint is located in $\{\lceil n\delta_{n}  \rceil,\ldots,\lfloor n(1-\delta_{n}  )\rfloor\}$ for some appropriately chosen $\delta_n \in (0,1)$. For asymptotic analysis, we make the following assumption: 

\begin{assumption}\label{Assumption-delta}
Let $\delta_n \in (0,1)$ for all $n \in \mathbb{N}$ with $\delta_{n}   \to 0$ and $n  \delta_{n}   \to \infty$ as $n \to \infty$.
\end{assumption}

Motivated by the oracle analysis in Section \ref{motivation}, we consider the following test statistic
\begin{align}\label{Tn-def}
    T_{n}   = \underset{\lceil n\delta_{n}  \rceil \leq t \leq \lfloor n(1-\delta_{n}  )\rfloor}{\max} \rho_{n}  (t/n),
\end{align}
where $\rho_n(t/n) = \rho(G_n(t/n),H_n(t/n))$. We reject $H_0$ in favor of $H_1$ if $T_n$ is large enough. The next two theorems describe the behavior of $T_n$ under $H_0$ and $H_1$.
\begin{theorem} \label{Tn under null}
If Assumption \ref{Assumption-delta} holds, then $T_{n} \xrightarrow[]{} 0$ $[\mathbb{P}]$ a.s. under $H_0$.
\end{theorem}

\noindent
Theorem \ref{Tn under null} states that in the absence of a true changepoint within the provided dataset, $T_n$ will converge to $0$ $[\mathbb{P}]$ almost surely.  To investigate the behavior of $T_n$ under $H_1$, we need to study the behavior of the function $\rho_n(\widetilde{\gamma})$ defined by
\begin{align}\label{rho_n_def}
	\rho_n(\widetilde{\gamma}) = \rho(G_n(\widetilde{\gamma}),H_n(\widetilde{\gamma})), 
\end{align}
where $\widetilde{\gamma} \in (0,1)$. The variable $\widetilde{\gamma}$ may be viewed as an arbitrary splitting fraction, which partitions the time-ordered data into two segments $\{Z_{1},\dots,Z_{\lfloor n\widetilde{\gamma}\rfloor}\}$ and $\{Z_{\lfloor n\widetilde{\gamma}\rfloor+1},\dots,Z_{n}\}$. One may now consider $T_n$ as a discrete maximization of the function $\rho_n(\cdot)$. The next result describes the almost sure uniform convergence of the empirical distance $d_n(\cdot)$ to its population counterpart and plays a pivotal role in the understanding of the behavior of $T_n$ where the function $d_{n}(\widetilde{\gamma})$ is defined by
\begin{align} \label{d_n_def}
    d_{n}(\widetilde{\gamma})=d \left(\widehat{P}_{G_{n}(\widetilde{\gamma})}, \widehat{P}_{H_{n}(\widetilde{\gamma})} \right),
\end{align}
where $\widehat{P}_{S}$ denotes the empirical distribution corresponding to the set $S$. 
\begin{lemma}\label{lma1}
If Assumption \ref{Assumption-delta} holds, then
\begin{align} \label{eq2}
	     \underset{\widetilde{\gamma} \in [\delta_n, 1-\delta_n]}{\sup} \left| d_{n}  (\widetilde{\gamma}) - d_*(\widetilde{\gamma})\right| \xrightarrow[]{}  0 \: \: \text{  $[\mathbb{P}]$ a.s.},
\end{align}
under $H_{1\gamma}$ for any prefixed $\gamma \in (0,1)$, where
\begin{align*}
	d_*(\widetilde{\gamma}) = d(\alpha F_{1} + (1-\alpha)F_2,\beta F_1 + (1-\beta)F_{2})
\end{align*}
with $\alpha = (\gamma/\widetilde{\gamma})\mathbb{I}(\widetilde{\gamma} > \gamma) + \mathbb{I}(\widetilde{\gamma} \leq \gamma)$ and $\beta = (\gamma - \widetilde{\gamma})/(1-\widetilde{\gamma})\mathbb{I}(\widetilde{\gamma} \leq \gamma)$. \\
Moreover,  
\begin{align} \label{eqq}
	d_*(\widetilde{\gamma}) = \left\{\frac{\gamma}{\widetilde{\gamma}}\mathbb{I}(\widetilde{\gamma} > \gamma) + \frac{1-\gamma}{1-\widetilde{\gamma}}\mathbb{I}(\widetilde{\gamma} \leq \gamma)\right\}^2 d(F_{1},F_{2}),
\end{align}
and the right hand side of \eqref{eqq} is uniquely maximized at $\widetilde{\gamma} = \gamma$. 
\end{lemma}

Note that the mixture distributions $\alpha F_{1} + (1-\alpha)F_2$ and $\beta F_1 + (1-\beta)F_{2}$ appearing in the above lemma are indeed the $\mathbb{P}$ almost sure weak limits of the representative distributions of the sets $\{Z_{1},\dots,Z_{\lfloor n\widetilde{\gamma}\rfloor}\}$ and $\{Z_{\lfloor n\widetilde{\gamma}\rfloor+1},\dots,Z_{n}\}$, respectively (see \citet{varadarajan1958convergence}). The above lemma shows that the asymptotic behavior of the empirical distance $d_n(\cdot)$ is governed by the representative distributions associated with the partition sets. Furthermore, equation \eqref{eqq} is obtained directly from \eqref{MMDproperty}.  The above result is now used to show that the behavior of the test statistic in the presence of a changepoint in the data is completely different from its behavior under the null hypothesis.
\begin{theorem} \label{Tn under alternative}
If Assumption \ref{Assumption-delta} holds, then $T_{n} \xrightarrow[]{} \gamma(1-\gamma)d(F_1,F_2)$ $[\mathbb{P}]$ a.s.  under $H_{1\gamma}$ for any prefixed $\gamma \in (0,1)$.
\end{theorem}
\noindent
Note that the almost sure limit obtained in the above theorem is positive since $F_1 \neq F_2$ implies  $d(F_1,F_2) > 0$  as the kernel is chosen to be a characteristic kernel. 

We will implement the test using the permutation distribution of $T_n$. This procedure will be valid due to the exchangeability of the data under $H_0$. More, specifically, we will generate $R$ independent permutations of the data. For the $r$th permutation, let us denote the permuted observations by $\{Z_1^{(r)},\ldots,Z_n^{(r)}\}$. Define 
\begin{align}\label{Tn-perm}
	T_n^{(r)} = \underset{\lceil n\delta_{n}  \rceil \leq t \leq \lfloor n(1-\delta_{n}  )\rfloor}{\max} \rho_n^{(r)}(t/n) \ 
	\text{where} \  \rho_n^{(r)}(\widetilde{\gamma}) = \rho( G_{n}^{(r)}(\widetilde{\gamma}), H_{n}^{(r)}(\widetilde{\gamma})) \\
    \text{with} \ G_{n}^{(r)}(\widetilde{\gamma}) = \left\{Z_1^{(r)},\ldots,Z_{\lfloor n\widetilde{\gamma}\rfloor}^{(r)}\right\},\ H_{n}^{(r)}(\widetilde{\gamma}) =\left\{Z_{\lfloor n\widetilde{\gamma}\rfloor+1}^{(r)},\ldots,Z_n^{(r)}\right\}. \nonumber
\end{align}
If the permutation $p$-value $p_n := \sum\limits_{r=1}^{R}\mathbb{I}(T_{n}^{(r)} > T_{n})/R$ is smaller than a specified significance level $\alpha$, we reject the null hypothesis $H_0$ implying the existence of a changepoint in the data.

We next investigate the consistency of the permutation test. Note that Theorem \ref{Tn under null} indicates that the permuted test statistic should also have a similar asymptotic degeneracy since the permutation mechanism mimics the null hypothesis. This fact along with Theorem \ref{Tn under alternative} paves a way of proving the consistency of the permutation test by proving that the permutation p-value converges to zero. To deal with the randomness of the data and that of the permutation procedure, we introduce the following technical framework. Let $(\Omega^{*},\mathcal{F}^{*},\mathbb{P}^*)$ be a measure space quantifying the distribution associated with permutation procedure (which is independent of data generating mechanism). For fixed $n \geq 1$, if $\Pi_n$ be a random variable denoting the permutation of the set $\{1,2,\dots,n\}$ then for any fixed permutation $\pi_0$ of the set $\{1,2,\dots,n\}$, $\mathbb{P}^*(\Pi_n = \pi_0) = \frac{1}{n!}$.  We will work on the product probability space $(\widetilde{\Omega},\widetilde{\mathcal{F}},\widetilde{\mathbb{P}})$, where $\widetilde{\Omega}=\Omega \times \Omega^*$,  $\widetilde{\mathcal{F}}=\mathcal{F} \otimes \mathcal{F}^*$ and $\widetilde{\mathbb{P}}$ is the probability measure on $(\widetilde{\Omega},\widetilde{\mathcal{F}})$ which is the unique extension of the probability measure $(\mathbb{P} \times \mathbb{P}^*)$.  The following theorem establishes the asymptotic degeneracy of the permuted test statistic and proves the consistency of the permutation test.
\begin{theorem} \label{consistency}
Suppose that Assumption \ref{Assumption-delta} is true. Then, the following hold:\\
(a) $T_{n}^{(r)} \xrightarrow[]{} 0\: \: \text{ $[\widetilde{\mathbb{P}}]$ a.s.}$ for all $r=1,2,\dots,R$ under $H_{1\gamma}$ for any prefixed $\gamma \in (0,1)$, \\
(b) $p_n \xrightarrow[]{} 0\: \: \text{ $[\widetilde{\mathbb{P}}]$ a.s.}$ under $H_1$, and the permutation test is consistent.        
\end{theorem}

When the permutation test if rejected, an estimate of the changepoint is given by
\begin{align} \label{eq1}
    \widehat{\tau}_{n}   = \underset{\lceil n\delta_{n}  \rceil \leq t \leq \lfloor n(1-\delta_{n}  )\rfloor}{\arg\max} \rho_{n}  (t/n).
\end{align}
Recall that under $H_{1\gamma}$, where $Z_{1},\dots,Z_{\lfloor n\gamma \rfloor} \stackrel{i.i.d.}{\sim} F_{1} \  \text{and} \ Z_{\lfloor n\gamma \rfloor+1},\dots,Z_{n}  \stackrel{i.i.d.}{\sim} F_2 \  \text{with} \ F_1 \neq F_2$, the parameter $\gamma$ corresponds to a break/change in the distribution of the time-ordered data. We denote $\gamma$ to be the true breakfraction in the data. The quantity $\widehat{\tau}_{n}/n$ is termed as the estimated breakfraction. The next result establishes its consistency.

\begin{theorem} \label{convergence of breakfraction in single changepoint setup}
Suppose that Assumption \ref{Assumption-delta} holds. Then, $\widehat{\tau}_{n}/n \xrightarrow[]{} \gamma \quad \text{$[\mathbb{P}]$ a.s.}$  under $H_{1\gamma}$ for any prefixed $\gamma \in (0,1)$.
\end{theorem}
Recall that the unique maximizer of the $d_*(\widetilde{\gamma})$ is $\widetilde{\gamma} = \gamma$ as obtained from Lemma \ref{lma1}. It can be shown that this is also the unique maximizer of the function $\rho_*(\widetilde{\gamma}) := \widetilde{\gamma}(1-\widetilde{\gamma})d_*(\widetilde{\gamma})$. Since $T_n$ may be viewed as a discrete maximizer of the function $\rho_n(\cdot)$, and $\rho_n(\cdot)$ converges uniformly to $\rho_*(\cdot)$ as a consequence of Lemma \ref{lma1}, the above theorem essentially shows that the maximizer $T_n$ (of  $\rho_n$) converges to the unique maximizer of its limit function $\rho_*(\cdot)$. Recall that our analysis of the oracle setting in Section \ref{motivation} revealed that the maximizer of the oracle discrepancy measure $\rho^*(\cdot,\cdot)$ is obtained at the true changepoint location. The above theorem states that the empirical measure $\rho_n(\cdot)$ mimics the oracle property.

 \section{Multiple Changepoint Detection: Unsupervised Scenario} \label{Method for unknown number of changepoints}
In this section , we consider the multiple changepoint setting. The dataset ${\cal D}_n = \{Z_1,Z_2,\ldots,Z_n\}$ consists of independent observations from $K_0+1$ distributions, denoted by $F_{1},F_{2},\dots,F_{K_0+1}$ respectively. Let $\gamma^{(1)},\gamma^{(2)},\dots,\gamma^{(K_0)}$ be the true breakfractions and $0 < \gamma^{(1)} < \gamma^{(2)} < \dots < \gamma^{(K_0)} < 1$. Also, define $\gamma^{(0)} = 0$ and $\gamma^{(K_0+1)} = 1$. Then for $i = 1,2,\dots,K_0+1$, we have $Z_{\lfloor n\gamma^{(i-1)} \rfloor+1},\dots,Z_{\lfloor n\gamma^{(i)} \rfloor} \mathop{\sim}\limits^{iid} F_{i}$, such that $F_{i-1} \neq F_{i}$. 

The oracle analysis in Section \ref{motivation} shows that the changepoint locations correspond to the local maxima of the $\rho^*$ curve. However, as demonstrated in that section, finding the local maxima of the $\rho^*$ curve may not yield all of the changepoints and as such an iterative procedure is needed. One approach would be to carry out a recursive binary segmentation approach where we detect the global maxima at each step. In the empirical setting, we will first assess the presence of any changepoint in the data using the permutation approach presented in the previous section. If the test is rejected, we split the data at the estimated changepoint location $\widehat{\tau}_n$ obtained in \eqref{eq1}. Next, the two subsets are tested for the presence of any more changepoints using the same permutation test now applied to the subsets. If neither of the two tests are rejected, then the procedure is stopped yielding a single estimated changepoint. Otherwise, we repeat the process on those segment(s) where the test has been rejected. This process is continued until the test is accepted in each segment. The following algorithm referred to as Algorithm DESC-U details the above procedure. 

\small{
\textbf{Algorithm DESC-U: Detection and Estimation of Structural Changes - Unsupervised}
\smallskip
\hrule
\smallskip
\noindent \textbf{Input}: Dataset ${\cal D}_n$ and the boundary parameter $\delta_n \in (0,1/2)$.
	\begin{enumerate}
    \item Apply the permutation test to ${\cal D}_n$.
		\item If the test is accepted, END, else, find a potential changepoint candidate $\widehat{\tau}_n$ using \eqref{eq1} and split ${\cal D}_n$ around $\widehat{\tau}_n$.
  \item Apply the permutation test to each of the output groups. If the test is accepted for a group, keep that group unchanged. Otherwise, find the estimated changepoint in that group as in \eqref{eq1} and split that group around the estimated changepoint. 
		\item Repeat Step 3 until the permutation test is accepted for each of the groups obtained.  
	\end{enumerate}
    \textbf{Output}: A partition $\mathcal{D}_n = \bigcup\limits_{i=1}^{\widehat{K}+1}C_{i}$ and the estimated breakfractions $\widehat{\gamma}_{n}^{(i)}=\sum\limits_{j=1}^{i}|C_{j}|/n$ for $i=1,2,\dots,\widehat{K}$, where $\widehat{K}$ is the estimated number of changepoints.}
    \smallskip
\hrule
\par Let us consider the setup of two changepoints in the data with the corresponding breakfractions denoted by $\gamma^{(1)}$ and $\gamma^{(2)}$. Consider an arbitrary splitting fraction $\widetilde{\gamma}$ and the associated empirical squared MMD distance $d_n(\widetilde{\gamma})$ as described in the previous section. If $\widetilde{\gamma} \in [\gamma^{(1)},\gamma^{(2)}]$, the $\mathbb{P}$ almost sure weak limits of the representative distributions associated with the partition induced by $\widetilde{\gamma}$ are given by 
$(\gamma^{(1)}/\widetilde{\gamma})F_1 + (1 - \gamma^{(1)}/\widetilde{\gamma})F_2$ and $(\gamma^{(2)} - \widetilde{\gamma})/(1-\widetilde{\gamma})F_2 + (1-\gamma^{(2)})/(1 - \widetilde{\gamma})F_3$. Consequently, analogous to Lemma \ref{lma1}, the asymptotic behavior of $d_n(\widetilde{\gamma})$ for $\widetilde{\gamma} \in [\gamma^{(1)},\gamma^{(2)}]$ will be determined by the MMD distance between the above mixture distributions. A similar analysis is also true when $\widetilde{\gamma} \in [\delta_n,\gamma^{(1)}]$ and $\widetilde{\gamma} \in [\gamma^{(2)},1-\delta_n]$.

The crucial observation to make is that the asymptotic analysis of $d_n(\widetilde{\gamma})$ in the multiple changepoint setup can be reduced to the analysis in the two changepoint setup. To this end, suppose $\widetilde{\gamma} \in [\delta_{n}, 1 - \delta_{n}]$ be a fraction such that $\gamma^{(i-1)} \leq \widetilde{\gamma} \leq \gamma^{(i)}$ for some $i \in \{2,3,\dots,K_0\}$. Then, the limiting representative distributions of the sets $\{Z_1,Z_2,\dots,Z_{\lfloor n\widetilde{\gamma}\rfloor}\}$ and $\{Z_{\lfloor n\widetilde{\gamma}\rfloor+1},\dots,Z_{n}\}$ are given by 
\begin{align*}
	\frac{\gamma^{(i-1)}}{\widetilde{\gamma}}Q_{1}^{i} + \frac{\widetilde{\gamma}-\gamma^{(i-1)}}{\widetilde{\gamma}}F_{i} \ \text{and} \ \frac{\gamma^{(i)}-\widetilde{\gamma}}{1-\widetilde{\gamma}}F_{i} + \frac{1-\gamma^{(i)}}{1-\widetilde{\gamma}}Q_{2}^{i}, \  \text{respectively, where} \\
	    Q_{1}^{i} := \sum\limits_{j=1}^{i-1} \frac{\gamma^{(j)}-\gamma^{(j-1)}}{\gamma^{(i-1)}} F_{j} \ \text{and} \ Q_{2}^{i} := \sum\limits_{j=i+1}^{K_0+1} \frac{\gamma^{(j)}-\gamma^{(j-1)}}{1-\gamma^{(i)}} F_{j}.
\end{align*}
Similar representations also hold for $\widetilde{\gamma} < \gamma^{(1)}$ and $\widetilde{\gamma} > \gamma^{(K_0)}$. The next result describes the asymptotic behavior of $d_n(\cdot)$. 

    \begin{lemma}\label{lma2}
      Under Assumption \ref{Assumption-delta} and the multiple changepoint setup described at the beginning of Section \ref{Method for unknown number of changepoints}, we have
      \begin{align*}
      	& \underset{\widetilde{\gamma} \in [\delta_n, 1-\delta_n]}{\sup} \left| d_{n}(\widetilde{\gamma}) - d_{**}(\widetilde{\gamma})\right|  \xrightarrow[]{} \: 0 \: \: \text{ $[\mathbb{P}]$ a.s.} , \ \text{where} \\
      	&d_{**}(\widetilde{\gamma}) =
      	\begin{cases}
      		d\left(F_{1},\frac{\gamma^{(1)}-\widetilde{\gamma}}{1-\widetilde{\gamma}}F_{1}+\sum\limits_{j=2}^{K_0+1} \frac{\gamma^{(j)}-\gamma^{(j-1)}}{1-\widetilde{\gamma}} F_{j}\right) & \text{if} \ \widetilde{\gamma} \in (0, \gamma^{(1)}], \\
      		d\left(\frac{\gamma^{(i-1)}}{\widetilde{\gamma}}Q_{1}^{i} + \frac{\widetilde{\gamma}-\gamma^{(i-1)}}{\widetilde{\gamma}}F_{i}, \frac{\gamma^{(i)}-\widetilde{\gamma}}{1-\widetilde{\gamma}}F_{i} + \frac{1-\gamma^{(i)}}{1-\widetilde{\gamma}}Q_{2}^{i}\right) & \text{if} \ \widetilde{\gamma} \in [\gamma^{(i-1)}, \gamma^{(i)}] \\  &\text{for} \ 2 \leq i \leq K_0, \\
      		d\left(\sum\limits_{j=1}^{K_0} \frac{\gamma^{(j)}-\gamma^{(j-1)}}{\widetilde{\gamma}}F_{j}+\frac{\widetilde{\gamma}-\gamma^{(K_0)}}{\widetilde{\gamma}}F_{K_0+1},F_{K_0+1}\right) & \text{if} \ \widetilde{\gamma} \in [\gamma^{(K_0)}, 1). 
      	\end{cases}
      \end{align*}
    \end{lemma}

 In the above result, the mixture distributions appearing in the definition of $d_{**}(\widetilde{\gamma})$ are indeed the limiting representative distributions of the partition sets obtained for the corresponding values of the splitting fraction $\widetilde{\gamma}$. Consequently, the asymptotic behavior of the empirical distance $d_n(\widetilde{\gamma})$ is completely determined by the representative distributions even in the multiple changepoint setting. 

\begin{lemma} \label{d_max}
    Under the assumption of Lemma \ref{lma2}, the continuous function $\widetilde{\gamma}(1-\widetilde{\gamma})d_{**}(\widetilde{\gamma})$ attains its unique maximum at $\gamma^{(1)}$ in the range  $\widetilde{\gamma} \in (0,\gamma^{(1)}]$, attains its maximum at either $\gamma^{(i-1)}$ or $\gamma^{(i)}$ in the range $\widetilde{\gamma} \in [\gamma^{(i-1)},\gamma^{(i)}]$, and attains its unique maximum at $\gamma^{(K_0)}$ in the range $\widetilde{\gamma} \in [\gamma^{(K_0)},1)$.
\end{lemma}

The above result shows that the limiting empirical distance $d_{**}(\cdot)$ potentially has multiple local maxima but these form a subset of the set of true breakfractions. In fact, it is part of the proof of Lemma \ref{d_max} that the function $\widetilde{\gamma} \mapsto \widetilde{\gamma}(1-\widetilde{\gamma})d_{**}(\widetilde{\gamma})$ is increasing in $(0,\gamma^{(1)}]$, is convex in $[\gamma^{(i-1)},\gamma^{(i)}]$ for each $i=2,3,\ldots,K_0$ and is decreasing in $[\gamma^{(K_0)},1)$. This behavior was also seen for the oracle MMD distance studied in Section \ref{motivation}. We next study the asymptotic behavior of the estimated breakfractions in the multiple changepoint setting. Define 
\begin{align}\label{estimated-breakfraction}
	\widehat{\gamma}_n = \widehat{\tau}_n/n, \  \text{where} \  \widehat{\tau}_n =  \underset{\lceil n\delta_{n}  \rceil \leq t \leq \lfloor n(1-\delta_{n}  )\rfloor}{\arg\max} \rho_{n}  (t/n) \  \text{and} \ \rho_n(\widetilde{\gamma}) = \frac{\lfloor n\widetilde{\gamma}\rfloor (n - \lfloor n\widetilde{\gamma}\rfloor)}{n^2}d_n(\widetilde{\gamma}).
\end{align}
Observe that $\widehat{\gamma}_n$ is the estimated breakfraction obtained in Step 2 of Algorithm DESC-U using the full dataset.  In terms of the outputs $\widehat{\gamma}_n^{(i)}$ of this algorithm, it is easily seen that $\widehat{\gamma}_n$ is one of these estimated breakfractions. Also, $\rho_n(\widehat{\gamma}_n) \geq \rho_n(\widehat{\gamma}_n^{(i)})$ for all $i=1,2,\ldots,\widehat{K}$ since $\widehat{\gamma}_n$ is the global maximizer obtained in Step 2 of the algorithm.

\begin{theorem} \label{testing and estimation in mcp}
Under the assumptions of Lemma \ref{lma2}, there exists a constant $c \in (0,\infty)$ such that 
\begin{align*}
  T_n \rightarrow c \quad \text{and} \quad \underset{1 \leq i \leq K_0}{\min} |\widehat{\gamma}_{n} - \gamma^{(i)}| \xrightarrow[]{} 0 \quad \text{$[\mathbb{P}]$ a.s.}
\end{align*}
\end{theorem}
\noindent
Theorem \ref{testing and estimation in mcp} shows that in the multiple changepoint setting, the estimated breakfraction obtained from the full dataset converges to one of the true breakfractions almost surely. Furthermore, it also implies that for a subset of the observations with at least one changepoint, the estimated breakfraction computed from that subset of the data will converge to one of the breakfractions present in that subset almost surely. Thus, the above theorem indicates that for all $i$, $\widehat{\gamma}_n^{(i)}$ obtained from a segment of the data should be a consistent estimator of one of the true breakfractions present in that segment. So, it is expected that Algorithm DESC-U will be able to consistently estimate all of the true breakfractions by virtue of its asymptotically correct recursive segmentation. Recall that in the oracle setting in Section \ref{motivation}, we could obtain all of the true changepoints by recursively splitting the data and finding the maxima of the oracle $\rho^*$-curve in those segments. The above discussion implies that the behavior of the proposed algorithm is similar to the oracle.

\section{Multiple Changepoint Detection: Supervised Scenario}
\label{methodology for prefixed number of changepoints}
Algorithm DESC-U, while effective in detecting changepoints, does not inherently accommodate prior information/specification regarding the number of changepoints. The specification of the desired number of changepoints to be obtained from a changepoint detection algorithm can be viewed in the same light as some of the clustering algorithms (e.g., $K$-means) where one has to specify apriori the desired number of clusters. Indeed, changepoint detection can be viewed as clustering of a time-ordered data sequence. The prior specification of the number of changepoints required in the output segmentation (denoted by $K$) may be available from domain knowledge or may be estimated from a previous study. For instance, if we want to detect changepoints in a time series of temperature curves, certain environmental factors and their times of occurrence (e.g., industrial revolution, world wars) that could potentially influence global temperature curves may be known in advance.  
%
%
%
In case the number of changepoints $K$ is specified in advance,  we can view the problem as one of changepoint estimation where we can do away with the assessment of significance of each estimated changepoint. One approach would be to aggressively split the data to obtain multiple changepoints and use a merging procedure to eliminate spurious segments (using MMD distance between adjacent segments) by taking into account the value of $K$. To this end, we introduce Algorithm DESC-S, where instead of evaluating the statistical significance of each estimated changepoint, we adopt a strategy of deliberate group splitting. Specifically, each segment is forcibly divided into two sub-groups. This process may result in the creation of spurious sub-groups within segments that contain no true changepoints. To rectify this, a subsequent merging step is employed. The merging decision is based on the $\rho$-distances between consecutive sub-groups, effectively quantifying the similarity between their underlying distributions. This merging process also serves as a mechanism to control the final number of estimated changepoints, thus ensuring that the algorithm adheres to the pre-specified value of $K$. The details of such an algorithm is given below.
\\

\small{
	\textbf{Algorithm DESC-S: Detection and Estimation of Structural Changes - Supervised}
	\smallskip
	\hrule
	\smallskip
	\noindent \textbf{Input}: Dataset ${\cal D}_n$, number of changepoints $K(\geq 1)$ and boundary parameter $\delta_n$ (a percentage of $n$).
	\begin{enumerate}
		\item Set $i = 0$. The input $C_{0}^{(1)}={\cal D}_n$.
		\item For each $j = 1,2,\dots,i+1$, find a potential changepoint candidate $\widehat{\tau}_n$ in $C_{i}^{(j)}$ as in \eqref{estimated-breakfraction} and split it around $\widehat{\tau}_n$ into the segments $D_{i}^{(2j-1)}$ and $D_{i}^{(2j)}$.
		\item Define $\rho^{i,j} = \rho\left( D_{i}^{(2j-1)}, D_{i}^{(2j)}\right)$ for $j=1,2,\dots,i+1$.
		\item If $i=0$, $C_{1}^{(1)}=D_{0}^{(1)}, C_{1}^{(2)}=D_{0}^{(2)}$. Else, arrange the $\rho^{i,j}$ values in increasing order of magnitude and let the ordered values be $\rho^{i,j_{1}} \leq \rho^{i,j_{2}} \leq \dots \leq \rho^{i,j_{i+1}}$. For $c=1,2,\dots,i$, merge $D_{i}^{(2j_{c}-1)}$ and $D_{i}^{(2j_{c})}$ and label the merged set as $C_{i+1}^{(c)}$. Label $D_{i}^{(2j_{i+1}-1)}$ and $D_{i}^{(2j_{i+1})}$ as $C_{i+1}^{(i+1)}$ and $C_{i+1}^{(i+2)}$ respectively.
		\item Update $i = i+1$. 
		\item Check if $i = K$.\\
		- If yes, stop.\\
		- Otherwise, repeat Step 2 to Step 6.
	\end{enumerate}
	
	\textbf{Output}: The estimated breakfractions $\widehat{\gamma}_{n,K}^{(l)}=\sum_{i=1}^{l}|C_{K}^{(i)}|/n$ for $l=1,2,\dots,K$.
	\smallskip
	\hrule
}
\vspace{0.1in}

\begin{figure}
	\centering	
	\includegraphics[scale=0.7]{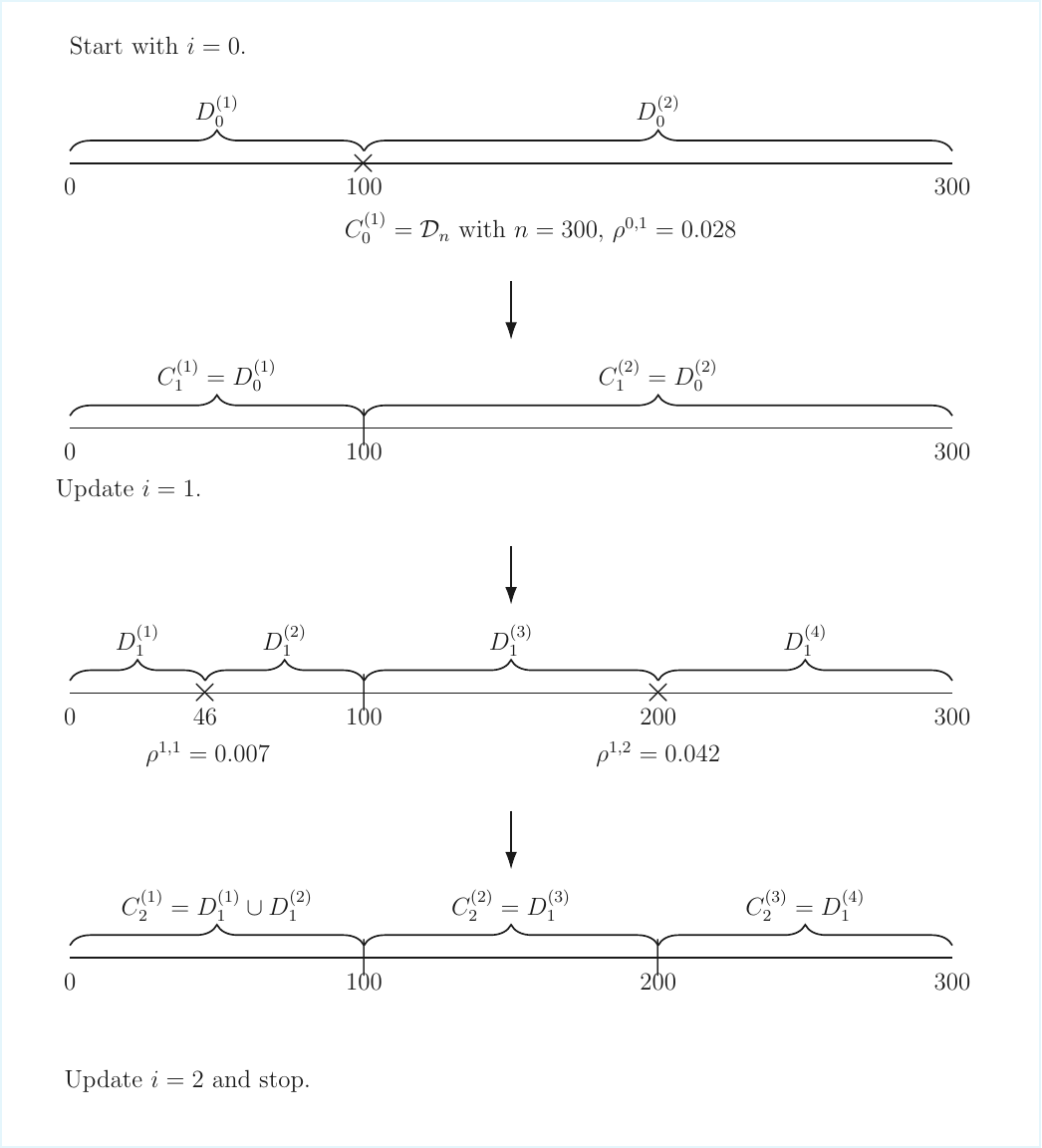}
	\caption{Pictorial representation of Algorithm DESC-S for $K_0 = 2$ and $n_1 = n_2 = n_3 = 100$}
	\label{DESC-S-picture}
\end{figure}

At the beginning of the $i$-th iteration ($i=0,1,\ldots,$) of the algorithm (comprising Steps $2$ through $5$), we have $i+1$ segments of the data. Once these are split into $2(i+1)$ segments using Step 2, the algorithm reduces the number of segments to $i+2$ using the merging step (Step 4) at the end of the $i$-th iteration. This is accomplished by merging $i$ pairs of consecutive segments that exhibit the minimum $\rho$ distance. A crucial aspect is to exclude pairs of segments to the immediate left and right of any existing (at the beginning of the $i$-th iteration) estimated changepoint. This ensures that we do not remove any estimated changepoint that was already existing up until that iteration. The algorithm thus ensures that after every iteration, the number of estimated changepoints increases by one. So, to obtain the specified number $K$ of changepoints, the algorithm will stop after the completion of the $(K-1)$-th iteration yielding $K+1$ segments. The previous algorithm is pictorially demonstrated in Fig \ref{DESC-S-picture}.

Algorithm DESC-S exhibits interesting theoretical properties depending on the relationship between $K$ and the true number of changepoints $K_0$. We start with the following definition that  states a desirable behavior of any changepoint detection algorithm. 
\begin{definition} \label{faithful}
    If $K_0$ and $K$ are the true number of changepoints and the number of changepoints obtained from a changepoint algorithm respectively, we call the algorithm ``Faithful'' if the following conditions hold.\\
    $(i)$ If $K=K_0$, the set of estimated breakfractions converges to the set of true breakfractions almost surely.\\
    $(ii)$ If $K<K_0$, the set of estimated breakfractions converges to a subset of the true breakfractions almost surely.\\
    $(iii)$ If $K>K_0$, the set of estimated breakfractions contains a subset which converges to the set of true breakfractions almost surely.    
\end{definition}
The following theorem provides an asymptotic behavior of Algorithm DESC-S.
\begin{theorem} \label{pop}
Under Assumption \ref{Assumption-delta} and the multiple changepoint setup described at the beginning of Section \ref{Method for unknown number of changepoints}, Algorithm DESC-S is ``Faithful''. More specifically, let $K$ be the specified number of changepoints, $K_0$ be the true number of changepoints, and $\widehat{\boldsymbol{\gamma}}_{[K]} := \{\widehat{\gamma}_{n,K}^{(1)}, \widehat{\gamma}_{n,K}^{(2)}, \dots, \widehat{\gamma}_{n,K}^{(K)}\}$ and $\boldsymbol{\gamma}_{[K_0]} := \{\gamma^{(1)}, \gamma^{(2)}, \dots, \gamma^{(K_0)}\}$ be the set of estimated breakfractions and the set of true breakfractions respectively. Then, the following statements are true. 
\begin{enumerate} [label=(\alph*)]
    \item When $K \leq K_0$,
    \begin{align*}
 \underset{1 \leq i \leq K}{\max} \: \underset{1 \leq i' \leq K_0}{\min} |\widehat{\gamma}_{n,K}^{(i)} - \gamma^{(i')}| \xrightarrow[]{} 0 \quad \text{$[\mathbb{P}]$ a.s.}
    \end{align*}
    \item When $K = K_0$,
    \begin{align*}
    d_H(\widehat{\boldsymbol{\gamma}}_{[K]},\boldsymbol{\gamma}_{[K_0]}) \xrightarrow[]{} 0 \quad \text{$[\mathbb{P}]$ a.s.},
    \end{align*}
    where $d_H(\cdot,\cdot)$ is the Hausdorff distance between subsets of $[0,1]$.
    \item When $K > K_0$,
    \begin{align*}
        \underset{1 \leq i' \leq K_0}{\max} \:\underset{1 \leq i \leq K}{\min} |\widehat{\gamma}_{n,K}^{(i)} - \gamma^{(i')}| \xrightarrow[]{} 0 \quad \text{$[\mathbb{P}]$ a.s.}
        \end{align*}
\end{enumerate}
\end{theorem}
\noindent
    Theorem \ref{pop} highlights the ability of Algorithm DESC-S to accurately estimate breakfractions when the number of changepoints is correctly specified. Moreover, when $K$ is misspecified, then the set of estimated breakfractions is asymptotically either contained in (if $K < K_0$) or contains (if $K > K_0$) the true set of breakfractions.

\section{Multiple Changepoint Detection: Semi-Supervised Scenario} \label{Methodology for specified bounds of number of changepoints}
In many practical applications, one may have prior knowledge regarding the true number of changepoints (denoted by $K_0$) in the form of bounds (upper or lower or both) on it. These bounds may be obtained from some previous studies involving similar data or from scientific information from the field of study. For example, if one is analyzing stock price index data (say, NIFTY 50 or S \& P 500) from the year 2000 onward, it is well-known that certain global socio-economic and demographic factors influenced the data including the 2008 recession and the COVID-19 outbreak. Thus, one can expect at least two structural changes in the data and one can also propose an upper bound on the maximum number of changepoints expected in the data. 

In this section, we will propose an algorithm which is designed to handle the case where we have knowledge of both the upper and the lower bounds on $K_0$, and these will be denoted by $K_u$ and $K_l$, respectively. We will later modify the algorithm to also include the cases where exactly one of these bounds are available.  When both $K_u$ and $K_l$ are given, we begin by employing Algorithm DESC-S with $K = K_u$ to initially obtain $K_u + 1$ segments of the data. Subsequently, we eliminate falsely detected changepoints using a backward elimination strategy. To this end, we compute the $p$-values obtained by using the permutation test as in Algorithm DESC-U to each of the adjacent pairs of segments and merge that pair of segments corresponding to the largest significant $p$-value while controlling a familywise error rate. These steps are repeated up until the number of detected changepoints equals the lower bound $K_l$ or all of the permutation $p$-values obtained are significant.  \citet{chiou2019identifying} also used a backward elimination strategy to discard insignificant changepoints. However, their elimination method is different in that they test the significance of the \textit{most unlikely changepoint}. Furthermore, their algorithm does not incorporate the knowledge of a lower bound on the number of changepoints unlike our method. The details of the algorithm are given below.
\\
\small{
\textbf{Algorithm DESC-SS: Detection and Estimation of Structural Changes -  Semi-Supervised}
\smallskip
\hrule
\smallskip
\noindent \textbf{Input}: Dataset ${\cal D}_n$, bounds $K_u$ and $K_l$, and boundary parameter $\delta_n$ (a percentage of $n$). 
	\begin{enumerate}
    \item Set $m=1$.
    \item Apply Algorithm DESC-S to ${\cal D}_n$ with $K = K_u-m+1$ and obtain the segments $C_1,\dots,C_{K_u-m+2}$.
		\item For each $i=1,2,\dots,K_{u}-m+1$, apply the permutation test as in Algorithm DESC-U to the set $\{C_{i} \cup C_{i+1}\}$ and obtain the p-value $p_i$ .
  \item If $\underset{i}{\max}\: p_{i} < \alpha/(K_{u}-m+1)$ or $m=K_u-K_l+1$, then define $\widehat{K} = K_u - m + 1$ and then END. Else, set $j=\underset{i}{\arg\max}  \: p_{i}$, merge $C_{j}$ and $C_{j+1}$ and define $\widehat{K} = K_u - m$.
  \item Update $m=m+1$ and label the sets obtained in the Step 4 as $C_1,C_2,\dots,C_{K_{u}-m+1}$. Then, go to Step 3.
  
	\end{enumerate}
    \textbf{Output}: The segments $C_1,C_2,\ldots,C_{\widehat{K}+1}$, where $K_{l} \leq \widehat{K} \leq K_{u}$.}
    \smallskip
\hrule
\vspace{0.1in}
\noindent In Step 3 of the above algorithm, the choice of significance level is done to ensure that the familywise error rate is controlled at the level $\alpha$ since there are $K_u-m+1$ many tests to be carried out in the $m$-th stage of the algorithm. 

\begin{remark}\label{remark-semi-supervised}
	Note that in Algorithm DESC-SS, we merge consecutive segments that are obtained from a $K_u + 1$ segmentation of the data using Algorithm DESC-S. If we assume that the true number of changepoints $K_0$ satisfies $K_0 \leq K_u$, then by the asymptotic behavior of Algorithm DESC-S in part (b) of Theorem \ref{pop}, no segment will have any true changepoint that is left undiscovered and there could potentially be some spurious changepoints which have been estimated. Consequently, if we consider the data as the union, say $S$, of two consecutive segments, we will always encounter the AMOC situation since the estimated changepoint in $S$ is either true or spurious. Thus, from part (b) of Theorem \ref{consistency}, it follows that we will (asymptotically) never merge these two consecutive segments if $S$ contains a true changepoint, that is, we will not commit any type II error while testing $H_{0,S}$: no changepoint in $S$ versus $H_{1,S}$: exactly one changepoint in $S$. \\
	\indent An alternate approach (forward selection) to the backward elimination strategy used in Algorithm DESC-SS would be to start with a $K_l + 1$ segmentation of the data using Algorithm DESC-S with $K = K_l$. Subsequently, we split each of the segments using Algorithm DESC-U until we reach the upper bound $K_u$ for the estimated number of changepoints or the permutation test is accepted for each of the segments. However, this approach has the following drawback compared to the backward elimination approach. Note that in this case, each segment can have no changepoints or multiple changepoints. So, while applying the permutation test on a segment $S$ with no changepoints, a type I error can be committed for $H_{0,S}$ versus $H_{1,S}$ (even asymptotically). On the other hand, if a segment $S$ has multiple changepoints, a type II error can be committed for the same hypotheses (even asymptotically). Thus, unlike the backward elimination strategy, here we can have both types of errors which may compromise the performance of this alternate forward selection approach.   
\end{remark}
We can modify Algorithm DESC-SS to handle the situations where exactly one of the bounds are specified. If only $K_u$ is provided, we can apply Algorithm DESC-SS with $K_l = 0$. On the other hand, if only $K_l$ is available, we can only adopt a forward selection approach where we first apply Algorithm DESC-S with $K = K_l$ and subsequently use Algorithm DESC-U with its stopping criterion. 

\section{Numerical Results} \label{simulation}
In this section, we investigate the numerical performance of the Algorithms DESC-U, DESC-S and DESC-SS. Algorithm DESC-U is also compared with some of the state of the art for changepoint detection in functional data, namely, (i) the \textbf{BGHK} method studied in \citet{berkes2009detecting} which is the standard CUSUM procedure on the functional principal component scores, (ii) the \textbf{change\textunderscore FF} method studied in \citet{aue2018detecting} which is the standard CUSUM method on the fully functional data, (iii) the \textbf{fmci} method proposed by \citet{harris2022scalable} where the authors introduced a pair of projections to represent the variability ``between'' and ``within'' the functional observations and applied the  univariate CUSUM statistic to identify the changepoints, (iv) the \textbf{EJS} method studied in \citet{44e75106888d4236b3935cb41db7bc34} which is based on a modified CUSUM procedure using a self-normalization technique, and (v) the \textbf{HK} method developed by \citet{hormann2010weakly} which uses a modified CUSUM approach based on estimation of the long-run variance of the functional time series. We also performed the method in \citet{padilla2022}, who proposed the functional binary seeded segmentation for detecting changes in the mean, using the codes provided by those authors. However, for all the simulated models considered in this paper, the performance of this method is found to be poor and so we did not report these results.

The method change\textunderscore FF is implemented for the multiple changepoint scenario by repeatedly applying the program \texttt{change\_FF} available in the R package \texttt{fChange}  (version 0.2.0 dated 16th February 2024) using binary segmentation since the program only returns a single estimated changepoint. The code for implementing the fmci method is available in \textit{https://github.com/trevor-harris/fmci}. The EJS method is the self-normalized CUSUM test for detecting changes in the mean function. We used the codes provided by \citet{padilla2022} for performing the EJS, HK and BGHK procedures. All of these methods use fully observed functional data and they convert discrete data to functional observations by using B-splines with 20 basis functions.

The algorithms proposed in this paper are implemented using the R software (version 4.5.1) on a standard computer equipped with a 12th Gen Intel(R) Core(TM) i7-12700 processor and 16 GB of RAM. The computation times for the three proposed algorithms are noted for typical simulation configurations. For instance, for datasets of size $n=600$ with a single changepoint, Algorithm DESC-U and Algorithm DESC-SS required a maximum of three minutes for each replication, whereas  Algorithm DESC-S required only a few seconds.

\subsection{Simulated Data}
For the simulation study, we generated functional data on a discrete grid of 128 equi-spaced points in the interval $[0,1]$. For implementation of the Algorithms DESC-U, DESC-S and DESC-SS, we use the Gaussian kernel $k(x,y) = \exp\{-\lVert x-y \rVert^2/2h^2\}$ in the definition of MMD, where the tuning parameter $h$ is chosen heuristically as $h = \mbox{Median}\{\lVert a-b \rVert : a,b \in \mathcal{D}, a \neq b\}$ (see Section 5 of \citet{NIPS2009_685ac8ca}). Here, $\mathcal{D}$ denotes the set of all observations and $\lVert \cdot \rVert$ denotes the usual norm on the space $L^2[0,1]$. We have also experimented with $h = 0.1,0.2,1,5,10$. However, these choices did not change the results significantly and thus we do not report them. Throughout this section, all performance metrics are reported based on $100$ Monte Carlo iterations. First, we consider the setup when all the observations are generated from the same distribution to investigate the performance of the procedures under the null hypothesis. Samples of size $n$ are drawn from the following models.
\vspace{0.025in}\\
\noindent\textbf{Model N1}:   $X(t) = \sum_{j=0}^{150}\sqrt{\theta_{j}}W_{j}\phi_{j}(t) + 0.5-100(t-0.1)(t-0.3)(t-0.5)(t-0.9) +0.8\sin{(1+10\pi t)}$, where $\theta_{j} = 0.7 \times 2^{-j}$, $W_{j} \overset{\mathrm{i.i.d}}{\sim} N(0,1)$, for $j = 0,1, 2, \dots, 150$, $\phi_{0}(t)=1$, $\phi_{j}(t) = \sqrt{2}\sin{(2\pi lt - \pi)}$ for $j=2l-1$ and $\phi_{j}(t) = \sqrt{2}\cos{(2\pi lt - \pi)}$ for $j=2l$ for $l=1,2,\dots,75$.
\vspace{0.075in}\\
\noindent\textbf{Model N2}: $X(t) = BB(t)$ where $BB(.)$ denotes the standard Brownian Bridge.
\vspace{0.075in}\\
\noindent \textbf{Model N3}:   $X(t) = \sum_{j=1}^{50}\sqrt{\theta_{j}}W_{j}\phi_{j}(t) + 2t$, where $\phi_{j}(t) = \sqrt{2}\sin{(jt\pi)}$, $\theta_{j} = exp(-j/3)$, $W_{j} \overset{\mathrm{i.i.d}}{\sim} N(0,1)$ for $j = 1, 2, \dots, 50$. 
\vspace{0.075in}\\
\noindent \textbf{Model N4}: $X(t) = \sum_{j=1}^{40}\sqrt{\theta_{j}}W_{j}\phi_{j}(t)$, where $\phi_{j}(t) = \sqrt{2}\sin{(jt\pi)}$, $\theta_{j} = j^{-2}$, $W_{j} \overset{\mathrm{i.i.d}}{\sim} N(0,1)$ for $j=1,2,\dots,40$.
\vspace{0.075in}\\
\noindent Table \ref{size} provides a summary of type-I errors for the test procedures in each of the competing methods for testing \eqref{amoc-hypothesis} across four simulation scenarios with different sample sizes. The permutation test in our method controls the type-I error in all scenarios and the same is also true for all the other competitors, with the exception of fmci. The test procedure in the fmci method maintains the nominal level only under Model N2 whereas the estimated size is significantly higher than the desired level for the other models. This discrepancy remains even when the sample size is increased from $100$ to $500$. 

 \begin{table}
 \centering
	\caption{Empirical size of the detection procedures at level $\alpha=0.05$}
        	\label{size}
	\begin{tabular} {@{}lcrcrrr@{}}
		\hline
		\multirow{2}{*}{$n$} &\multirow{2}{*}{\textbf{Methods}} &\multicolumn{4}{c}{\textbf{Estimated probability of Type-I error}} \\
		\cline{3-6}
		& &\textbf{Model N1}  &\textbf{Model N2}  &\textbf{Model N3}  &\textbf{Model N4}
		\\
		\hline
		\multirow{6}{*}
		{100} & DESC-U& 0.06& 0.06& 0.08& 0.02\\
		
		& BGHK& 0.04& 0.08& 0.06& 0.03\\
		
		& change\textunderscore FF& 0.03& 0.07& 0.05& 0.06\\
		
		& fmci& 0.21& 0.04& 0.14& 0.21\\
		
		& EJS& 0.05& 0.06& 0.02& 0.06\\
		
		& HK& 0.06& 0.07& 0.06& 0.08\\
		[10pt]
		\multirow{6}{*}
		{500} & DESC-U& 0.03& 0.03& 0.02& 0.04\\
		
		& BGHK& 0.07& 0.02& 0.05& 0.04\\
		
		& change\textunderscore FF& 0.05& 0.06& 0.07& 0.04\\
		
		& fmci& 0.16& 0.03& 0.19& 0.16\\
		
		& EJS& 0.05& 0.05& 0.03& 0.02\\
		
		& HK& 0.06& 0.04& 0.02& 0.05\\
		\hline
	\end{tabular}

\end{table}

We will now explore the performance of the procedures in the following models that employ a single changepoint configuration. The pre-change process is denoted by $X_1$ and the post-change process is denoted by $X_2$. The data $Z_1,\ldots,Z_{n_1}$ are i.i.d. copies of $X_1$ while $Z_{n_1+1},\ldots,Z_n$ are i.i.d. copies of $X_2$.
\vspace{0.075in}\\
\noindent \textbf{Model 1} [Location change]:  $X_{l}(t) = \sum_{j=1}^{50}\sqrt{\theta_{jl}}W_{j}\phi_{j}(t) + \mu_{l}(t)$, $l = 1, 2$,
where for each $j$, $\phi_{j}(t) = \sqrt{2}\sin{(jt\pi)}$. $\theta_{j1} = \theta_{j2} = exp(-j/3)$, $W_{j} \overset{\mathrm{i.i.d}}{\sim} N(0,1)$ for $j = 1, 2, \dots, 50$, $\mu_1(t) = 2t$ and $\mu_2(t) = 6t(1-t)$.
\vspace{0.075in}\\
\noindent \textbf{Model 2} [Location change]:    $X_{l}(t) = \sum_{j=1}^{40}(\sqrt{\theta_{jl}}W_{j} + \mu_{jl})\phi_{j}(t)$, $l = 1, 2$,
where for $j=1,2,\dots,40$, $\phi_{j}(t) = \sqrt{2}\sin{(jt\pi)}$, $\theta_{j1} = \theta_{j2} = j^{-2}$, $W_{j} \overset{\mathrm{i.i.d}}{\sim} t_{3}/\sqrt{3}$, $\mu_{j1} = 0$ and $\mu_{j2} = 0.75(-1)^{j+1} I\{1 \leq j \leq 3\}$. 
\vspace{0.075in}\\
\noindent \textbf{Model 3} [Location change]:   $X_{l}(t) = \sum_{j=0}^{150}\sqrt{\theta_{j}}W_{j}\phi_{j}(t) + \mu_{l}(t)$,
where $\theta_{j} = 0.7 \times 2^{-j}$, $W_{j} \overset{\mathrm{i.i.d}}{\sim} N(0,1)$, for $j = 0,1, 2, \dots, 150$, $\phi_{0}(t)=1$, $\phi_{j}(t) = \sqrt{2}\sin{(2\pi lt - \pi)}$ for $j=2l-1$ and $\phi_{j}(t) = \sqrt{2}\cos{(2\pi lt - \pi)}$ for $j=2l$ for $l=1,2,\dots,75$, $\mu_{1}(t)=0.5-100(t-0.1)(t-0.3)(t-0.5)(t-0.9)+0.8\sin{(1+10\pi t)}$ and $\mu_{2}(t)=1+3t^2-5t^3+0.6\sin{(1+10\pi t)}$.
\vspace{0.075in}\\
\noindent \textbf{Model 4} [Location change]: $X_{1}(t)=BB(t)$ and $X_{2}(t)=BB(t)+\sin{(t)}$.
\vspace{0.075in}\\
\noindent \textbf{Model 5} [Scale change]: $X_{l}(t) = \sum_{j=1}^{40}\sqrt{\theta_{jl}}W_{j}\phi_{j}(t)$, $l = 1, 2$,
where for $j=1,2,\dots,40$, $\phi_{j}(t) = \sqrt{2}\sin{(jt\pi)}$, $W_{j} \overset{\mathrm{i.i.d}}{\sim} N(0,1)$, $\theta_{j1} = j^{-2}$, $\theta_{j2} = 3j^{-2}$.
\vspace{0.075in}\\
\noindent \textbf{Model 6} [Change in eigenvalues]:   $X_{l}(t) = \sum_{j=1}^{50}\sqrt{\theta_{jl}}W_{j}\phi_{j}(t)$, $l = 1, 2$,
where for each $j$, $\phi_{j}(t) = \sqrt{2}\sin{(jt\pi)}$. $\theta_{j1} = j^{-2}$, $\theta_{j2} = exp(-j)$, $W_{j} \overset{\mathrm{i.i.d}}{\sim} N(0,1)$ for $j = 1, 2, \dots, 50$.
\vspace{0.075in}\\
\noindent \textbf{Model 7} [Change in eigenfunctions]: $X_{l}(t) = \sum_{j=1}^{40}\sqrt{\theta_{jl}}W_{j}\phi_{jl}(t)$, $l = 1, 2$,
where for $j=1,2,\dots,40$, $\phi_{j1}(t) = \sqrt{2}\sin{(jt\pi)}$, $\phi_{j2}$ is the $j$-th term of Fourier Basis, $\theta_{j1} = \theta_{j2} = j^{-2}$, $W_{j} \overset{\mathrm{i.i.d}}{\sim} N(0,1)$.
\vspace{0.075in}\\
We denote the set of true changepoints by $L = \{\tau_1,\ldots,\tau_{K_0}\}$ (ordered) and the set of estimated changepoints obtained by using any changepoint detection algorithm by $\widehat{L} = \{\widehat{\tau}_1,\ldots,\widehat{\tau}_{\widehat{K}}\}$ (ordered). We will define $\widehat{L} \simeq L$ if 
\begin{align}\label{success}
	\widehat{K} = K_0 \quad \text{and} \quad \max_{1 \leq i \leq K_0} |\widehat{\tau}_i  - \tau_i| \leq 1,
\end{align}
that is, each of the estimated changepoints lies within one neighborhood of a distinct true changepoint. Similarly, we define $\widehat{L} \supsetsim L$ if 
\begin{align}\label{success-super}
	\widehat{K} > K_0 \quad \text{and} \quad  \underset{1 \leq i' \leq K_0}{\max} \:\underset{1 \leq i \leq \widehat{K}}{\min} |\widehat{\tau}_i - \tau_{i'}| \leq 1,
\end{align}
that is, for every true changepoint, there exists an estimated changepoint which differs from it by at most one unit of time. In Table \ref{changepoint detection in AMOC}, we report the performance of Algorithm DESC-U and its competitors for the above models under a variety of changepoint locations and sample sizes. 

\begin{table}
          \centering  \caption{Performance of different algorithms in single changepoint setup ($K_0 = 1$).  The three numbers in each cell are (i) $\mathbb{P}(\widehat{K}=1) \times 100\%$, (ii)  $\mathbb{P}(\widehat{L} \simeq L) \times 100\%$ and (iii) $\mathbb{P}(\widehat{L} \supsetsim L) \times 100\%$, respectively.}
	\label{changepoint detection in AMOC}
	\scalebox{1}{\begin{tabular}{@{}lcrrrrrr@{}}
		\hline
\multirow{2}{*}{\textbf{Model}} &\multirow{2}{*}{$n_1$} &\multicolumn{6}{c}{\textbf{Methods}} \\
\cline{3-8}
		 & &\textbf{DESC-U}  &\textbf{BGHK}  &\textbf{change\textunderscore FF}  &\textbf{fmci} &\textbf{EJS} &\textbf{HK} \\
		\hline
         1& 45& 92\:;\:92\:;\:8& 89\:;\:80\:;\:10& 85\:;\:75\:;\:12& 89\:;\:83\:;\:10& 92\:;\:76\:;\:6& 91\:;\:81\:;\:8\\
        
         ($n = 300$)& 150& 88\:;\:88\:;\:12& 82\:;\:82\:;\:18& 95\:;\:95\:;\:5& 89\:;\:83\:;\:10& 89\:;\:81\:;\:10& 83\:;\:83\:;\:17\\
        
         & 240& 81\:;\:81\:;\:19& 95\:;\:89\:;\:5& 92\:;\:89\:;\:7& 85\:;\:80\:;\:13& 92\:;\:77\:;\:8& 97\:;\:91\:;\:3\\
        [6pt]
         1        & 90& 91\:;\:91\:;\:9& 90\:;\:80\:;\:9& 89\:;\:81\:;\:9& 86\:;\:81\:;\:14& 93\:;\:67\:;\:5& 91\:;\:81\:;\:8\\
        
         ($n = 600$)& 300& 95\:;\:95\:;\:5& 92\:;\:92\:;\:8& 93\:;\:93\:;\:7& 81\:;\:78\:;\:19& 93\:;\:67\:;\:5& 93\:;\:93\:;\:7\\
        
         & 480& 91\:;\:91\:;\:9& 91\:;\:89\:;\:9& 87\:;\:81\:;\:13& 83\:;\:82\:;\:16& 85\:;\:65\:;\:13& 92\:;\:90\:;\:8\\
        [6pt]
         2& 45& 91\:;\:91\:;\:9& 85\:;\:84\:;\:15& 94\:;\:92\:;\:5& 76\:;\:72\:;\:22& 89\:;\:86\:;\:10& 85\:;\:83\:;\:13\\
        
         ($n = 300$)& 150& 90\:;\:90\:;\:10& 92\:;\:92\:;\:8& 94\:;\:94\:;\:6& 79\:;\:71\:;\:19& 94\:;\:94\:;\:6& 94\:;\:94\:;\:6\\
        
         & 240& 93\:;\:93\:;\:7& 88\:;\:88\:;\:12& 88\:;\:88\:;\:12& 85\:;\:84\:;\:13& 92\:;\:88\:;\:8& 86\:;\:86\:;\:8\\
        [6pt]
           2      & 90& 88\:;\:88\:;\:12& 86\:;\:84\:;\:13& 96\:;\:94\:;\:3& 84\:;\:71\:;\:14& 87\:;\:78\:;\:12& 86\:;\:84\:;\:13\\
        
         ($n = 600$)& 300& 92\:;\:92\:;\:8& 92\:;\:92\:;\:8& 94\:;\:94\:;\:6& 78\:;\:73\:;\:19& 92\:;\:82\:;\:7& 93\:;\:92\:;\:7\\
        
         & 480& 88\:;\:88\:;\:12& 86\:;\:86\:;\:14& 94\:;\:93\:;\:6& 82\:;\:78\:;\:16& 83\:;\:71\:;\:14& 86\:;\:86\:;\:14\\
        [6pt]
        3& 45& 90\:;\:90\:;\:10& 89\:;\:81\:;\:10& 89\:;\:79\:;\:7& 88\:;\:77\:;\:11& 93\:;\:79\:;\:7& 93\:;\:81\:;\:7\\
        
         ($n = 300$)& 150& 89\:;\:89\:;\:11& 85\:;\:85\:;\:15& 94\:;\:94\:;\:6& 83\:;\:79\:;\:16& 92\:;\:72\:;\:8& 85\:;\:85\:;\:15\\
        
         & 240& 89\:;\:89\:;\:11& 94\:;\:87\:;\:5& 97\:;\:89\:;\:3& 79\:;\:72\:;\:20& 91\:;\:73\:;\:9& 94\:;\:87\:;\:5\\
        [6pt]
           3      & 90& 88\:;\:88\:;\:12& 93\:;\:82\:;\:6& 90\:;\:74\:;\:9& 86\:;\:66\:;\:13& 95\:;\:60\:;\:2& 95\:;\:83\:;\:4\\
        
         ($n = 600$)& 300& 94\:;\:93\:;\:6& 93\:;\:93\:;\:7& 93\:;\:93\:;\:7& 82\:;\:75\:;\:15& 96\:;\:65\:;\:3& 92\:;\:92\:;\:8\\
        
         & 480& 91\:;\:91\:;\:9& 90\:;\:85\:;\:10& 93\:;\:90\:;\:6& 81\:;\:79\:;\:18& 95\:;\:55\:;\:2& 91\:;\:86\:;\:9\\
        [6pt]
         4& 45& 82\:;\:77\:;\:17& 68\:;\:30\:;\:18& 65\:;\:37\:;\:13& 82\:;\:54\:;\:7& 91\:;\:27\:;\:1& 70\:;\:28\:;\:18\\
        
         ($n = 300$)& 150& 86\:;\:77\:;\:14& 92\:;\:64\:;\:7& 92\:;\:73\:;\:7& 83\:;\:51\:;\:12& 94\:;\:34\:;\:2& 93\:;\:65\:;\:6\\
        
         & 240& 88\:;\:82\:;\:11& 79\:;\:50\:;\:11& 77\:;\:56\:;\:13& 81\:;\:60\:;\:15& 94\:;\:34\:;\:6& 82\:;\:50\:;\:10\\
        [6pt]
           4      & 90& 93\:;\:84\:;\:7& 69\:;\:25\:;\:22& 68\:;\:37\:;\:26& 86\:;\:51\:;\:12& 89\:;\:24\:;\:4& 69\:;\:25\:;\:20\\
        
         ($n = 600$)& 300& 89\:;\:79\:;\:11& 92\:;\:71\:;\:6& 93\:;\:78\:;\:6& 85\:;\:61\:;\:11& 97\:;\:29\:;\:2& 92\:;\:71\:;\:6\\
        
         & 480& 91\:;\:86\:;\:8& 78\:;\:45\:;\:8& 85\:;\:53\:;\:7& 82\:;\:65\:;\:13& 94\:;\:30\:;\:1& 80\:;\:46\:;\:8\\
        [6pt]
         5& 45& 91\:;\:91\:;\:9& 4\:;\:0\:;\:0& 4\:;\:0\:;\:0& 90\:;\:87\:;\:10& 2\:;\:0\:;\:0& 4\:;\:0\:;\:0\\
        
         ($n = 300$)& 150& 91\:;\:91\:;\:9& 3\:;\:0\:;\:0& 6\:;\:0\:;\:0& 77\:;\:71\:;\:19& 4\:;\:0\:;\:0& 3\:;\:0\:;\:0\\
        
         & 240& 82\:;\:82\:;\:18& 10\:;\:0\:;\:0& 8\:;\:0\:;\:0& 87\:;\:78\:;\:11& 6\:;\:0\:;\:0& 8\:;\:0\:;\:0\\
        [6pt]
           5      & 90& 92\:;\:92\:;\:8& 3\:;\:0\:;\:0& 5\:;\:0\:;\:0& 90\:;\:83\:;\:10& 3\:;\:0\:;\:0& 3\:;\:0\:;\:0\\
        
         ($n = 600$)& 300& 91\:;\:91\:;\:9& 3\:;\:0\:;\:0& 6\:;\:0\:;\:0& 78\:;\:74\:;\:22& 5\:;\:0\:;\:0& 4\:;\:0\:;\:0\\
        
         & 480& 91\:;\:91\:;\:9& 7\:;\:1\:;\:0& 18\:;\:2\:;\:0& 80\:;\:71\:;\:18& 10\:;\:1\:;\:0& 8\:;\:1\:;\:0\\
        [6pt]
         6& 45& 91\:;\:91\:;\:9& 10\:;\:0\:;\:0& 8\:;\:0\:;\:0& 86\:;\:80\:;\:12& 10\:;\:0\:;\:0& 10\:;\:0\:;\:0\\
        
         ($n = 300$)& 150& 90\:;\:90\:;\:10& 5\:;\:0\:;\:0& 7\:;\:0\:;\:0& 86\:;\:83\:;\:14& 8\:;\:0\:;\:0& 5\:;\:0\:;\:0\\
        
         & 240& 85\:;\:85\:;\:15& 6\:;\:0\:;\:0& 5\:;\:0\:;\:0& 84\:;\:80\:;\:16& 6\:;\:0\:;\:0& 6\:;\:0\:;\:0\\
        [6pt]
           6      & 90& 94\:;\:94\:;\:6& 11\:;\:1\:;\:0& 10\:;\:0\:;\:0& 94\:;\:85\:;\:3& 8\:;\:0\:;\:0& 10\:;\:1\:;\:0\\
        
         ($n = 600$)& 300& 91\:;\:91\:;\:9& 4\:;\:0\:;\:0& 7\:;\:0\:;\:0& 87\:;\:81\:;\:10& 8\:;\:0\:;\:0& 3\:;\:0\:;\:0\\
        
         & 480& 94\:;\:94\:;\:6& 3\:;\:0\:;\:0& 5\:;\:0\:;\:0& 86\:;\:83\:;\:11& 3\:;\:0\:;\:0& 3\:;\:0\:;\:0\\
        [6pt]

         7& 45& 89\:;\:89\:;\:11& 2\:;\:0\:;\:0& 3\:;\:0\:;\:0& 96\:;\:90\:;\:4& 4\:;\:0\:;\:0& 2\:;\:0\:;\:0\\
        
         ($n = 300$)& 150& 86\:;\:86\:;\:14& 7\:;\:0\:;\:0& 5\:;\:0\:;\:0& 90\:;\:89\:;\:10& 3\:;\:0\:;\:0& 8\:;\:0\:;\:0\\
        
         & 240& 91\:;\:91\:;\:9& 4\:;\:0\:;\:0& 5\:;\:1\:;\:0& 83\:;\:72\:;\:17& 4\:;\:0\:;\:0& 4\:;\:0\:;\:0\\
        [6pt]
         7 & 90& 85\:;\:85\:;\:15& 4\:;\:0\:;\:0& 2\:;\:0\:;\:0& 98\:;\:84\:;\:2& 5\:;\:0\:;\:0& 4\:;\:0\:;\:0\\
        
         ($n = 600$)& 300& 93\:;\:93\:;\:7& 6\:;\:0\:;\:0& 7\:;\:0\:;\:0& 92\:;\:85\:;\:8& 8\:;\:0\:;\:0& 5\:;\:0\:;\:0\\
        
         & 480& 90\:;\:90\:;\:10& 3\:;\:0\:;\:0& 11\:;\:0\:;\:0& 77\:;\:75\:;\:19& 4\:;\:1\:;\:0& 3\:;\:0\:;\:0\\
        \hline
        \end{tabular}
}
        \end{table}

Table \ref{changepoint detection in AMOC} demonstrates that Algorithm DESC-U correctly rejects the null hypothesis in \eqref{amoc-hypothesis} in all the models, which corroborates Theorem \ref{consistency} about the consistency of the permutation test. Consequently, it correctly detects the presence of a changepoint in the data. Additionally, it estimates the changepoint location accurately in all of the models except perhaps Model 4. However, Algorithm DESC-U significantly outperforms all of its competitors with respect to the metric $\mathbb{P}(\widehat{L} \simeq L)$ for this model. The third metric $\mathbb{P}(\widehat{L} \supsetsim L)$, which signifies overestimation while containing the true changepoints, also indicates that the performance of the proposed algorithm is comparable with its competitors. Among the other methods, the performance of BGHK, change\textunderscore FF and HK is comparable to our method for the location change models (Model 1-4) except Model 4 where their performance deteriorates. The EJS method compares well with our method in terms of $\mathbb{P}(\widehat{K}=1)$ for all the location-change models, however, it significantly underperforms with respect to the metric $\mathbb{P}(\widehat{L} \simeq L)$. All of these competitors fail completely for the covariance-change models (Model 5-7). Algorithm DESC-U exhibits slightly better performance than the fmci method in most of the models under the metric $\mathbb{P}(\widehat{K}=1)$. However, we note from Table \ref{size} that the fmci method fails to control the type-I error at the desired level in many situations, which may attribute to the inflated values of $\mathbb{P}(\widehat{K}=1)$ in Table \ref{changepoint detection in AMOC}. Algorithm DESC-U outperforms the fmci method with regards to the metric $\mathbb{P}(\widehat{L} \simeq L)$ for all the models.

We next investigate the performance of the test procedures in the various algorithms considered in this section as well as the probability of ``success'' of any algorithm (defined by $\mathbb{P}(\widehat{L} \simeq L)$) when we vary either the signal strength or the breakfraction $\gamma$.  To this end, we consider the following models. As earlier, the pre-change process is denoted by $X_1$ and the post-change process is denoted by $X_2$. The data $Z_1,\ldots,Z_{\lfloor n\gamma\rfloor}$ are i.i.d. copies of $X_1$ while $Z_{\lfloor n\gamma\rfloor+1},\ldots,Z_n$ are i.i.d. copies of $X_2$ for a breakfraction $\gamma \in (0,1)$.
\vspace{0.075in}\\
\noindent \textbf{Model M1}: $X_1(t)=BB(t)$ and $X_{2}(t)=BB(t)+c\sin{(t)}$.
\vspace{0.075in}\\
\noindent \textbf{Model M2}: $X_{l}(t) = \sum_{j=1}^{40}\sqrt{\theta_{jl}}W_{j}\phi_{j}(t)$, $l = 1, 2$, where for $j=1,2,\dots,40$, $\phi_{j}(t) = \sqrt{2}\sin{(jt\pi)}$, $\theta_{j1} = j^{-2}$, $\theta_{j2} = cj^{-2}$.
\vspace{0.075in}\\
\begin{figure}
\centering
\includegraphics[scale=0.35]{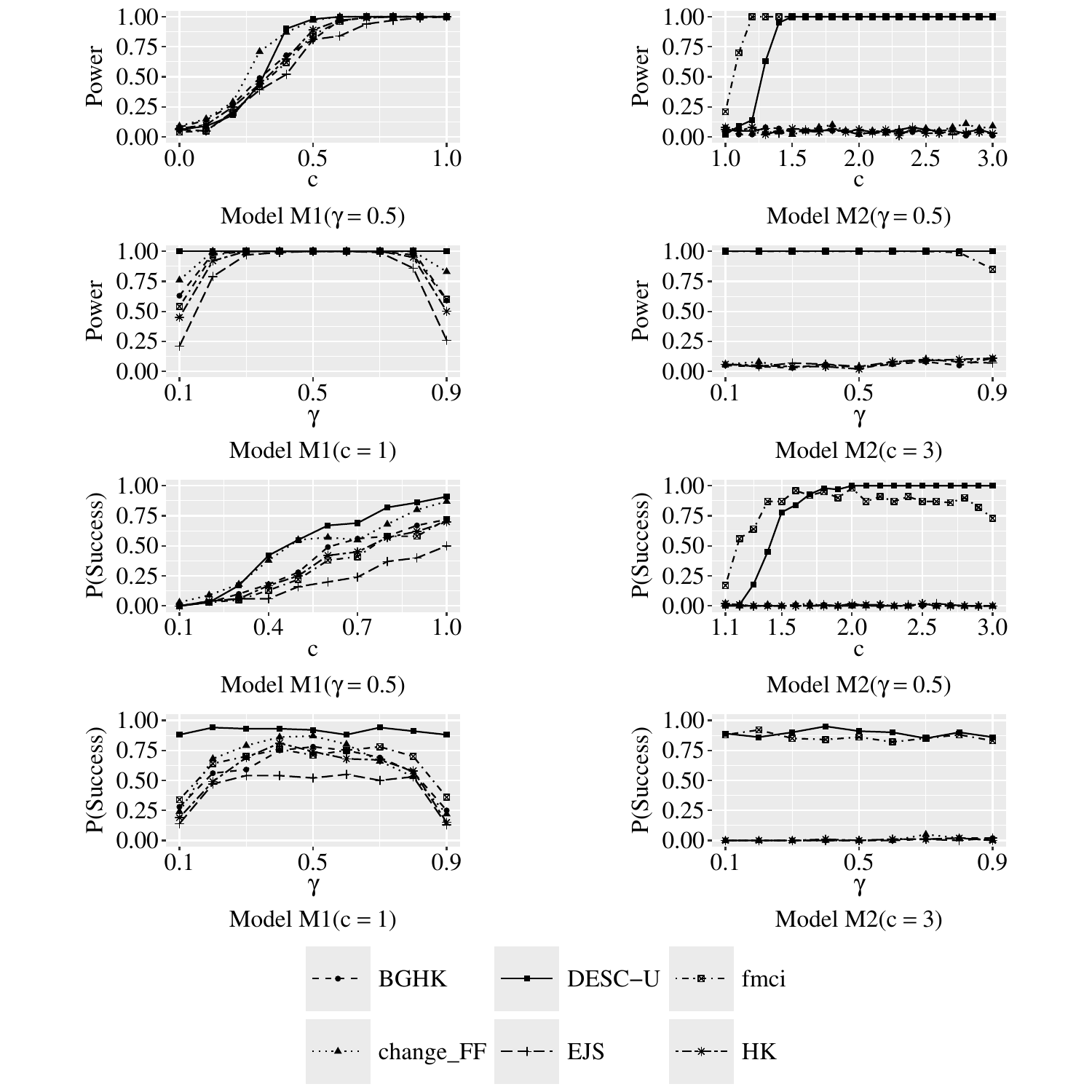}
	\caption{Powers and probabilities of success ($\mathbb{P}(\widehat{L} \simeq L)$) under Models M1 and M2 when $n = 100$.}
	\label{fig:power_detectability_curves}
\end{figure}
Figure \ref{fig:power_detectability_curves} displays the power curves and the probabilities of success when exactly one of $c$ or $\gamma$ is held fixed.
 Note that the values $c = 1$ in Model M1 and $c = 3$ in Model M2 correspond to Models 4 and 5 in the earlier simulation study in the single changepoint setting. Also, the values $c = 0$ in Model M1 and $c = 1$ in Model M2 correspond to the null models N2 and N4 considered earlier, which are the pre-change processes for Models 4 and 5 respectively. It is observed that for the location-change model M1, the test procedure in the change\textunderscore FF method has the best power ($\gamma$ fixed with varying $c$) performance followed closely by the permutation test in Algorithm DESC-U. The other test procedures have a slower rate to increase to 1 (as $c$ grows) than these two tests. On the other hand, when the signal is fixed and $\gamma$ varies, all of the competing test procedures show a significant drop in power when the change happens towards either boundary of the data set since they are based on asymptotic approximations. The permutation test in Algorithm DESC-U remains unaffected being an exact test and its power remains at 1 throughout. Under the scale-change model M2, all of the tests except the permutation test and that in the fmci procedure fail completely. Furthermore, the test in the fmci procedure is not comparable since it has a significantly inflated size although its power reaches 1 the fastest ($\gamma$ fixed with varying $c$) and does not suffer from any boundary problems ($c$ fixed with varying $\gamma$). The power of permutation test quickly reaches 1 as the signal strength grows while its power stays at 1 for varying $\gamma$. In terms of the probability of success (representing accuracy of the estimation technique in the changepoint method), Algorithm DESC-U is the best for Model M1 followed closely by the change\textunderscore FF method when the signal strength varies. However, when $\gamma$ varies keeping $c$ fixed, the drop in the probability of success for the competing methods is significant even for non-boundary values $\gamma = 0.3$ and $\gamma = 0.7$. Algorithm DESC-U, on the other hand, does not show any drop in performance and has a significantly higher probability of success.  In terms of the probability of success (keeping $\gamma$ fixed) under Model M2, all the competitors of Algorithm DESC-U except fmci fail completely while the fmci method is not comparable due to its elevated size. Also, the probability of success for the fmci method seems to drop off after $c = 2$ while it quickly reaches 1 for Algorithm DESC-U. When $c$ is fixed, the probabilities of success for Algorithm DESC-U and the fmci method are similar while it is almost zero for the other methods. However, since the fmci method does not control the probability of type I error, its performance should not be compared with that of the proposed algorithm.


We next consider the multiple changepoint setting $(K_0 > 1)$ and investigate the performance of the algorithms. We will confine our analysis to the setup with $K_0 = 2$ where we will investigate the effect of both epidemic and non-epidemic changes in either location or scale. In each of the following models, the data $Z_1,\ldots,Z_{n_1}$ are i.i.d. copies of $X_1$, the data $Z_{n_1+1},\ldots,Z_{n_1+n_2}$ are i.i.d. copies of $X_2$ and $Z_{n_1+n_2+1},\ldots,Z_n$ are i.i.d. copies of $X_3$. 
\vspace{0.075in}\\
\noindent \textbf{Model 8} [Location change]:   $X_{l}(t) = \sum_{j=0}^{150}\sqrt{\theta_{j}}W_{j}\phi_{j}(t) + \mu_{l}(t)$,
where $\theta_{j} = 0.7 \times 2^{-j}$, $W_{j} \overset{\mathrm{i.i.d}}{\sim} N(0,1)$, for $j = 0,1, 2, \dots, 150$, $\phi_{0}(t)=1$, $\phi_{j}(t) = \sqrt{2}\sin{(2\pi rt - \pi)}$ for $j=2r-1$ and $\phi_{j}(t) = \sqrt{2}\cos{(2\pi rt - \pi)}$ for $j=2r$ with $r=1,2,\dots,75$, $\mu_{1}(t)=0.5-100(t-0.1)(t-0.3)(t-0.5)(t-0.9)$, $\mu_{2}(t)=1+3t^2-5t^3+1.5\sin{(1+10\pi t)}$ and $\mu_{3}(t)=1 + 3t^2 - 5t^3$.
\vspace{0.075in}\\
\noindent \textbf{Model 9} [Epidemic change in location]: $X_{1}(t)=BB(t)$, $X_{2}(t)=BB(t)+t$ and $X_{3}(t)=BB(t)$.
\vspace{0.075in}\\
\noindent \textbf{Model 10} [Change in eigenvalues]:   $X_{l}(t) = \sum_{j=1}^{50}\sqrt{\theta_{jl}}W_{j}\phi_{j}(t)$, $l = 1, 2, 3$,
where for each $j$, $\phi_{j}(t) = \sqrt{2}\sin{(jt\pi)}$, $W_{j} \overset{\mathrm{i.i.d}}{\sim} N(0,1)$, $\theta_{j1} = j^{-2}$, $\theta_{j2}=j^{-1.05}$, and $\theta_{j3} = exp(-j)$ for $j = 1, 2, \dots, 50$.
\vspace{0.075in}\\
\noindent \textbf{Model 11} [Epidemic change in scale]: $X_{l}(t) = \sum_{j=1}^{40}\sqrt{\theta_{jl}}W_{j}\phi_{j}(t)$, $l = 1, 2, 3$,
where for $j=1,2,\dots,40$, $\phi_{j}(t) = \sqrt{2}\sin{(jt\pi)}$, $W_{j} \overset{\mathrm{i.i.d}}{\sim} t_{3}/\sqrt{3}$, $\theta_{j1} = j^{-2}$, $\theta_{j2} = 3j^{-2}$ and $\theta_{j3} = j^{-2}$.
\vspace{0.075in}\\
\noindent \textbf{Model 12} [Epidemic change in eigenfunctions]: $X_{l}(t) = \sum_{j=1}^{40}\sqrt{\theta_{jl}}W_{j}\phi_{jl}(t)$, $l = 1, 2$,
where for $j=1,2,\dots,40$, $\phi_{j1}(t) = \sqrt{2}\sin{(jt\pi)}$, $\phi_{j2}$ is the $j$-th term of Fourier Basis, $\phi_{j3}(t) = \sqrt{2}\sin{(jt\pi)}$ $\theta_{j1} = \theta_{j2} =\theta_{j3} = exp(-j/3)$, $W_{j} \overset{\mathrm{i.i.d}}{\sim} N(0,1)$.

\begin{table}
    \centering        \caption{Performance of different algorithms in the multiple changepoint setup ($K_0 = 2$).  The three numbers in each cell are (i) $\mathbb{P}(\widehat{K}=2) \times 100\%$, (ii)  $\mathbb{P}(\widehat{L} \simeq L) \times 100\%$ and (iii) $\mathbb{P}(\widehat{L} \supsetsim L) \times 100\%$, respectively.}
	\label{changepoint detection under mcp}
    \renewcommand{\arraystretch}{1.4}
\scalebox{0.85}{
    \begin{tabular} {@{}lcrrrrrr@{}}
		\hline
\multirow{2}{*}{\textbf{Model}} &\multirow{2}{*}{$(n_1,n_2)$} &\multicolumn{6}{c}{\textbf{Methods}} \\
\cline{3-8}
		 & &\textbf{DESC-U}  &\textbf{BGHK}  &\textbf{change\textunderscore FF}  &\textbf{fmci} &\textbf{EJS} &\textbf{HK}\\
		\cline{1-8}
         8& 45,\:75& 85\:;\:85\:;\:15& 88\:;\:87\:;\:11& 95\:;\:95\:;\:5& 80\:;\:72\:;\:18& 0\:;\:0\:;\:0& 0\:;\:0\:;\:0\\
        
         ($n=300$)& 100,\:100& 85\:;\:85\:;\:15& 79\:;\:79\:;\:21& 93\:;\:93\:;\:7& 79\:;\:64\:;\:15& 0\:;\:0\:;\:0& 0\:;\:0\:;\:0\\
        
        & 180,\:45& 82\:;\:82\:;\:18& 84\:;\:84\:;\:16& 90\:;\:90\:;\:10& 81\:;\:56\:;\:5& 4\:;\:0\:;\:0& 50\:;\:46\:;\:3\\
        [6pt]
         8        & 90,\:150& 81\:;\:81\:;\:19& 82\:;\:80\:;\:18& 87\:;\:87\:;\:13& 80\:;\:64\:;\:17& 0\:;\:0\:;\:0& 1\:;\:0\:;\:0\\
        
         ($n=600$)& 200,\:200& 84\:;\:84\:;\:16& 88\:;\:88\:;\:12& 88\:;\:88\:;\:12& 72\:;\:63\:;\:24& 0\:;\:0\:;\:0& 0\:;\:0\:;\:0\\
        
         & 360,\:90& 82\:;\:72\:;\:13& 93\:;\:93\:;\:7& 78\:;\:52\:;\:12& 79\:;\:64\:;\:12& 3\:;\:0\:;\:0& 93\:;\:93\:;\:7\\
        [6pt]
        9& 45,\:75& 77\:;\:68\:;\:19& 76\:;\:34\:;\:9& 72\:;\:42\:;\:17& 84\:;\:45\:;\:9& 0\:;\:0\:;\:0& 77\:;\:33\:;\:10\\
        
        ($n=300$)& 100,\:100& 84\:;\:67\:;\:15& 83\:;\:45\:;\:12& 84\:;\:61\:;\:14& 86\:;\:49\:;\:11& 0\:;\:0\:;\:0& 86\:;\:48\:;\:8\\
        
        & 180,\:45& 75\:;\:57\:;\:21& 64\:;\:31\:;\:14& 70\:;\:45\:;\:20& 87\:;\:48\:;\:6& 0\:;\:0\:;\:0& 60\:;\:23\:;\:10\\
        [6pt]     
         9  & 90,\:150& 82\:;\:72\:;\:13& 75\:;\:40\:;\:11& 78\:;\:52\:;\:12& 82\:;\:46\:;\:9& 0\:;\:0\:;\:0& 79\:;\:40\:;\:11\\
        
        ($n=600$)& 200,\:200& 87\:;\:70\:;\:11& 84\:;\:48\:;\:7& 85\:;\:64\:;\:8& 84\:;\:46\:;\:9& 0\:;\:0\:;\:0& 85\:;\:49\:;\:6\\
        
        & 360,\:90& 77\:;\:60\:;\:17& 62\:;\:28\:;\:17& 62\:;\:34\:;\:23& 82\:;\:45\:;\:7& 0\:;\:0\:;\:0& 62\:;\:28\:;\:13\\
        [6pt]
        10& 45,\:75& 81\:;\:81\:;\:19& 0\:;\:0\:;\:0& 5\:;\:0\:;\:0& 79\:;\:65\:;\:12& 0\:;\:0\:;\:0& 0\:;\:0\:;\:0\\
        
        ($n=300$)& 100,\:100& 86\:;\:86\:;\:14& 0\:;\:0\:;\:0& 2\:;\:0\:;\:0& 69\:;\:53\:;\:22& 0\:;\:0\:;\:0& 0\:;\:0\:;\:0\\
        
        & 180,\:45& 87\:;\:87\:;\:13& 0\:;\:0\:;\:0& 1\:;\:0\:;\:0& 85\:;\:66\:;\:12& 0\:;\:0\:;\:0& 0\:;\:0\:;\:0\\
        [6pt]     
         10  & 90,\:150& 87\:;\:87\:;\:13& 0\:;\:0\:;\:0& 2\:;\:0\:;\:0& 63\:;\:54\:;\:28& 0\:;\:0\:;\:0& 0\:;\:0\:;\:0\\
        
        ($n=600$)& 200,\:200& 85\:;\:85\:;\:15& 0\:;\:0\:;\:0& 6\:;\:0\:;\:0& 79\:;\:66\:;\:15& 0\:;\:0\:;\:0& 0\:;\:0\:;\:0\\
        
        & 360,\:90& 82\:;\:82\:;\:17& 0\:;\:0\:;\:0& 4\:;\:0\:;\:0& 79\:;\:62\:;\:8& 0\:;\:0\:;\:0& 0\:;\:0\:;\:0\\
        [6pt]
        11& 45,\:75& 90\:;\:84\:;\:9& 2\:;\:0\:;\:0& 0\:;\:0\:;\:0& 87\:;\:76\:;\:8& 0\:;\:0\:;\:0& 1\:;\:0\:;\:0\\
        
        ($n=300$)& 100,\:100& 90\:;\:79\:;\:8& 1\:;\:0\:;\:0& 1\:;\:0\:;\:0& 81\:;\:69\:;\:11& 1\:;\:0\:;\:0& 1\:;\:0\:;\:0\\
        
         & 180,\:45& 92\:;\:72\:;\:4& 0\:;\:0\:;\:0& 1\:;\:0\:;\:0 & 90\:;\:64\:;\:7& 0\:;\:0\:;\:0& 0\:;\:0\:;\:0\\
        [6pt]       
         11 & 90,\:150& 86\:;\:77\:;\:12& 1\:;\:0\:;\:0& 2\:;\:0\:;\:0& 80\:;\:66\:;\:16& 0\:;\:0\:;\:0& 1\:;\:0\:;\:0\\
        
        ($n=600$)& 200,\:200& 90\:;\:80\:;\:9& 0\:;\:0\:;\:0& 1\:;\:0\:;\:0& 83\:;\:72\:;\:9& 0\:;\:0\:;\:0& 0\:;\:0\:;\:0\\
        
        & 360,\:90& 89\:;\:80\:;\:8& 1\:;\:0\:;\:0& 1\:;\:0\:;\:0& 82\:;\:67\:;\:14& 0\:;\:0\:;\:0& 1\:;\:0\:;\:0\\
        [6pt]
        12& 45,\:75& 82\:;\:81\:;\:18& 0\:;\:0\:;\:0& 0\:;\:0\:;\:0& 86\:;\:79\:;\:11& 0\:;\:0\:;\:0& 0\:;\:0\:;\:0\\
        
        ($n=300$)& 100,\:100& 76\:;\:74\:;\:20& 0\:;\:0\:;\:0& 0\:;\:0\:;\:0& 86\:;\:81\:;\:11& 1\:;\:0\:;\:0& 0\:;\:0\:;\:0\\
        
         & 180,\:45& 84\:;\:72\:;\:14& 0\:;\:0\:;\:0& 0\:;\:0\:;\:0& 85\:;\:73\:;\:12& 0\:;\:0\:;\:0& 0\:;\:0\:;\:0\\
        [6pt]     
        12 & 90,\:150& 83\:;\:81\:;\:17& 0\:;\:0\:;\:0& 1\:;\:0\:;\:0& 88\:;\:81\:;\:10& 0\:;\:0\:;\:0& 0\:;\:0\:;\:0\\
        
        ($n=600$)& 200,\:200& 89\:;\:83\:;\:11& 1\:;\:0\:;\:0& 1\:;\:0\:;\:0& 86\:;\:81\:;\:13& 0\:;\:0\:;\:0& 0\:;\:0\:;\:0\\
        
        & 360,\:90& 85\:;\:80\:;\:13& 1\:;\:0\:;\:0& 1\:;\:0\:;\:0& 81\:;\:73\:;\:14& 0\:;\:0\:;\:0& 1\:;\:0\:;\:0\\
        \hline
        \end{tabular}
        }
        \end{table}
        
The performance of different changepoint detection methods are summarized in Table \ref{changepoint detection under mcp}. It is seen from this table that in the location-change models 8 and 9, Algorithm DESC-U compares favorably with the BGHK, change\textunderscore FF and fmci methods in terms of the accurately detecting the number of changepoints in the data. With regards to the estimation of the changepoint locations ($\mathbb{P}(\widehat{L} \simeq L)$), it is observed that our method have similar performance to the BGHK and change\textunderscore FF methods for Model 8 while the fmci method underperforms. On the other hand, for the epidemic location-change model (Model 9), our method shows significantly larger values of $\mathbb{P}(\widehat{L} \simeq L)$ than all of these three competitors. The EJS method completely fails for both the location-change models since it almost always underestimates the number of changepoints. The performance of the HK method is satisfactory in terms of $\mathbb{P}(\widehat{K}=2)$ for Model 9 while it underperforms with regards to $\mathbb{P}(\widehat{L} \simeq L)$. However, under Model 8, it shows a very erratic behavior. For Models 10, 11 and 12 exhibiting changes in the covariance, all the methods except DESC-U and fmci fail completely. Our method outperforms the fmci method in terms of both $\mathbb{P}(\widehat{K}=2)$ and $\mathbb{P}(\widehat{L} \simeq L)$ for Models 10 and 11, while their performance is comparable for Model 12. 


Next, we analyze the performance of Algorithm DESC-S when the number of changepoints is predetermined for models 1 through 12. We have documented the frequency with which the estimated set of changepoints matches the actual set of changepoints or falls within one unit of the true changepoints, specifically when the specified number of changepoints ($K$) equals the actual number of changepoints ($K_0$). In cases where $K$ is less than $K_0$ (applicable to models 8 to 12), we have indicated how often each estimated changepoint corresponds to one of the true changepoints. Conversely, when $K$ exceeds $K_0$, we have noted how frequently the estimated set of changepoints includes all of the true changepoints.
The performance of Algorithm DESC-S are summarized in Tables 1 and 2 in the Supplementary Material. These two tables illustrate that the changepoint locations identified by Algorithm DESC-S are in close agreement with the actual changepoint locations, or are found within one neighborhood of them, in almost all cases where the specified and actual number of changepoints match. In situations of underspecification, the estimated changepoints mainly represent a subset of the true changepoints, while in cases of overspecification, the estimated changepoints generally form a superset of the true changepoints. These findings corroborate Theorem \ref{pop} on the ``faithful'' property of the algorithm.

Finally, we have analyzed the performance of Algorithm DESC-SS under specific constraints related to the number of changepoints and the results are reported in Tables \ref{DESC-SS in AMOC} and \ref{DESC-SS in mcp}. 
It is clearly seen from these tables that Algorithm DESC-SS has an improvement in the performance over that of Algorithm DESC-U (reported in Tables \ref{changepoint detection in AMOC} and \ref{changepoint detection under mcp}), particularly when the upper bound of the number of changepoints is provided. Furthermore, the performance of Algorithm DESC-SS remains unaffected by additional information regarding the lower bound of the number of changepoints. When only the lower bound is provided, the performance of Algorithm DESC-SS degrades a bit and is similar to the performance of Algorithm DESC-U. The reason for this slight degradation in the performance is discussed in the second paragraph in Remark \ref{remark-semi-supervised}. 

\begin{table}
\centering
\caption{Performance of Algorithm DESC-SS in the single changepoint scenario: The three numbers in each cell are (i) $\mathbb{P}(\widehat{K}=1) \times 100\%$, (ii) $\mathbb{P}(\widehat{L} \simeq L) \times 100\%$ and (iii) $\mathbb{P}(\widehat{L} \supsetsim L) \times 100\%$.}
	\label{DESC-SS in AMOC}
    \renewcommand{\arraystretch}{1.25}
	\scalebox{0.75}{\begin{tabular} {@{}lcrrrrr@{}}
		\hline
\textbf{Model} &$n_1$ &$K_u= 2$ & $K_u= 3$ &$K_l=1, K_u= 2$ &$K_l=1, K_u= 3$ &$K_l=1$\\
		\hline
         1& 45 & 95\:;\:95\:;\:5 & 95\:;\:90\:;\:4& 95\:;\:95\:;\:5& 95\:;\:90\:;\:4& 89\:;\:89\:;\:11\\
        
        ($n = 300$) & 150 & 94\:;\:94\:;\:6& 95\:;\:95\:;\:5& 94\:;\:94\:;\:6& 95\:;\:95\:;\:5& 84\:;\:84\:;\:16\\
        
        & 240 & 95\:;\:95\:;\:5& 93\:;\:93\:;\:7& 95\:;\:95\:;\:5& 93\:;\:93\:;\:7& 87\:;\:87\:;\:13\\
        [6pt]         
        1& 90 & 99\:;\:99\:;\:1 & 94\:;\:94\:;\:4& 99\:;\:99\:;\:1 & 94\:;\:94\:;\:4& 92\:;\:92\:;\:8\\
        
        ($n = 600$) & 300 & 98\:;\:98\:;\:2& 97\:;\:96\:;\:3& 98\:;\:98\:;\:2& 97\:;\:96\:;\:3& 92\:;\:92\:;\:8\\
        
        & 480 & 99\:;\:99\:;\:1& 100\:;\:100\:;\:0& 99\:;\:99\:;\:1& 100\:;\:100\:;\:0& 91\:;\:91\:;\:9\\
        [6pt]
        2& 45 & 100\:;\:100\:;\:0& 100\:;\:99\:;\:0& 100\:;\:100\:;\:0& 100\:;\:99\:;\:0& 92\:;\:92\:;\:8\\
        
        ($n = 300$) & 150 & 100\:;\:100\:;\:0& 100\:;\:100\:;\:0& 100\:;\:100\:;\:0& 100\:;\:100\:;\:0& 89\:;\:89\:;\:11\\
        
        & 240 & 100\:;\:100\:;\:0& 100\:;\:100\:;\:0& 100\:;\:100\:;\:0& 100\:;\:100\:;\:0& 88\:;\:88\:;\:12\\
        [6pt]        
        2& 90 & 100\:;\:100\:;\:0& 100\:;\:100\:;\:0& 100\:;\:100\:;\:0& 100\:;\:100\:;\:0& 89\:;\:89\:;\:11\\
        
        ($n = 600$) & 300 & 100\:;\:100\:;\:0& 100\:;\:100\:;\:0& 100\:;\:100\:;\:0& 100\:;\:100\:;\:0& 93\:;\:93\:;\:7\\
        
        & 480 & 100\:;\:100\:;\:0& 100\:;\:100\:;\:0& 100\:;\:100\:;\:0& 100\:;\:100\:;\:0& 89\:;\:89\:;\:11\\
        [6pt]
        3& 45 & 96\:;\:96\:;\:4& 96\:;\:93\:;\:3& 96\:;\:96\:;\:4& 96\:;\:93\:;\:3& 88\:;\:88\:;\:12\\
        
        ($n = 300$) & 150 & 98\:;\:98\:;\:2& 98\:;\:97\:;\:2& 98\:;\:98\:;\:2& 98\:;\:97\:;\:2& 92\:;\:92\:;\:8\\
        
        & 240 & 97\:;\:97\:;\:3& 96\:;\:93\:;\:3& 97\:;\:97\:;\:3& 96\:;\:93\:;\:3& 89\:;\:89\:;\:11\\
        [6pt]        
        3& 90 & 96\:;\:96\:;\:4& 95\:;\:95\:;\:5& 96\:;\:96\:;\:4& 95\:;\:95\:;\:5& 90\:;\:90\:;\:10\\
        
        ($n = 600$) & 300 & 96\:;\:95\:;\:4& 96\:;\:92\:;\:4& 96\:;\:95\:;\:4& 96\:;\:92\:;\:4& 91\:;\:91\:;\:9\\
        
        & 480 & 97\:;\:97\:;\:3& 96\:;\:94\:;\:4& 97\:;\:97\:;\:3& 96\:;\:94\:;\:4& 88\:;\:88\:;\:12\\
        [6pt]
        4& 45 & 97\:;\:91\:;\:3& 95\:;\:85\:;\:4& 97\:;\:91\:;\:3& 95\:;\:85\:;\:4& 83\:;\:78\:;\:16\\
        
        ($n = 300$) & 150 & 96\:;\:87\:;\:4& 93\:;\:82\:;\:7& 96\:;\:87\:;\:4& 93\:;\:82\:;\:7& 88\:;\:80\:;\:11\\
        
        & 240 & 97\:;\:90\:;\:3& 94\:;\:85\:;\:6& 97\:;\:90\:;\:3& 94\:;\:85\:;\:6& 89\:;\:83\:;\:10\\
        [6pt]        
        4& 90 & 96\:;\:86\:;\:4& 94\:;\:84\:;\:6& 96\:;\:86\:;\:4& 94\:;\:84\:;\:6& 90\:;\:81\:;\:10\\
        
        ($n = 600$) & 300 & 91\:;\:81\:;\:9& 91\:;\:81\:;\:8& 91\:;\:81\:;\:9& 91\:;\:81\:;\:8& 87\:;\:77\:;\:13\\
        
        & 480 & 100\:;\:94\:;\:0& 96\:;\:88\:;\:3& 100\:;\:94\:;\:0& 96\:;\:88\:;\:3& 89\:;\:84\:;\:10\\
        [6pt]
        5& 45 & 94\:;\:93\:;\:5& 94\:;\:76\:;\:4& 94\:;\:93\:;\:5& 94\:;\:76\:;\:4& 90\:;\:90\:;\:10\\
        
        ($n = 300$) & 150 & 93\:;\:93\:;\:7& 93\:;\:89\:;\:6& 93\:;\:93\:;\:7& 93\:;\:89\:;\:6& 89\:;\:89\:;\:11\\
        
        & 240 & 93\:;\:93\:;\:7& 96\:;\:96\:;\:4& 93\:;\:93\:;\:7& 96\:;\:96\:;\:4& 84\:;\:84\:;\:16\\
        [6pt]        
        5& 90 & 97\:;\:95\:;\:3& 97\:;\:86\:;\:3& 97\:;\:95\:;\:3& 97\:;\:86\:;\:3& 94\:;\:94\:;\:6\\
        
        ($n = 600$) & 300 & 96\:;\:96\:;\:4& 94\:;\:90\:;\:6& 96\:;\:96\:;\:4& 94\:;\:90\:;\:6& 91\:;\:91\:;\:9\\
        
        & 480 & 99\:;\:99\:;\:1& 99\:;\:99\:;\:1& 99\:;\:99\:;\:1& 99\:;\:99\:;\:1& 87\:;\:87\:;\:13\\
        [6pt]
        6& 45 & 98\:;\:98\:;\:2& 98\:;\:98\:;\:2& 98\:;\:98\:;\:2& 98\:;\:98\:;\:2& 85\:;\:85\:;\:15\\
        
        ($n = 300$) & 150 & 96\:;\:94\:;\:4& 98\:;\:93\:;\:2& 96\:;\:94\:;\:4& 98\:;\:93\:;\:2& 87\:;\:87\:;\:13\\
        
        & 240 & 98\:;\:94\:;\:2& 96\:;\:85\:;\:2& 98\:;\:94\:;\:2& 96\:;\:85\:;\:2& 90\:;\:90\:;\:10\\
        [6pt]        
        6& 90 & 98\:;\:98\:;\:2& 98\:;\:98\:;\:2& 98\:;\:98\:;\:2& 98\:;\:98\:;\:2& 91\:;\:91\:;\:9\\
        
        ($n = 600$) & 300 & 91\:;\:90\:;\:9& 95\:;\:93\:;\:5& 91\:;\:90\:;\:9& 95\:;\:93\:;\:5& 89\:;\:89\:;\:11\\
        
        & 480 & 97\:;\:94\:;\:3& 96\:;\:85\:;\:4& 97\:;\:94\:;\:3& 96\:;\:85\:;\:4& 91\:;\:91\:;\:9\\
        [6pt]
        7& 45 & 95\:;\:92\:;\:5& 95\:;\:66\:;\:5& 95\:;\:92\:;\:5& 95\:;\:66\:;\:5& 93\:;\:84\:;\:7\\
        
        ($n = 300$) & 150 & 94\:;\:92\:;\:6& 94\:;\:86\:;\:6& 94\:;\:92\:;\:6& 94\:;\:86\:;\:6& 92\:;\:92\:;\:8\\
        
        & 240 & 94\:;\:94\:;\:6& 95\:;\:95\:;\:5& 94\:;\:94\:;\:6& 95\:;\:95\:;\:5& 94\:;\:94\:;\:6\\
        [6pt] 
        7& 90 & 96\:;\:92\:;\:4& 95\:;\:80\:;\:5& 96\:;\:92\:;\:4& 95\:;\:80\:;\:5& 95\:;\:93\:;\:5\\
        
        ($n = 600$) & 300 & 98\:;\:96\:;\:2& 95\:;\:87\:;\:5& 98\:;\:96\:;\:2& 95\:;\:87\:;\:5& 94\:;\:94\:;\:6\\
        
        & 480 & 99\:;\:99\:;\:1& 100\:;\:100\:;\:0& 99\:;\:99\:;\:1& 100\:;\:100\:;\:0& 91\:;\:88\:;\:9\\
        \hline
        \end{tabular}
            }
\end{table}

\begin{table}
           \centering \caption{Performance of Algorithm DESC-SS in the multiple changepoint scenario ($K_0 = 2$): The three numbers in each cell are (i) $\mathbb{P}(\widehat{K}=1) \times 100\%$, (ii) $\mathbb{P}(\widehat{L} \simeq L) \times 100\%$ and (iii) $\mathbb{P}(\widehat{L} \supsetsim L) \times 100\%$.}
	\label{DESC-SS in mcp}
    \renewcommand{\arraystretch}{1.5}
    \scalebox{0.75}{
	\begin{tabular} {@{}lcrrrrrrr@{}}
		\hline
\textbf{Model} &$(n_1,n_2)$ &$K_l=1,K_u = 3$ &$K_l=1,K_u = 4$ & $K_u = 2$ &$K_u = 3$ &$K_u = 4$ &$K_l = 1$ &$K_l = 2$\\
		\hline
        8& 45,\:75 & 97\:;\:97\:;\:3& 96\:;\:96\:;\:4& 100\:;\:100\:;\:0&97\:;\:97\:;\:3& 96\:;\:96\:;\:4& 79\:;\:79\:;\:21& 84\:;\:84\:;\:16\\
        
        ($n = 300$) & 100,\:100 & 92\:;\:92\:;\:8& 94\:;\:90\:;\:6& 100\:;\:100\:;\:0& 92\:;\:92,8& 94\:;\:90\:;\:6& 86\:;\:86\:;\:14& 85\:;\:85\:;\:15\\
        
        & 180,\:45 & 97\:;\:97\:;\:3& 99\:;\:95\:;\:1& 100\:;\:100\:;\:0& 97\:;\:97\:;\:3& 99\:;\:95\:;\:1& 79\:;\:79\:;\:21& 81\:;\:81\:;\:19\\
        [6pt]         
        8& 90,\:150 & 97\:;\:97\:;\:3& 97\:;\:97\:;\:3& 100\:;\:100\:;\:0& 97\:;\:97\:;\:3& 97\:;\:97\:;\:3& 84\:;\:84\:;\:16& 84\:;\:84\:;\:16\\
        
        ($n = 600$) & 200,\:200 & 91\:;\:91\:;\:9& 94\:;\:94\:;\:6& 100\:;\:100\:;\:0& 91\:;\:91\:;\:9& 94\:;\:94\:;\:6& 83\:;\:83\:;\:17& 80\:;\:80\:;\:20\\
        
        & 360,\:90 & 99\:;\:99\:;\:1& 99\:;\:98\:;\:1& 100\:;\:100\:;\:0& 99\:;\:99\:;\:1& 99\:;\:98\:;\:1& 83\:;\:83\:;\:17& 86\:;\:86\:;\:14\\
        [6pt]
        9& 45,\:75 & 97\:;\:85\:;\:2& 93\:;\:80\:;\:4& 100\:;\:87\:;\:0& 97\:;\:85\:;\:2& 93\:;\:80\:;\:4& 73\:;\:64\:;\:23& 77\:;\:67\:;\:20\\
        
        ($n = 300$) & 100,\:100 & 92\:;\:74\:;\:8& 90\:;\:73\:;\:8& 100\:;\:79\:;\:0& 92\:;\:74\:;\:8& 90\:;\:73\:;\:8& 81\:;\:64\:;\:19& 81\:;\:67\:;\:16\\
        
        & 180,\:45 & 97\:;\:77\:;\:3& 92\:;\:67\:;\:7& 100\:;\:70\:;\:0& 97\:;\:77\:;\:3& 92\:;\:67\:;\:7& 76\:;\:59\:;\:19& 75\:;\:58\:;\:20\\
        [6pt]        
        9& 90,\:150 & 93\:;\:80\:;\:5& 95\:;\:81\:;\:4& 100\:;\:85\:;\:0& 93\:;\:80\:;\:5& 95\:;\:81\:;\:4& 87\:;\:75\:;\:10& 81\:;\:69\:;\:16\\
        
        ($n = 600$) & 200,\:200 & 99\:;\:82\:;\:0& 97\:;\:79\:;\:1& 100\:;\:76\:;\:0& 99\:;\:82\:;\:0& 97\:;\:79\:;\:1& 87\:;\:71\:;\:10& 76\:;\:61\:;\:16\\
        
        & 360,\:90 & 96\:;\:74\:;\:2& 92\:;\:71\:;\:5& 100\:;\:71\:;\:0& 96\:;\:74\:;\:2& 92\:;\:71\:;\:5& 74\:;\:58\:;\:19& 74\:;\:58\:;\:19
        \\
        [6pt]
        10& 45,\:75 & 97\:;\:97\:;\:3& 97\:;\:95\:;\:3& 100\:;\:100\:;\:0& 97\:;\:97\:;\:3& 97\:;\:95\:;\:3& 80\:;\:80\:;\:20& 83\:;\:83\:;\:17\\
        
        ($n = 300$) & 100,\:100 & 95\:;\:95\:;\:5& 98\:;\:98\:;\:2& 100\:;\:99\:;\:0& 95\:;\:95\:;\:5& 98\:;\:98\:;\:2& 83\:;\:83\:;\:17& 76\:;\:76\:;\:24\\
        
        & 180,\:45 & 96\:;\:96\:;\:4& 99\:;\:99\:;\:1& 100\:;\:98\:;\:0& 96\:;\:96\:;\:4& 99\:;\:99\:;\:1& 85\:;\:85\:;\:15& 85\:;\:85\:;\:15\\
        [6pt]        
        10& 90,\:150 & 98\:;\:98\:;\:2& 99\:;\:99\:;\:1& 100\:;\:100\:;\:0& 98\:;\:98\:;\:2& 99\:;\:99\:;\:1& 85\:;\:85\:;\:15& 84\:;\:84\:;\:16\\
        
        ($n = 600$) & 200,\:200 & 95\:;\:95\:;\:5& 94\:;\:94\:;\:6& 100\:;\:100\:;\:0& 95\:;\:95\:;\:5& 94\:;\:94\:;\:6& 89\:;\:89\:;\:11& 85\:;\:85\:;\:15\\
        
        & 360,\:90 & 95\:;\:94\:;\:5& 98\:;\:98\:;\:2& 100\:;\:99\:;\:0& 95\:;\:94\:;\:5& 98\:;\:98\:;\:2& 86\:;\:86\:;\:14& 87\:;\:87\:;\:13\\
        [6pt]
        11& 45,\:75 & 98\:;\:98\:;\:2& 99\:;\:97\:;\:1& 100\:;\:93\:;\:0& 98\:;\:98\:;\:2& 99\:;\:97\:;\:1& 83\:;\:83\:;\:17& 82\:;\:82\:;\:18\\
        
        ($n = 300$) & 100,\:100 & 98\:;\:98\:;\:2& 96\:;\:96\:;\:4& 100\:;\:100\:;\:0& 98\:;\:98\:;\:2& 96\:;\:96\:;\:4& 84\:;\:84\:;\:16& 84\:;\:84\:;\:16\\
        
        & 180,\:45 & 98\:;\:96\:;\:2& 98\:;\:96\:;\:2& 100\:;\:99\:;\:0& 98\:;\:96\:;\:2& 98\:;\:96\:;\:2& 84\:;\:83\:;\:16& 86\:;\:85\:;\:14\\
        [6pt]        
        11& 90,\:150 & 100\:;\:100\:;\:0& 100\:;\:100\:;\:0& 100\:;\:100\:;\:0& 100\:;\:100\:;\:0& 100\:;\:100\:;\:0& 91\:;\:91\:;\:9& 90\:;\:90\:;\:10\\
        
        ($n = 600$) & 200,\:200 & 94\:;\:94\:;\:6& 94\:;\:94\:;\:6& 100\:;\:100\:;\:0& 100\:;\:83\:;\:0& 100\:;\:81\:;\:0& 85\:;\:85\:;\:15& 85\:;\:85\:;\:15\\
        
        & 360,\:90 & 98\:;\:96\:;\:1& 98\:;\:94\:;\:2& 100\:;\:97\:;\:0& 98\:;\:96\:;\:1& 98\:;\:94\:;\:2& 85\:;\:83\:;\:14& 82\:;\:80\:;\:17\\
        [6pt]
        12& 45,\:75 & 98\:;\:96\:;\:2& 97\:;\:75\:;\:3& 100\:;\:99\:;\:0& 98\:;\:96\:;\:2& 97\:;\:75\:;\:3& 88\:;\:75\:;\:11& 88\:;\:76\:;\:10\\
        
        ($n = 300$) & 100,\:100 & 94\:;\:72\:;\:6& 90\:;\:62\:;\:10& 100\:;\:79\:;\:0& 94\:;\:72\:;\:6& 90\:;\:62\:;\:10& 86\:;\:68\:;\:13& 68\:;\:53\:;\:28\\
        
        & 180,\:45 & 98\:;\:84\:;\:2& 98\:;\:84\:;\:2& 100\:;\:86\:;\:0& 98\:;\:84\:;\:2& 98\:;\:84\:;\:2& 83\:;\:52\:;\:9& 77\:;\:59\:;\:18\\
        [6pt]       
        12& 90,\:150 & 96\:;\:89\:;\:4& 98\:;\:78\:;\:2& 100\:;\:98\:;\:0& 96\:;\:89\:;\:4& 98\:;\:78\:;\:2& 94\:;\:84\:;\:10& 87\:;\:81\:;\:13\\
        
        ($n = 600$) & 200,\:200 & 96\:;\:78\:;\:4& 90\:;\:71\:;\:10& 100\:;\:85\:;\:0& 96\:;\:78\:;\:4& 90\:;\:71\:;\:10& 86\:;\:72\:;\:12& 84\:;\:71\:;\:13\\
        
        & 360,\:90 & 97\:;\:90\:;\:3& 97\:;\:89\:;\:3& 100\:;\:93\:;\:0& 97\:;\:90\:;\:3& 97\:;\:89\:;\:3& 84\:;\:66\:;\:12& 85\:;\:68\:;\:10\\
        \hline
        \end{tabular}
        }

\end{table}

\subsection{Real Data Analysis}
In this section, we apply the proposed algorithms and the competing algorithms considered in the previous section on a real dataset. We consider the Central England Temperature (CET) data, which consists of 251 years of average daily temperatures in Central England spanning from the years 1772 to 2022. To ensure a balanced sample, data corresponding to February 29th in the leap years has been excluded and no additional transformations have been made to the dataset. This dataset was kindly provided to us by the authors of \citet{BHT2025}.  The data can be viewed as $251$ curves with $365$ measurements on each curve. \citet{berkes2009detecting} and \citet{chen2023greedy} also analyzed this dataset in their papers. \citet{berkes2009detecting} detects changes at the years $1808$, $1850$, $1926$ and $1992$ while \citet{chen2023greedy} identifies changes at the years $1896$ and $1987$. The method in the latter paper is denoted by \textbf{GS} in this section. We first applied Algorithm DESC-U on the full data and it identifies two changepoints at the years $1897$ and $1988$. This finding aligns with the results reported by \citet{chen2023greedy}. Figure 1 in the Supplementary Material shows the median temperature curves for the three segments in the full data obtained by Algorithm DESC-U, namely, $1772-1896$, $1897-1987$ and $1988-2022$. The value of the median curve at each day is obtained by taking the median of the temperatures on that day for all years in a given segment. The plot shows an increasing trend  in the overall annual temperature. We have also applied the other competing algorithms on this dataset and the estimated changepoints of all of the above methods considered are shown in Table \ref{competitors on CET-v1}. It is seen that the HK method estimates the same set of changepoints as Algorithm DESC-U. This table also illustrates that all methodologies have an estimated changepoint around the year $1990$ (ranging from $1987$ to $1993$), which can perhaps to be attributed to the rapid global warming observed during the 1990's. Indeed, this supports the assertion made in the ``State of the World $1989$'' report by WorldWatch (see \citet{koutaissoff1989state}), which suggested that the $90$'s would be a crucial decade for climate change. 



\begin{table}	\caption{Changepoints detected by different methods on the full dataset.}
	\label{competitors on CET-v1}    	
    \renewcommand{\arraystretch}{1.5}
	\resizebox{\linewidth}{!}{
		\begin{tabular} {@{}lccccccc@{}}
			\hline
			\textbf{Methods}  &\textbf{DESC-U}  &\textbf{GS} &\textbf{BGHK} &\textbf{change\textunderscore FF} &\textbf{fmci} &\textbf{EJS} &\textbf{HK}
			\\
			\hline
			\textbf{Segments} & 1897,1988 & 1896,1987 & 1808, 1850, 1926, 1992 & 1843, 1920, 1989 & 1845, 1987 & 1928, 1993 & 1897, 1988
			\\
			\hline
		\end{tabular}
	}

\end{table}

Temperature variations are not consistent throughout the seasons. Elements such as solar radiation, atmospheric circulation, and land surface dynamics can affect each season in distinct ways. Human activities, including alterations in land use or localized/seasonal pollution, may exert a more significant influence during certain seasons. Focusing solely on annual data may overlook alterations in the seasonal cycle itself, such as an extension of summer or a reduction of winter, even if the overall annual temperature change is minimal. By examining seasonal data, it becomes possible to identify changepoints that may be unique to a specific season, which could be obscured when analyzing annual averages. Consequently, we conduct a changepoint analysis on seasonal temperature data from central England to provide a more comprehensive and nuanced insight into the evolving temperature patterns over time. The other (more technical) advantage of doing such an analysis is that one can treat the seasonal curves across years as nearly independent since they are quite separated in time as opposed to the annual temperature curves which are better treated as a functional time series. Note that our algorithm and its theoretical properties hinge on the independence of the time-ordered data and hence splitting the data as above would help in justifying the performance of our method. 
\par 
Central England experiences four distinct seasons: Spring (March to May), Summer (June to August), Autumn (September to November), and Winter (December to February). Spring brings sudden showers and blossoming plants, while summer is the warmest, featuring long sunny days and occasional heatwaves. Autumn can be mild and dry or wet and windy, with leaves changing color. Winter is the coldest, with freezing temperatures, icy conditions, and sometimes snow. Although there are officially four seasons in Central England, one year can be partitioned into two parts : one consisting of Summer and Autumn, the other consisting of Winter and Spring (See \url{https://weather.metoffice.gov.uk/learn-about/met-office-for-schools/other-content/other-resources/our-seasons}).  
\par 
For our analysis, we first divide every year in the data into two groups: one comprising Summer and Autumn, and the other combining Winter and Spring. We then examine whether there are significant temperature variations within these two groups individually. In the first group, which includes Summer and Autumn, Algorithm DESC-U detects a changepoint at $1928$ which is consistent with the results presented by \citet{berkes2009detecting}, where the changepoint is identified around $1926$. This estimated year aligns with the Early Twentieth Century Warming phenomenon (see \citet{hegerl2018early}) which “\textit{still defies full explanation}” as mentioned in \citet{bronnimann2009early}. This early warming phenomenon is a well-documented occurrence that partially overlaps with the notable warming of the Arctic (see \citet{bengtsson2004early}). In this same group, Algorithm DESC-U detects one more changepoint at the year $1993$ which corresponds to the advent of rapid global warming. For the second group comprising of Winter and Spring, Algorithm DESC-U identifies two changepoints. The first estimated changepoint is $1851$, which can be attributed to the human impact of the Industrial Revolution, aligning closely with the $1850$ estimate presented by \citet{berkes2009detecting}. Additionally, there is significant evidence of a changepoint occurring in $1988$ which is also estimated by Algorithm DESC-U, marking the onset of rapid global warming. 
\begin{figure}[h]
\centering
  \begin{subfigure}{.45\linewidth}
  \hspace*{-4cm}
\includegraphics[scale=0.25]{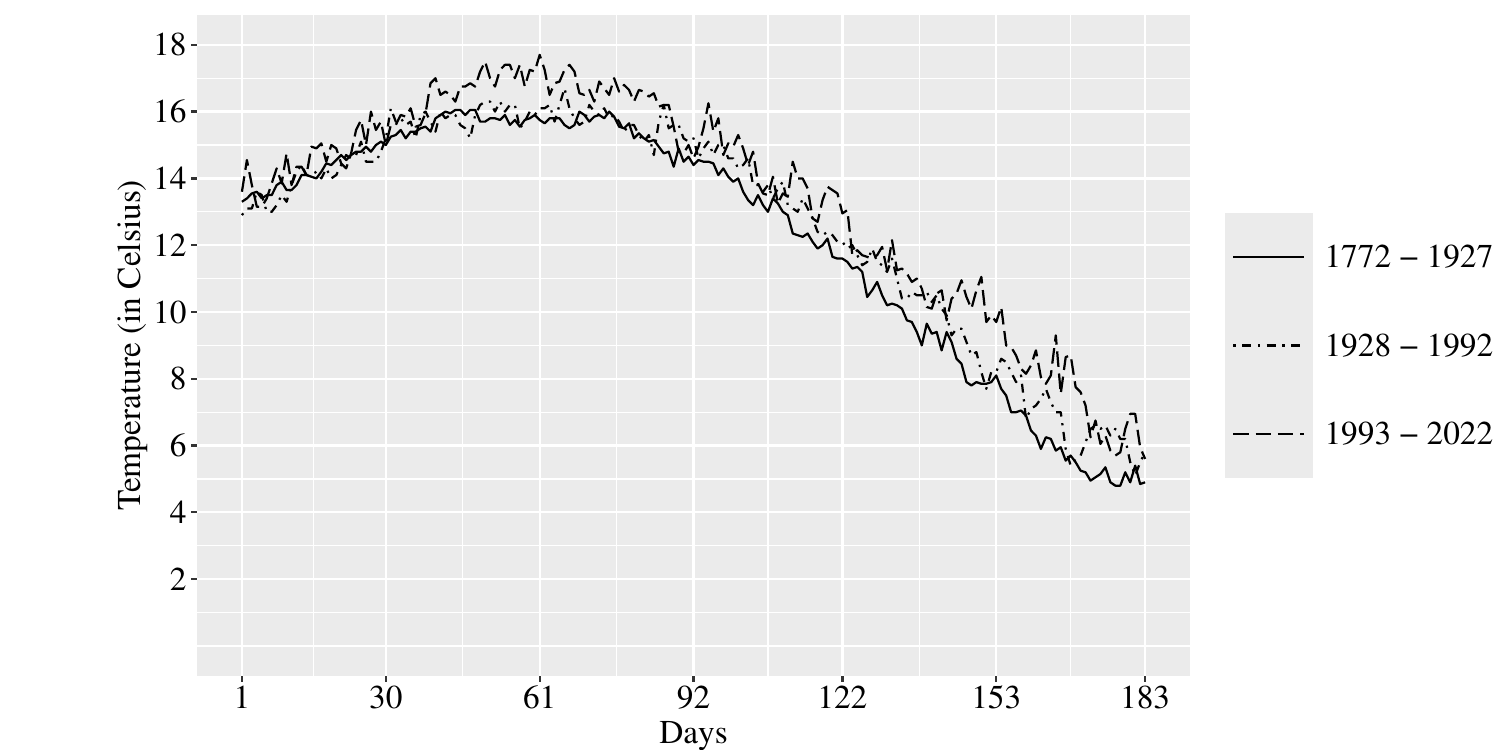}
  \end{subfigure}
  \hspace*{-3cm}
  \begin{subfigure}{.25\linewidth}
\includegraphics[scale=0.25]{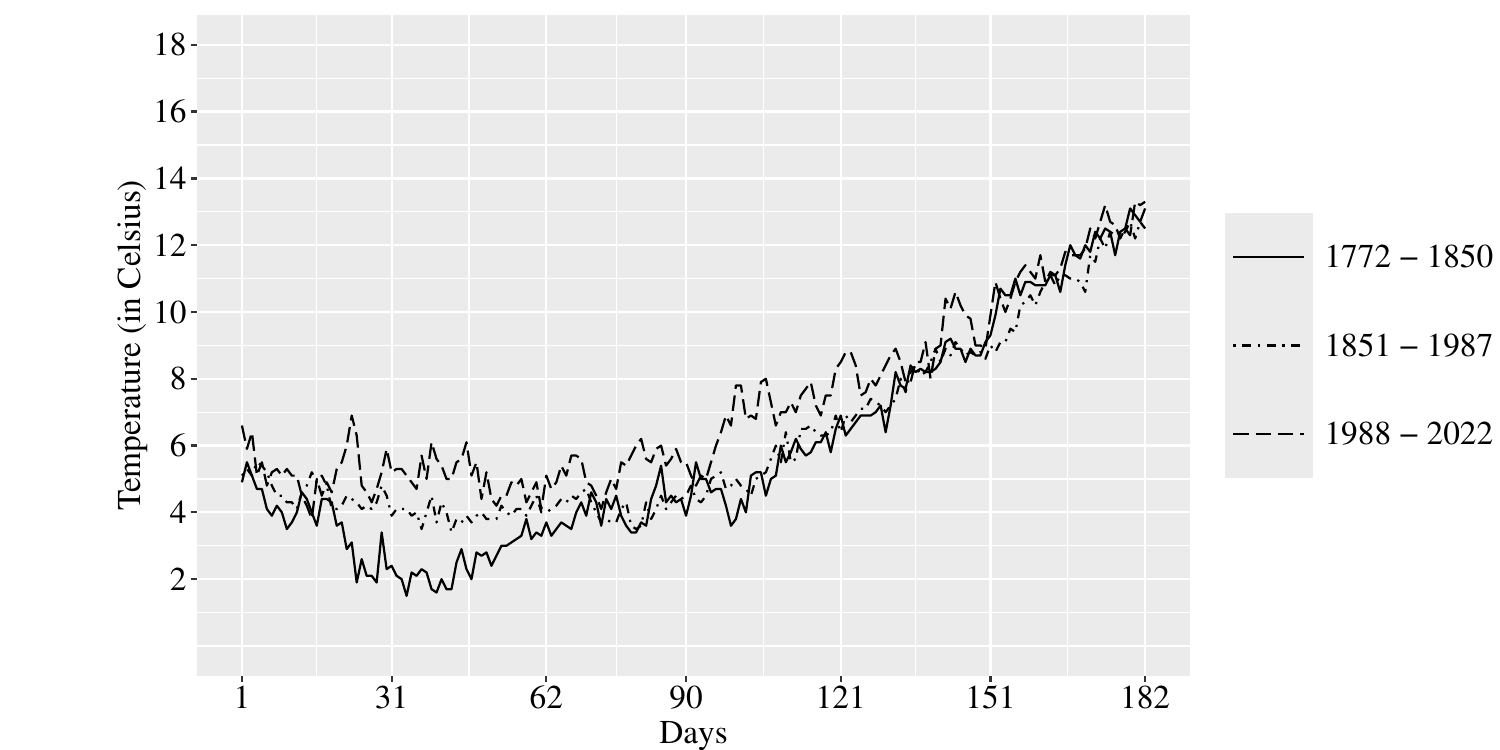}
  \end{subfigure}
  \caption{Median temperature curves for the segments in the CET data obtained by Algorithm DESC-U for each of the groups Summer + Autumn (left panel) and Winter + Spring (right panel).}
  \label{fig:DESC-U_two_parts}
  \end{figure}
Figure \ref{fig:DESC-U_two_parts} shows the median temperature curves for the segments obtained by Algorithm DESC-U and clearly demonstrates the warming trend in the two parts of data separately. We have also carried out the analysis for each of the four seasons separately. The estimated changepoints are reported in Table \ref{competitors on CET} (see the last four columns). It is observed that the estimated changepoints, with the exception of the year $1937$, match with those obtained either from the analysis of the full data or the analysis of the seasonal grouped data done earlier (see the first three columns in Table \ref{competitors on CET}). In Figure \ref{fig:DESC-U_every_season}, we plot the year-wise median curves for the segments obtained by Algorithm DESC-U for each of the four seasons separately. Figure \ref{fig:DESC-U_every_season} indicates that all of the estimated seasonal changepoints are significant. This is due to the fact that the plots in Figure \ref{fig:DESC-U_every_season} clearly show an increase in the median temperature in each post-change segment. Furthermore, this increase in the median temperature occurs on almost all days of every season. We also note in passing that our deep-dive analysis (both using the full dataset or the seasonally segmented data) does not find any evidence of a changepoint around the year $1808$, which was reported in \citet{berkes2009detecting}.  
  
\begin{table}
	\caption{Changepoints detected by Algorithm DESC-U.}
	\label{competitors on CET}	\resizebox{\linewidth}{!}{
		\begin{tabular} {@{}ccccccc@{}}
			\hline
			\textbf{Full data}  &\textbf{Summer + Autumn}  &\textbf{Winter + Spring} &\textbf{Summer} &\textbf{Autumn} &\textbf{Winter} &\textbf{Spring}
			\\
			\hline
			1897, 1988& 1928, 1993& 1851, 1988&  1988& 1937, 1993& 1897& 1988
			\\
			\hline
		\end{tabular}
	}

\end{table}

\begin{figure}[h!] 
    \centering
   \includegraphics[scale=0.3]{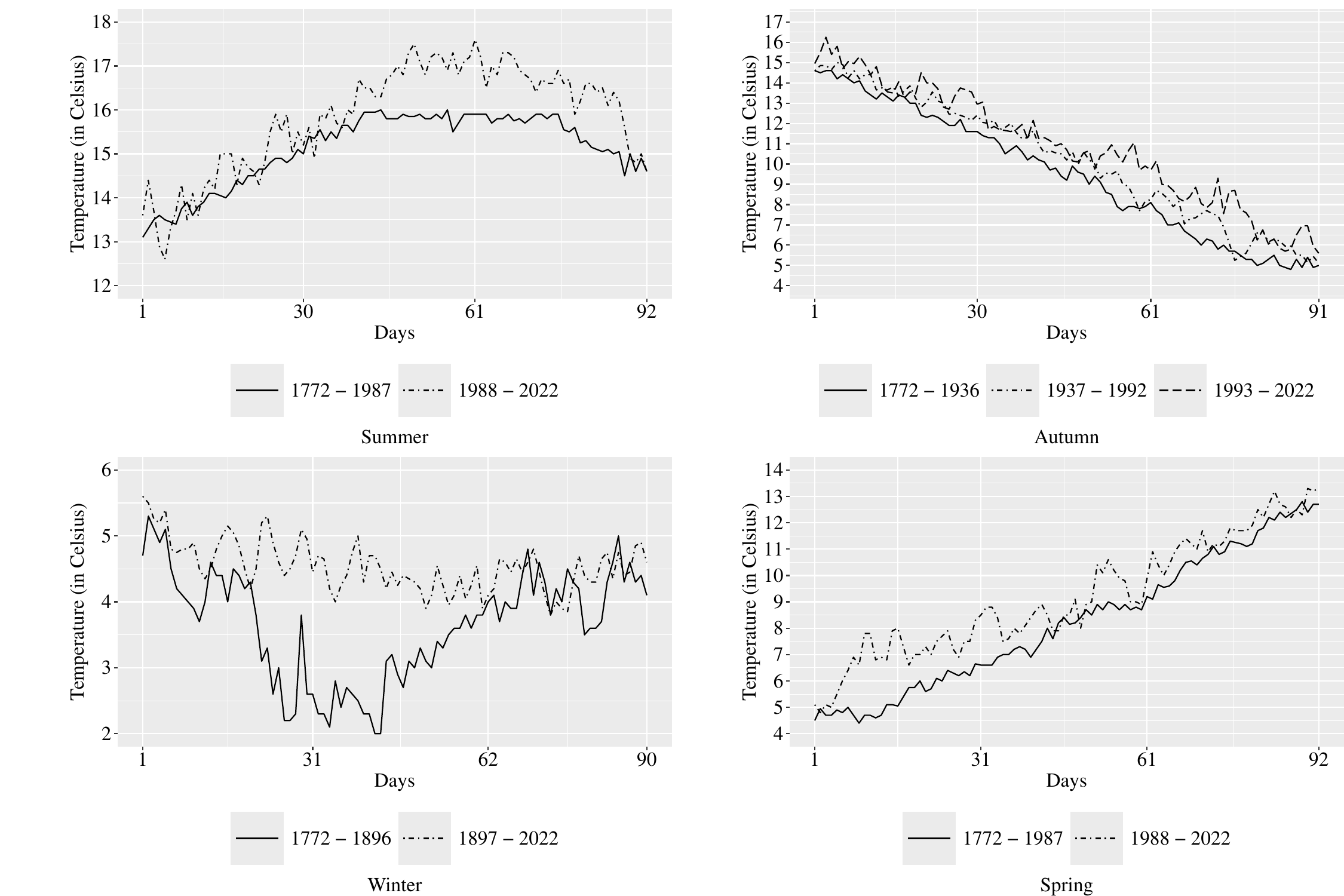}
    \caption{Median temperature curves for each of the segments obtained by Algorithm DESC-U for the four seasons in the CET data.}
    \label{fig:DESC-U_every_season}
\end{figure}

We have also implemented Algorithm DESC-SS on the above dataset with the bounds $K_l=2$ and $K_u=4$ on the number of changepoints. We observed that Algorithm DESC-SS identifies changes in the same years as Algorithm DESC-U for both the full data as well as the seasonally segmented data. This similarity is also observed even when exactly one of the two bounds is used.

\section{Concluding Remarks} \label{conclusion}
In this paper, we develop novel nonparametric changepoint algorithms for detecting distributional change in each of the three scenarios: (i) when the number of changepoints is unknown, (ii) when it is pre-specified, and (iii) when upper and/or lower bounds are specified. The last two scenarios have hardly been considered in the literature on functional changepoint detection.
In addition, we provide theoretical guarantees for the two algorithms in these scenarios, including a new notion of efficacy (the "faithful" property) applicable to any changepoint detection algorithm.

Although the numerical analysis focuses on univariate functional data, the algorithms proposed in the paper can also be applied to multivariate functional data. Indeed, these changepoint detection algorithms are suited for data in any metric space. The theoretical framework that ensures consistent changepoint detection only requires the kernel to be characteristic with respect to that metric space. This is unlike the classical CUSUM-type methods reviewed in the Section \ref{Introduction} which need the linear structure of the space to be even defined and thus cannot be used for data in non-linear metric spaces.

The permutation test in the proposed algorithms may not be appropriate when there is lack of exchangeability in the data, for instance, if there is some form of temporal dependence. Although the numerical study in this paper uses dense regular grids for the functional data, we have observed that in case of moderately sparse regular grids, the proposed algorithms are also quite effective. However, the performance deteriorates if the grid is very sparse. In case the functional data is observed over an irregular grid or is partially observed across the domain, appropriate modifications of the proposed algorithms will be necessary. These topics are earmarked for future research.

\section*{Acknowledments}
The authors are grateful to Prof. Lorenzo Trapani, Prof. Lajos Horvath and Dr. Cooper Boniece for kindly providing access to the real dataset used in this study.    


\section*{Funding}
The research of Anirvan Chakraborty is supported by the MATRICS grant bearing number MTR/2021/000211 provided by the Science and Engineering Research Board, Government of India.

\bibliographystyle{plainnat}

\bibliography{Ref}

@inproceedings{NIPS2008_d07e70ef,
 author = {Fukumizu, Kenji and Gretton, Arthur and Sch\"{o}lkopf, Bernhard and Sriperumbudur, Bharath K.},
 booktitle = {Advances in Neural Information Processing Systems},
 editor = {D. Koller and D. Schuurmans and Y. Bengio and L. Bottou},
 pages = {},
 publisher = {Curran Associates, Inc.},
 title = {Characteristic Kernels on Groups and Semigroups},
 url = {https://proceedings.neurips.cc/paper_files/paper/2008/file/d07e70efcfab08731a97e7b91be644de-Paper.pdf},
 volume = {21},
 year = {2008}
}

@article{sriperumbudur2010hilbert,
  title={Hilbert space embeddings and metrics on probability measures},
  author={Sriperumbudur, Bharath K and Gretton, Arthur and Fukumizu, Kenji and Sch{\"o}lkopf, Bernhard and Lanckriet, Gert RG},
  journal={The Journal of Machine Learning Research},
  volume={11},
  pages={1517--1561},
  year={2010},
  publisher={JMLR. org}
}

@article{gretton2006kernel,
  title={A kernel method for the two-sample-problem},
  author={Gretton, Arthur and Borgwardt, Karsten and Rasch, Malte and Sch{\"o}lkopf, Bernhard and Smola, Alex},
  journal={Advances in neural information processing systems},
  volume={19},
  year={2006}
}

@article{ziegel2024characteristic,
  title={Characteristic kernels on Hilbert spaces, Banach spaces, and on sets of measures},
  author={Ziegel, Johanna and Ginsbourger, David and D{\"u}mbgen, Lutz},
  journal={Bernoulli},
  volume={30},
  number={2},
  pages={1441--1457},
  year={2024},
  publisher={Bernoulli Society for Mathematical Statistics and Probability}
}

@inproceedings{NIPS2009_685ac8ca,
 author = {Fukumizu, Kenji and Gretton, Arthur and Lanckriet, Gert and Sch\"{o}lkopf, Bernhard and Sriperumbudur, Bharath K.},
 booktitle = {Advances in Neural Information Processing Systems},
 editor = {Y. Bengio and D. Schuurmans and J. Lafferty and C. Williams and A. Culotta},
 pages = {},
 publisher = {Curran Associates, Inc.},
 title = {Kernel Choice and Classifiability for RKHS Embeddings of Probability Distributions},
 url = {https://proceedings.neurips.cc/paper_files/paper/2009/file/685ac8cadc1be5ac98da9556bc1c8d9e-Paper.pdf},
 volume = {22},
 year = {2009}
}

@article{berkes2009detecting,
  title={Detecting changes in the mean of functional observations},
  author={Berkes, Istv{\'a}n and Gabrys, Robertas and Horv{\'a}th, Lajos and Kokoszka, Piotr},
  journal={Journal of the Royal Statistical Society Series B: Statistical Methodology},
  volume={71},
  number={5},
  pages={927--946},
  year={2009},
  publisher={Oxford University Press}
}

@article{aue2009estimation,
  title={Estimation of a change-point in the mean function of functional data},
  author={Aue, Alexander and Gabrys, Robertas and Horv{\'a}th, Lajos and Kokoszka, Piotr},
  journal={Journal of Multivariate Analysis},
  volume={100},
  number={10},
  pages={2254--2269},
  year={2009},
  publisher={Elsevier}
}

@article{aston2012detecting,
  title={Detecting and estimating changes in dependent functional data},
  author={Aston, John AD and Kirch, Claudia},
  journal={Journal of Multivariate Analysis},
  volume={109},
  pages={204--220},
  year={2012},
  publisher={Elsevier}
}

@article{44e75106888d4236b3935cb41db7bc34,
title = {Testing the structural stability of temporally dependent functional observations and application to climate projections},
author = {Xianyang Zhang and Xiaofeng Shao and Katharine Hayhoe and Wuebbles, Donald J.},
year = {2011},
volume = {5},
pages = {1765--1796},
journal = {Electronic Journal of Statistics},
issn = {1935-7524},
publisher = {Institute of Mathematical Statistics},
}

@article{hormann2010weakly,
  title={Weakly Dependent Functional Data},
  author={H{\"o}rmann, Siegfried and Kokoszka, Piotr},
  journal={The Annals of Statistics},
  volume={38},
  number={3},
  pages={1845--1884},
  year={2010}
}

@article{sharipov2016sequential,
  title={Sequential block bootstrap in a Hilbert space with application to change point analysis},
  author={Sharipov, Olimjon and Tewes, Johannes and Wendler, Martin},
  journal={Canadian Journal of Statistics},
  volume={44},
  number={3},
  pages={300--322},
  year={2016},
  publisher={Wiley Online Library}
}

@article{gromenko2017detection,
  title={Detection of change in the spatiotemporal mean function},
  author={Gromenko, Oleksandr and Kokoszka, Piotr and Reimherr, Matthew},
  journal={Journal of the Royal Statistical Society Series B: Statistical Methodology},
  volume={79},
  number={1},
  pages={29--50},
  year={2017},
  publisher={Oxford University Press}
}

@article{aue2018detecting,
  title={Detecting and dating structural breaks in functional data without dimension reduction},
  author={Aue, Alexander and Rice, Gregory and S{\"o}nmez, Ozan},
  journal={Journal of the Royal Statistical Society Series B: Statistical Methodology},
  volume={80},
  number={3},
  pages={509--529},
  year={2018},
  publisher={Oxford University Press}
}

@article{li2021bayesian,
  title={Bayesian change point detection for functional data},
  author={Li, Xiuqi and Ghosal, Subhashis},
  journal={Journal of Statistical Planning and Inference},
  volume={213},
  pages={193--205},
  year={2021},
  publisher={Elsevier}
}

@article{chiou2019identifying,
  title={Identifying multiple changes for a functional data sequence with application to freeway traffic segmentation},
  author={Chiou, Jeng-Min and Chen, Yu-Ting and Hsing, Tailen},
  journal={The Annals of Applied Statistics},
  volume={13},
  number={3},
  pages={1430--1463},
  year={2019},
  publisher={JSTOR}
}

@article{harris2022scalable,
  title={Scalable multiple changepoint detection for functional data sequences},
  author={Harris, Trevor and Li, Bo and Tucker, J Derek},
  journal={Environmetrics},
  volume={33},
  number={2},
  pages={e2710},
  year={2022},
  publisher={Wiley Online Library}
}

@article{chen2023greedy,
  title={Greedy segmentation for a functional data sequence},
  author={Chen, Yu-Ting and Chiou, Jeng-Min and Huang, Tzee-Ming},
  journal={Journal of the American Statistical Association},
  volume={118},
  number={542},
  pages={959--971},
  year={2023},
  publisher={Taylor \& Francis}
}

@article{aue2020structural,
  title={Structural break analysis for spectrum and trace of covariance operators},
  author={Aue, Alexander and Rice, Gregory and S{\"o}nmez, Ozan},
  journal={Environmetrics},
  volume={31},
  number={1},
  pages={e2617},
  year={2020},
  publisher={Wiley Online Library}
}

@article{horvath2022change,
  title={Change point analysis of covariance functions: A weighted cumulative sum approach},
  author={Horv{\'a}th, Lajos and Rice, Gregory and Zhao, Yuqian},
  journal={Journal of Multivariate Analysis},
  volume={189},
  pages={104877},
  year={2022},
  publisher={Elsevier}
}

@article{hoeffding1963probability,
  title={Probability inequalities for sums of bounded random variables},
  author={Hoeffding, Wassily},
  journal={Journal of the American statistical association},
  volume={58},
  number={301},
  pages={13--30},
  year={1963},
  publisher={Taylor \& Francis}
}

@article{dette2021detecting,
  title={Detecting structural breaks in eigensystems of functional time series},
  author={Dette, Holger and Kutta, Tim},
  journal={Electronic Journal of Statistics},
  volume={15},
  pages={944--983},
  year={2021}
}

@article{bastian2024multiple,
  title={Multiple change point detection in functional data with applications to biomechanical fatigue data},
  author={Bastian, Patrick and Basu, Rupsa and Dette, Holger},
  journal={The Annals of Applied Statistics},
  volume={18},
  number={4},
  pages={3109--3129},
  year={2024},
  publisher={Institute of Mathematical Statistics}
}

@misc{hegerl2018early,
  title={The early 20th century warming: anomalies, causes, and consequences. WIREs Climate Change 9 (4): e522},
  author={Hegerl, GC and Br{\"o}nnimann, S and Schurer, A and Cowan, T},
  year={2018}
}

@article{bengtsson2004early,
  title={The early twentieth-century warming in the Arctic—A possible mechanism},
  author={Bengtsson, Lennart and Semenov, Vladimir A and Johannessen, Ola M},
  journal={Journal of Climate},
  volume={17},
  number={20},
  pages={4045--4057},
  year={2004}
}

@article{bronnimann2009early,
  title={Early twentieth-century warming},
  author={Br{\"o}nnimann, Stefan},
  journal={Nature Geoscience},
  volume={2},
  number={11},
  pages={735--736},
  year={2009},
  publisher={Nature Publishing Group UK London}
}

@article{koutaissoff1989state,
  title={The State of the World 1989, by Lester Brown et al.},
  author={Koutaissoff, Elisabeth},
  journal={Environmental Conservation},
  volume={16},
  number={2},
  pages={190--190},
  year={1989},
  publisher={Cambridge University Press}
}

@inproceedings{padilla2022,
author = {Padilla, Carlos Misael Madrid and Zhao, Zifeng and Wang, Daren and Yu, Yi},
title = {Change-point detection for sparse and dense functional data in general dimensions},
year = {2022},
isbn = {9781713871088},
publisher = {Curran Associates Inc.},
address = {Red Hook, NY, USA},
booktitle = {Proceedings of the 36th International Conference on Neural Information Processing Systems},
articleno = {2690},
numpages = {13},
location = {New Orleans, LA, USA},
series = {NIPS '22}
}

@article{BHT2025,
	title = {On changepoint detection in functional data using empirical energy distance},
	journal = {Journal of Econometrics},
	volume = {250},
	pages = {106023},
	year = {2025},
	issn = {0304-4076},
	url = {https://www.sciencedirect.com/science/article/pii/S0304407625000776},
	author = {B. Cooper Boniece and Lajos Horváth and Lorenzo Trapani}
}

@article{matteson2014nonparametric,
  title={A nonparametric approach for multiple change point analysis of multivariate data},
  author={Matteson, David S and James, Nicholas A},
  journal={Journal of the American Statistical Association},
  volume={109},
  number={505},
  pages={334--345},
  year={2014},
  publisher={Taylor \& Francis}
}

@article{varadarajan1958convergence,
  title={On the convergence of sample probability distributions},
  author={Varadarajan, Veeravalli S},
  journal={Sankhy{\=a}: The Indian Journal of Statistics (1933-1960)},
  volume={19},
  number={1/2},
  pages={23--26},
  year={1958},
  publisher={JSTOR}
}

@article{harchaoui2008kernel,
  title={Kernel change-point analysis},
  author={Harchaoui, Zaid and Moulines, Eric and Bach, Francis},
  journal={Advances in neural information processing systems},
  volume={21},
  year={2008}
}

@article{zhang2025change,
  title={Change-Point Detection for Object-valued Time Series},
  author={Zhang, Yi and Zhu, Changbo and Shao, Xiaofeng},
  journal={Journal of Business \& Economic Statistics},
  number={accepted},
  pages={1--23},
  year={2025},
  publisher={Taylor \& Francis}
}

@article{jiang2024two,
  title={Two-sample and change-point inference for non-Euclidean valued time series},
  author={Jiang, Feiyu and Zhu, Changbo and Shao, Xiaofeng},
  journal={Electronic Journal of Statistics},
  volume={18},
  number={1},
  pages={848--894},
  year={2024},
  publisher={The Institute of Mathematical Statistics and the Bernoulli Society}
}

@article{dubey2020frechet,
  title={Fr{\'e}chet change-point detection},
  author={Dubey, Paromita and M{\"u}ller, Hans-Georg},
  journal={The Annals of Statistics},
  volume={48},
  number={6},
  pages={3312--3335},
  year={2020},
  publisher={JSTOR}
}

@article{song2024practical,
  title={Practical and powerful kernel-based change-point detection},
  author={Song, Hoseung and Chen, Hao},
  journal={IEEE Transactions on Signal Processing},
  year={2024},
  publisher={IEEE}
}

@inproceedings{harchaoui2007retrospective,
  title={Retrospective mutiple change-point estimation with kernels},
  author={Harchaoui, Zaid and Capp{\'e}, Olivier},
  booktitle={2007 IEEE/SP 14th Workshop on Statistical Signal Processing},
  pages={768--772},
  year={2007},
  organization={IEEE}
}

@article{truong2020selective,
  title={Selective review of offline change point detection methods},
  author={Truong, Charles and Oudre, Laurent and Vayatis, Nicolas},
  journal={Signal Processing},
  volume={167},
  pages={107299},
  year={2020},
  publisher={Elsevier}
}

@article{xu2025change,
  title={Change-point detection with deep learning: A review},
  author={Xu, Ruiyu and Song, Zheren and Wu, Jianguo and Wang, Chao and Zhou, Shiyu},
  journal={Frontiers of Engineering Management},
  volume={12},
  number={1},
  pages={154--176},
  year={2025},
  publisher={Springer}
}

@article{chen2023graph,
  title={Graph-based change-point analysis},
  author={Chen, Hao and Chu, Lynna},
  journal={Annual Review of Statistics and Its Application},
  volume={10},
  number={1},
  pages={475--499},
  year={2023},
  publisher={Annual Reviews}
}

@article{arlot2019kernel,
  title={A kernel multiple change-point algorithm via model selection},
  author={Arlot, Sylvain and Celisse, Alain and Harchaoui, Zaid},
  journal={Journal of machine learning research},
  volume={20},
  number={162},
  pages={1--56},
  year={2019}
}

@article{garreau2018consistent,
  title={Consistent change-point detection with kernels},
  author={Garreau, Damien and Arlot, Sylvain},
  journal={Electronic Journal of Statistics},
  volume={12},
  number={2},
  pages={4440--4486},
  year={2018}
}

@article{horvath2025change,
  title={Change point analysis for functional data using empirical characteristic functionals},
  author={Horv{\'a}th, Lajos and Rice, Gregory and VanderDoes, Jeremy},
  journal={Journal of Time Series Analysis},
  year={2025},
  publisher={Wiley Online Library}
}

@article{ramsay2025robust,
  title={Robust changepoint detection in the variability of multivariate functional data},
  author={Ramsay, Kelly and Chenouri, Shoja'eddin},
  journal={Journal of Nonparametric Statistics},
  pages={1--22},
  year={2025},
  publisher={Taylor \& Francis}
}

@article{chu2019asymptotic,
  title={Asymptotic distribution-free change-point detection for multivariate and non-Euclidean data},
  author={Chu, Lynna and Chen, Hao},
  journal={The Annals of Statistics},
  volume={47},
  number={1},
  pages={382--414},
  year={2019},
  publisher={JSTOR}
}

@article{dai2019discovering,
  title={Discovering common change-point patterns in functional connectivity across subjects},
  author={Dai, Mengyu and Zhang, Zhengwu and Srivastava, Anuj},
  journal={Medical image analysis},
  volume={58},
  pages={101532},
  year={2019},
  publisher={Elsevier}
}

@inproceedings{kostic2025change,
  title={Change Point Detection in Hadamard Spaces by Alternating Minimization},
  author={Kostic, Anica and Runge, Vincent and Truong, Charles},
  booktitle={International Conference on Artificial Intelligence and Statistics},
  pages={1963--1971},
  year={2025},
  organization={PMLR}
}

@inproceedings{wang2024non,
  title={Non-parametric online change point detection on Riemannian manifolds},
  author={Wang, Xiuheng and Borsoi, Ricardo Augusto and Richard, C{\'e}dric},
  booktitle={41st International Conference on Machine Learning, ICML 2024},
  year={2024}
}

@inproceedings{wang2023online,
  title={Online change point detection on Riemannian manifolds with Karcher mean estimates},
  author={Wang, Xiuheng and Borsoi, Ricardo Augusto and Richard, C{\'e}dric},
  booktitle={2023 31st European Signal Processing Conference (EUSIPCO)},
  pages={2033--2037},
  year={2023},
  organization={IEEE}
}

\section*{Supplementary Material} \label{supplementary}

The supplementary material contains the proofs of all the technical results and some figures and tables from the data analyses.

\section*{ Proofs of Mathematical Results}

\subsection*{ Proof of Theorem 3.1}
Suppose \{$Z_1,Z_2,\dots,Z_n$\} is a set of i.i.d. observations from distribution $F$. Let $\widetilde \gamma \in [\delta_{n}, 1 - \delta_{n}]$ where \{$\delta_n$\} is a sequence of positive numbers such that $\delta_n \to 0$ and $n\delta_n \to \infty$ as $n \to \infty$. Split the observations into two sets 
\begin{equation*}
	G_{n}(\widetilde{\gamma}) = \left\{Z_1,Z_2,\dots,Z_{\lfloor n\widetilde{\gamma} \rfloor}\right\} \: \: \text{and} \: \: H_{n}(\widetilde{\gamma}) = \left\{Z_{\lfloor n\widetilde{\gamma} \rfloor+1},Z_{\lfloor n\widetilde{\gamma} \rfloor+2},\dots,Z_{n}\right\}. 
\end{equation*}
Let $\widetilde{r} = \lfloor n\widetilde{\gamma} \rfloor$. Then
\begin{ceqn}
	\begin{center}
		$d_{n}(\widetilde{\gamma}) = \frac{1}{\widetilde{r}^2}\sum\limits_{i=1}^{\widetilde{r}}\sum\limits_{i'=1}^{\widetilde{r}}k(Z_{i},Z_{i'}) + \frac{1}{(n - \widetilde{r})^2}\sum\limits_{j=\widetilde{r}+1}^{n}\sum\limits_{j'=\widetilde{r}+1}^{n}k(Z_{j},Z_{j'}) - \frac{2}{\widetilde{r}(n - \widetilde{r})}\sum\limits_{i=1}^{\widetilde{r}}\sum\limits_{j=\widetilde{r}+1}^{n}k(Z_{i},Z_{j})$.
	\end{center}
\end{ceqn}
{Let $\mathbb{N}$ denote the set of natural numbers. Using the  strong law of large numbers for $V$-statistics, for every $\epsilon > 0$, there exists $N_1=N_1(\epsilon) \in \mathbb{N}$ such that
\begin{ceqn}
	\begin{center}
		$\left| \frac{1}{{r}^{2}}\sum\limits_{i=1}^{r}\sum\limits_{i'=1}^{r}k(Z_{i},Z_{i'}) - \mu_{k}^{X} \right| < \epsilon$    $[\mathbb{P}]~ a.s.$ whenever $r \geq N_1$,
	\end{center}
\end{ceqn}
where $\mu_{k}^{X}=\mathbb{E}[k(Z,Z')]$ such that $Z,Z' \stackrel{i.i.d.}{\sim} F$.
Since $n\delta_{n} \to \infty$, there exists $N_1^{*}=N_1^{*}(\epsilon) \in \mathbb{N}$ such that $\lfloor n\delta_{n} \rfloor \geq N_1$ whenever $n \geq N_1^{*}$. Since $\widetilde{\gamma} > \delta_{n}$,  $\widetilde{r} \geq \lfloor n\delta_{n} \rfloor \geq N_1$ for $n \geq N_1^{*}$ (a similar line of argument is used in \citet{matteson2014nonparametric}). Thus,
\begin{align} \label{eq28}
	\left| \frac{1}{\widetilde{r}^2}\sum\limits_{i=1}^{\widetilde{r}}\sum\limits_{i'=1}^{\widetilde{r}}k(Z_{i},Z_{i'}) - \mu_{k}^{X} \right| < \epsilon   \: \:  [\mathbb{P}]~a.s. \: \: \text{whenever} \: \: n \geq N_1^{*}.
\end{align}
Analogously, there exist integers $N_2^{*}, N_3^{*} \in \mathbb{N}$ such that
\begin{align} \label{eq29}
	\left| \frac{1}{(n - \widetilde{r})^2}\sum\limits_{j=\widetilde{r}+1}^{n}\sum\limits_{j'=\widetilde{r}+1}^{n}k(Z_{j},Z_{j'}) - \mu_{k}^{X} \right| < \epsilon   \: \:  [\mathbb{P}]~a.s.\: \: \text{whenever} \: \: n \geq N_2^{*}.
\end{align}
\begin{center}
	and
\end{center}
\begin{align} \label{eq3}
	\left| \frac{1}{\widetilde{r}(n - \widetilde{r})}\sum\limits_{i=1}^{\widetilde{r}}\sum\limits_{j=\widetilde{r}+1}^{n}k(Z_{i},Z_{j}) - \mu_{k}^{X} \right| < \epsilon   \: \:  [\mathbb{P}]~a.s.\: \: \text{whenever} \: \: n \geq N_3^{*}.
\end{align}
Now,
\begin{align*}
	d_{n}(\widetilde{\gamma}) & =  \frac{1}{\widetilde{r}^2}\sum\limits_{i=1}^{\widetilde{r}}\sum\limits_{i'=1}^{\widetilde{r}}k(Z_{i},Z_{i'}) + \frac{1}{(n - \widetilde{r})^2}\sum\limits_{j=\widetilde{r}+1}^{n}\sum\limits_{j'=\widetilde{r}+1}^{n}k(Z_{j},Z_{j'})  - \frac{2}{\widetilde{r}(n - \widetilde{r})}\sum\limits_{i=1}^{\widetilde{r}}\sum\limits_{j=\widetilde{r}+1}^{n}k(Z_{i},Z_{j})  \\
    & = \left\{ \frac{1}{\widetilde{r}^2}\sum\limits_{i=1}^{\widetilde{r}}\sum\limits_{i'=1}^{\widetilde{r}}k(Z_{i},Z_{i'}) - \mu_{k}^{X} \right\} + \left\{ \frac{1}{(n - \widetilde{r})^2}\sum\limits_{j=\widetilde{r}+1}^{n}\sum\limits_{j'=\widetilde{r}+1}^{n}k(Z_{j},Z_{j'}) - \mu_{k}^{X} \right\} \\ & \quad - 2\left\{ \frac{1}{\widetilde{r}(n - \widetilde{r})}\sum\limits_{i=1}^{\widetilde{r}}\sum\limits_{j=\widetilde{r}+1}^{n}k(Z_{i},Z_{j}) - \mu_{k}^{X} \right\} \\
	& \leq \left| \frac{1}{\widetilde{r}^2}\sum\limits_{i=1}^{\widetilde{r}}\sum\limits_{i'=1}^{\widetilde{r}}k(Z_{i},Z_{i'}) - \mu_{k}^{X} \right| + \left| \frac{1}{(n - \widetilde{r})^2}\sum\limits_{j=\widetilde{r}+1}^{n}\sum\limits_{j'=\widetilde{r}+1}^{n}k(Z_{j},Z_{j'}) - \mu_{k}^{X} \right| \\ & \quad + 2\left| \frac{1}{\widetilde{r}(n - \widetilde{r})}\sum\limits_{i=1}^{\widetilde{r}}\sum\limits_{j=\widetilde{r}+1}^{n}k(Z_{i},Z_{j}) - \mu_{k}^{X} \right|. 
\end{align*}
Let $N^{*} = \max\{N_1^{*},N_2^{*},N_3^{*}\}$. Then, by equations (\ref{eq28}), (\ref{eq29}) and (\ref{eq3}), we obtain $d_{n}(\widetilde{\gamma}) < 4\epsilon$, $[\mathbb{P}]~a.s.$ for all $n \geq N^{*}$.
Since $\epsilon$ is an arbitrary positive number and $N^{*}$ does not depend on $\widetilde{\gamma}$,
\begin{align} \label{convergence of d under null}
	\underset{{\widetilde{\gamma}} \in [\delta_n, 1-\delta_n]}{\sup} d_{n}(\widetilde{\gamma}) \xrightarrow[]{} 0 \quad [\mathbb{P}]~a.s.
\end{align}
Also, $0 \leq \frac{\lfloor n\widetilde{\gamma} \rfloor(n - \lfloor n\widetilde{\gamma} \rfloor)}{n^{2}} \leq 1$. Hence,
\begin{align} \label{convergence of rho under null}
	\underset{\lceil n\delta_{n} \rceil \leq t \leq \lfloor n(1 - \delta_{n}) \rfloor}{\max} \rho_{n}(t/n) \leq \underset{\widetilde{\gamma} \in [\delta_n, 1-\delta_n]}{\sup}\rho_{n}(\widetilde{\gamma}) &\leq \underset{\widetilde{\gamma} \in [\delta_n, 1-\delta_n]}{\sup}\frac{\lfloor n\widetilde{\gamma} \rfloor(n - \lfloor n\widetilde{\gamma} \rfloor)}{n^{2}}\underset{{\widetilde{\gamma}} \in [\delta_n, 1-\delta_n]}{\sup} d_{n}(\widetilde{\gamma}) \nonumber\\&\leq \underset{{\widetilde{\gamma}} \in [\delta_n, 1-\delta_n]}{\sup} d_{n}(\widetilde{\gamma}).
\end{align}
Therefore, under $H_0$,
\begin{align*} 
	T_{n} = \underset{\lceil n\delta_{n} \rceil \leq t \leq \lfloor n(1 - \delta_{n}) \rfloor}{\max} \rho_{n}(t/n) \xrightarrow[]{} 0 \quad [\mathbb{P}]~a.s.
\end{align*}

\subsection*{ Proof of Lemma 3.2}
Suppose $H_{1\gamma}$ is true for some $\gamma \in (0,1)$, i.e., $Z_{1},\dots,Z_{\lfloor n\gamma \rfloor} \stackrel{i.i.d.}{\sim} F_{1} \  \text{and} \ Z_{\lfloor n\gamma \rfloor+1},\dots,Z_{n}  \stackrel{i.i.d.}{\sim} F_2 \  \text{with} \ F_1 \neq F_2$. Choose any $\widetilde{\gamma} \in [\delta_n, 1 - \delta_n]$ and denote $\lfloor n\widetilde{\gamma}\rfloor$ by $\widetilde{s}$.
Then, $G_{n}(\widetilde{\gamma}) = \left\{Z_1,Z_2,\dots,Z_{\widetilde{s}}\right\}$ and $H_{n}(\widetilde{\gamma}) = \left\{Z_{\widetilde{s}+1},Z_{\widetilde{s}+2},\dots,Z_{n}\right\}$. Thus,
\begin{align*}
	d_{n}(\widetilde{\gamma}) =& \frac{1}{\widetilde{s}^2}\sum\limits_{i=1}^{\widetilde{s}}\sum\limits_{i'=1}^{\widetilde{s}}k(Z_{i},Z_{i'}) + \frac{1}{(n - \widetilde{s})^2}\sum\limits_{j=\widetilde{s} + 1}^{n}\sum\limits_{j'=\widetilde{s} + 1}^{n}k(Z_{j},Z_{j'}) \\&- \frac{2}{\widetilde{s}(n - \widetilde{s})}\sum\limits_{i=1}^{\widetilde{s}}\sum\limits_{j=\widetilde{s}+1}^{n}k(Z_{i},Z_{j}).
\end{align*}
Suppose $\widetilde{\gamma} \leq \gamma$. Then $\widetilde{s} \leq s$ where $s=\lfloor n\gamma \rfloor$. By the strong law of large numbers for $V$-statistics, for every $\epsilon^{*} > 0$, there exists $N_1^{(1)} = N_1^{(1)}(\epsilon^{*})\in \mathbb{N}$ such that
\begin{align*}
	\left|\frac{1}{r^2}\sum\limits_{i=1}^{r}\sum\limits_{i'=1}^{r}k(Z_{i},Z_{i'}) - \mu_{k}^{X}\right| < \epsilon^{*} \: \: \text{whenever}\:\: r > N_1^{(1)} \: \:  [\mathbb{P}]~a.s., 
\end{align*}
where $Z_{1},Z_{2},\dots,Z_{r} \stackrel{i.i.d}{\sim} F_{1}$ and $\mu_{k}^{X}=\mathbb{E}[k(X,X')]$ with $X,X' \stackrel{i.i.d.}{\sim} F_{1}$. Since $n\delta_{n} \xrightarrow[]{} \infty$, there exists $M_1^{(1)} \in \mathbb{N}$ such that $\lfloor n\delta_n  \rfloor > N_1^{(1)}$ whenever $n > M_1^{(1)}$. As $\widetilde{s} \geq \lfloor n\delta_{n}\rfloor$, for $n > M_1^{(1)}$, we have
\begin{align} \label{eq4}
	\left|\frac{1}{\widetilde{s}^2}\sum\limits_{i=1}^{\widetilde{s}}\sum\limits_{i'=1}^{\widetilde{s}}k(Z_{i},Z_{i'}) - \mu_{k}^{X}\right| < \epsilon \: \:  [\mathbb{P}] \: \: a.s. 
\end{align}
Now, suppose $\widetilde{\gamma} > \gamma$. Then, $\widetilde{s} \geq s$. In $G_{n}(\widetilde{\gamma})$, first $s$ observations are from distribution $F_1$, while the remaining ($\widetilde{s} - s$) observations are from distribution $F_2$. 
 By the strong law of large numbers, for any $\epsilon>0$, there exists $N_{2}^{(1)} \in \mathbb{N}$ such that
\begin{align*}
	\left|\frac{1}{s^2}\sum\limits_{i=1}^{s}\sum\limits_{i'=1}^{s}k(Z_{i},Z_{i'}) - \mu_{k}^{X}\right| < \epsilon \:  \: \:  [\mathbb{P}]~a.s. \:\: \text{whenever $s>N_{2}^{(1)}$.} 
\end{align*} 
 Since $\gamma \in (0,1)$ is a fixed quantity, there exists $M_2^{(1)} \in \mathbb{N}$ such that $s = \lfloor n\gamma \rfloor > N_{2}^{(1)}$ whenever $n \geq M_{2}^{(1)}$. Hence, for $n \geq M_{2}^{(1)}$,
\begin{align} \label{eq15}
	\left|\frac{1}{s^2}\sum\limits_{i=1}^{s}\sum\limits_{i'=1}^{s}k(Z_{i},Z_{i'}) - \mu_{k}^{X}\right| < \epsilon \:  \: \:  [\mathbb{P}]~a.s. 
\end{align}

Note that
\begin{align*}
	\left|\left(\frac{s}{\widetilde{s}}\right)^2 - \left(\frac{\gamma}{\widetilde{\gamma}}\right)^2\right| = \left|\left(\frac{s}{\widetilde{s}} - \frac{\gamma}{\widetilde{\gamma}}\right)^2 + \frac{2\gamma}{\widetilde{\gamma}}\left(\frac{s}{\widetilde{s}} - \frac{\gamma}{\widetilde{\gamma}}\right)\right| \leq \left(\frac{s}{\widetilde{s}} - \frac{\gamma}{\widetilde{\gamma}}\right)^{2} + \frac{2\gamma}{\widetilde{\gamma}}\left|\frac{s}{\widetilde{s}} - \frac{\gamma}{\widetilde{\gamma}}\right| \leq \left(\frac{s}{\widetilde{s}} - \frac{\gamma}{\widetilde{\gamma}}\right)^{2} + 2\left|\frac{s}{\widetilde{s}} - \frac{\gamma}{\widetilde{\gamma}}\right|.
\end{align*}
Now, 
\begin{align*}
	\left|\frac{s}{\widetilde{s}} - \frac{\gamma}{\widetilde{\gamma}}\right|  \leq \left|\frac{s}{\widetilde{s}} - \frac{n\gamma}{\widetilde{s}}\right| + \left|\frac{n\gamma}{\widetilde{s}} - \frac{\gamma}{\widetilde{\gamma}}\right|  = \frac{|s - n\gamma|}{\widetilde{s}} + \frac{\gamma}{\widetilde{\gamma} \widetilde{s}}|n\widetilde{\gamma} - \widetilde{s}| \leq \frac{2}{\widetilde{s}}.
\end{align*}
Therefore,
\begin{align*}
	\left|\left(\frac{s}{\widetilde{s}}\right)^2 - \left(\frac{\gamma}{\widetilde{\gamma}}\right)^2\right| \leq 4\left(\frac{1}{\widetilde{s}^2} + \frac{1}{\widetilde{s}}\right) \leq \frac{8}{\widetilde{s}} \leq \frac{8}{\lfloor n\delta_{n}\rfloor}.
\end{align*}
Note that there exists $N_{3}^{(1)} \in \mathbb{N}$ such that $8/r<\epsilon$ whenever $r \geq N_{3}^{(1)}$. Since $n\delta_{n} \xrightarrow[]{} \infty$, there exists $M_2^{(2)} \in \mathbb{N}$ such that $\lfloor n\delta_{n}\rfloor \geq N_{3}^{(1)}$ whenever $n \geq M_{2}^{(2)}$. Hence for $n \geq M_{2}^{(2)}$, 
\begin{align} \label{eq19}
    \left|\left(\frac{s}{\widetilde{s}}\right)^2 - \left(\frac{\gamma}{\widetilde{\gamma}}\right)^2\right| < \epsilon.
\end{align}
Equations \eqref{eq15} and \eqref{eq19} imply that for $n > \max \{M_{2}^{(1)},M_{2}^{(2)}\}$,
\begin{align*}
	\left|\frac{1}{\widetilde{s}^2}\sum\limits_{i=1}^{s}\sum\limits_{i'=1}^{s}k(Z_{i},Z_{i'}) - \left(\frac{\gamma}{\widetilde{\gamma}}\right)^{2}\mu_{k}^{X}\right| &\leq  \left|\left(\frac{s}{\widetilde{s}}\right)^{2} - \left(\frac{\gamma}{\widetilde{\gamma}}\right)^{2}\right|\left|\frac{1}{s^2}\sum\limits_{i=1}^{s}\sum\limits_{i'=1}^{s}k(Z_{i},Z_{i'}) - \mu_{k}^{X}\right| + \mu_{k}^{X}\left|\left(\frac{s}{\widetilde{s}}\right)^{2} - \left(\frac{\gamma}{\widetilde{\gamma}}\right)^{2}\right| \\& \quad + \left(\frac{\gamma}{\widetilde{\gamma}}\right)^{2}\left|\frac{1}{s^2}\sum\limits_{i=1}^{s}\sum\limits_{i'=1}^{s}k(Z_{i},Z_{i'}) - \mu_{k}^{X}\right|\\
    &\leq \epsilon^{2} + \epsilon \mu_{k}^{X} + \epsilon\\
    & < \epsilon^{2}+2\epsilon.
\end{align*}
The last inequality follows due to the boundedness of Gaussian kernel (see Section 7.1 of the main paper). Note that $M_{2}^{(1)}$ and $M_{2}^{(2)}$ are independent of $\widetilde{\gamma}$.
Since $\epsilon$ is arbitrary positive number, for every $\epsilon^{*}>0$, there exists $N_{1}^{*} \in \mathbb{N}$ (which also does not depend on $\widetilde{\gamma}$) such that for $n > N_{1}^{*}$, 
\begin{align} \label{eq5}
	\left|\frac{1}{\widetilde{s}^2}\sum\limits_{i=1}^{\widetilde{s}}\sum\limits_{i'=1}^{\widetilde{s}}k(Z_{i},Z_{i'}) - \left(\frac{\gamma}{\widetilde{\gamma}}\right)^{2}\mu_{k}^{X}\right| < \frac{\epsilon^{*}}{3} \: \: [\mathbb{P}]~a.s.
\end{align}
Similarly, we can find two positive integers $N_2^{*}$ and $N_3^{*}$ such that for $n > N_2^{*}$,
\begin{align} \label{eq6}
	\left|\frac{1}{\widetilde{s}^2}\sum\limits_{i=s+1}^{\widetilde{s}}\sum\limits_{i'=s+1}^{\widetilde{s}}k(Z_{i},Z_{i'}) - \left(\frac{\widetilde{\gamma}-\gamma}{\widetilde{\gamma}}\right)^{2}\mu_{k}^{Y}\right| < \frac{\epsilon^{*}}{3} \: \: [\mathbb{P}]~a.s.
\end{align}
	and for $n > N_3^{*}$,
\begin{align} \label{eq7}
	\left|\frac{1}{\widetilde{s}^2}\sum\limits_{i=1}^{s}\sum\limits_{i'=s+1}^{\widetilde{s}}k(Z_{i},Z_{i'}) - \frac{\gamma(\widetilde{\gamma}-\gamma)}{\widetilde{\gamma}^{2}}\mu_{k}^{XY}\right| < \frac{\epsilon^{*}}{6} \: \: [\mathbb{P}]~a.s.
\end{align}
where $\mu_{k}^{Y}=\mathbb{E}[k(Y,Y')]$ and $\mu_{k}^{XY}=\mathbb{E}[k(X,Y)]$ with $Y,Y' \stackrel{i.i.d.}{\sim} F_{2}$ and $X$ and $Y$ are independent.
Now,
\begin{align*}
	& \left|\frac{1}{\widetilde{s}^{2}}\sum\limits_{i=1}^{\widetilde{s}}\sum\limits_{i'=1}^{\widetilde{s}}k(Z_{i},Z_{i'}) - \left\{\left(\frac{\gamma}{\widetilde{\gamma}}\right)^{2}\mu_{k}^{X} + \left(\frac{\widetilde{\gamma}-\gamma}{\widetilde{\gamma}}\right)^{2}\mu_{k}^{Y} + \frac{2\gamma(\widetilde{\gamma}-\gamma)}{\widetilde{\gamma}^{2}}\mu_{k}^{XY}\right\} \right| \\
	& \leq \left|\frac{1}{\widetilde{s}^{2}}\sum\limits_{i=1}^{s}\sum\limits_{i'=1}^{s}k(Z_{i},Z_{i'}) - \left(\frac{\gamma}{\widetilde{\gamma}}\right)^{2}\mu_{k}^{X}\right| + \left|\frac{1}{\widetilde{s}^{2}}\sum\limits_{i=s+1}^{\widetilde{s}}\sum\limits_{i'=s+1}^{\widetilde{s}}k(Z_{i},Z_{i'}) - \left(\frac{\widetilde{\gamma}-\gamma}{\widetilde{\gamma}}\right)^{2}\mu_{k}^{Y}\right| \\
	& \quad + 2\left|\frac{1}{\widetilde{s}^{2}}\sum\limits_{i=1}^{s}\sum\limits_{i'=s+1}^{\widetilde{s}}k(Z_{i},Z_{i'}) - \frac{\gamma(\widetilde{\gamma}-\gamma)}{\widetilde{\gamma}^2}\mu_{k}^{XY}\right|.
\end{align*}
So, by equations (\ref{eq5}), (\ref{eq6}) and (\ref{eq7}), for $n > \max\{N_1^{*},N_2^{*},N_3^{*}\}$,
\begin{align} \label{eq8}
	\left|\frac{1}{\widetilde{s}^{2}}\sum\limits_{i=1}^{\widetilde{s}}\sum\limits_{i'=1}^{\widetilde{s}}k(Z_{i},Z_{i'}) - \left\{\left(\frac{\gamma}{\widetilde{\gamma}}\right)^{2}\mu_{k}^{X} + \left(\frac{\widetilde{\gamma}-\gamma}{\widetilde{\gamma}}\right)^{2}\mu_{k}^{Y} + \frac{2\gamma(\widetilde{\gamma}-\gamma)}{\widetilde{\gamma}^{2}}\mu_{k}^{XY}\right\} \right| < \epsilon^{*} \:\: [\mathbb{P}]~a.s.
\end{align}
Using equations (\ref{eq4}) and (\ref{eq8}), we have for $n > \max\{M_1^{(1)},\max\{N_1^{*},N_2^{*},N_3^{*}\}\}$,
\begin{align} \label{eq9}
	\left|\frac{1}{\widetilde{s}^{2}}\sum\limits_{i=1}^{\widetilde{s}}\sum\limits_{i'=1}^{\widetilde{s}}k(Z_{i},Z_{i'}) - \mu_{k}^{X}\mathbb{I}(\widetilde{\gamma} \leq \gamma)  - \left\{\left(\frac{\gamma}{\widetilde{\gamma}}\right)^{2}\mu_{k}^{X} + \left(\frac{\widetilde{\gamma}-\gamma}{\widetilde{\gamma}}\right)^{2}\mu_{k}^{Y} + \frac{2\gamma(\widetilde{\gamma}-\gamma)}{\widetilde{\gamma}^{2}}\mu_{k}^{XY}\right\}\mathbb{I}(\widetilde{\gamma} > \gamma) \right| < \epsilon^{*} \:\:[\mathbb{P}]~a.s. 
\end{align}
Since $\epsilon^{*}$ is arbitrary positive number and $M_1^{(1)},N_1^{*},N_2^{*},N_3^{*}$ are independent of $\widetilde{\gamma}$, 
\begin{align} \label{eq10}
	\underset{\widetilde{\gamma} \in [\delta_n, 1-\delta_n]}{\sup} \left|\frac{1}{\widetilde{s}^{2}}\sum\limits_{i=1}^{\widetilde{s}}\sum\limits_{i'=1}^{\widetilde{s}}k(Z_{i},Z_{i'}) - \mu_{k}^{X}\mathbb{I}(\widetilde{\gamma} \leq \gamma) - \left\{\left(\frac{\gamma}{\widetilde{\gamma}}\right)^{2}\mu_{k}^{X} + \left(\frac{\widetilde{\gamma}-\gamma}{\widetilde{\gamma}}\right)^{2}\mu_{k}^{Y} +  \frac{2\gamma(\widetilde{\gamma}-\gamma)}{\widetilde{\gamma}^{2}}\mu_{k}^{XY}\right\}\mathbb{I}(\widetilde{\gamma} > \gamma) \right| \xrightarrow[]{} \: 0 \quad [\mathbb{P}] ~ a.s. 
\end{align}
Similarly, it can be shown that
\begin{align} \label{eq11}
	\underset{\widetilde{\gamma} \in [\delta_n, 1-\delta_n]}{\sup} &\left|\frac{1}{(n-\widetilde{s})^{2}}\sum\limits_{i=\widetilde{s}+1}^{n}\sum\limits_{i'=\widetilde{s}+1}^{n}k(Z_{i},Z_{i'}) - \mu_{k}^{Y}\mathbb{I}(\widetilde{\gamma} > \gamma) - \left\{\left(\frac{\gamma-\widetilde{\gamma}}{1-\widetilde{\gamma}}\right)^{2}\mu_{k}^{X} + \left(\frac{1-\gamma}{1-\widetilde{\gamma}}\right)^{2}\mu_{k}^{Y} \right. \right. \nonumber\\  &\left. \left. + \frac{2(\gamma-\widetilde{\gamma})(1-\gamma)}{(1-\widetilde{\gamma})^{2}}\mu_{k}^{XY}\right\}\mathbb{I}(\widetilde{\gamma} \leq \gamma) \right| 
	\xrightarrow[]{} \: 0 \quad [\mathbb{P}]~a.s. 
\end{align}
\begin{center}
	and
\end{center}
\begin{align} \label{eq12}
	\underset{\widetilde{\gamma} \in [\delta_n, 1-\delta_n]}{\sup} &\left|\frac{1}{\widetilde{s}(n-\widetilde{s})}\sum\limits_{i=1}^{\widetilde{s}}\sum\limits_{i'=\widetilde{s}+1}^{n}k(Z_{i},Z_{i'}) - \left\{\frac{\gamma}{\widetilde{\gamma}}\mu_{k}^{XY} + \frac{\widetilde{\gamma}-\gamma}{\widetilde{\gamma}}\mu_{k}^{Y}\right\}\mathbb{I}(\widetilde{\gamma} > \gamma) - \left\{\frac{\gamma-\widetilde{\gamma}}{1-\widetilde{\gamma}}\mu_{k}^{X} \right.\right. \nonumber\\ &\left. \left. + \frac{1-\gamma}{1-\widetilde{\gamma}}\mu_{k}^{XY}\right\}\mathbb{I}(\widetilde{\gamma} \leq \gamma) \right| \xrightarrow[]{} \: 0 \quad [\mathbb{P}] ~ a.s. 
\end{align}
Combining equations~(\ref{eq10})--(\ref{eq12}) and applying the triangle inequality, we obtain
\begin{align} 
\label{eq25}
	\underset{\widetilde{\gamma} \in [\delta_n, 1-\delta_n]}{\sup} \left| d_{n}(\widetilde{\gamma}) - \left\{\frac{\gamma}{\widetilde{\gamma}}\mathbb{I}(\widetilde{\gamma} > \gamma) + \frac{1-\gamma}{1-\widetilde{\gamma}}\mathbb{I}(\widetilde{\gamma} \leq \gamma)\right\}^2 d(F_{1},F_{2}) \right| \xrightarrow[]{} \: 0 \quad [\mathbb{P}] ~ a.s.
\end{align}
Now,
\begin{align} 
\label{eq26}
    d_*(\widetilde{\gamma}) &= d(\alpha F_{1} + (1-\alpha)F_2,\beta F_1 + (1-\beta)F_{2}) \nonumber\\&=(\alpha - \beta)^{2}d(F_1,F_2)\:\: [\text{follows from equation (1) of the main paper}]\nonumber\\
    &=\left\{\frac{\gamma}{\widetilde{\gamma}}\mathbb{I}(\widetilde{\gamma} > \gamma) + \mathbb{I}(\widetilde{\gamma} \leq \gamma)-\frac{\gamma-\widetilde{\gamma}}{1-\widetilde{\gamma}}\mathbb{I}(\widetilde{\gamma} \leq \gamma)\right\}^{2}d(F_1,F_2)\:\: [\text{putting the values of $\alpha$ and $\beta$}]\nonumber\\
    & =\left\{\frac{\gamma}{\widetilde{\gamma}}\mathbb{I}(\widetilde{\gamma} > \gamma) + \frac{1-\gamma}{1-\widetilde{\gamma}}\mathbb{I}(\widetilde{\gamma} \leq \gamma)\right\}^2 d(F_{1},F_{2}). 
\end{align}
Equations \eqref{eq25} and \eqref{eq26} together imply that
\begin{align*}
	     \underset{\widetilde{\gamma} \in [\delta_n, 1-\delta_n]}{\sup} \left| d_{n}  (\widetilde{\gamma}) - d_*(\widetilde{\gamma})\right| \xrightarrow[]{}  0 \: \: [\mathbb{P}]~a.s.,
\end{align*}
under $H_{1\gamma}$ for any prefixed $\gamma \in (0,1)$.\\
From equation \eqref{eq26}, it is clear that $d_{*}(\widetilde{\gamma})$ is an increasing function of $\widetilde{\gamma}$ in the interval $[0,\gamma]$ and is decreasing in the interval $[\gamma,1]$. Therefore, $d_{*}(\widetilde{\gamma})$ achieves its maximum value at $\widetilde{\gamma}=\gamma$. This completes the proof of Lemma 3.2.}

\subsection*{ Proof of Theorem 3.3}
Let $H_{1\gamma}$ be true for some prefixed $\gamma \in (0,1)$. Define 
\begin{align} \label{S_gamma}
S(\widetilde{\gamma}) = \left\{\frac{\gamma}{\widetilde{\gamma}}\mathbb{I}(\widetilde{\gamma} > \gamma) + \frac{1-\gamma}{1-\widetilde{\gamma}}\mathbb{I}(\widetilde{\gamma} \leq \gamma)\right\}^2.
\end{align}
By Lemma 3.2, $d_{n}(\widetilde{\gamma})-S(\widetilde{\gamma})d(F_{1},F_{2}) \xrightarrow[]{} 0$ $[\mathbb{P}]~a.s.$, uniformly in $\widetilde{\gamma} \in [\delta_{n}, 1 - \delta_{n}]$.\\
Note that 
$$
\widetilde{\gamma}(1 - \widetilde{\gamma})S(\widetilde{\gamma})=\begin{cases}
	\frac{\widetilde{\gamma}(1-\gamma)^{2}}{1-\widetilde{\gamma}} & \text{if $\widetilde{\gamma} \leq \gamma$},\\
	\frac{\gamma^{2}(1-\widetilde{\gamma})}{\widetilde{\gamma}} & \text{if $\widetilde{\gamma} > \gamma$}.
\end{cases}
$$
Since the function $\widetilde{\gamma}(1 - \widetilde{\gamma})S(\widetilde{\gamma})$ is increasing in the interval $[0,\gamma]$ and is decreasing in the interval $[\gamma,1]$, $\widetilde{\gamma}(1 - \widetilde{\gamma})S(\widetilde{\gamma})$ attains its maximum value at $\widetilde{\gamma} = \gamma$.\\
The changepoint for a dataset $\{Z_1, Z_2, \ldots, Z_n\}$ is estimated by
\begin{align*}
	\widehat{\tau}_{n} = \underset{\lceil n\delta_{n} \rceil \leq t \leq \lfloor n(1-\delta_{n}) \rfloor}{arg\,max} \rho_{n}(t/n), 
\end{align*}
where $\rho_n(\widetilde{\gamma}) = \rho \left(\widehat{P}_{G_{n}(\widetilde{\gamma})},\widehat{P}_{H_{n}(\widetilde{\gamma})}\right)$ with $G_{n}(\widetilde{\gamma})=\left\{Z_{1},\dots,Z_{\lfloor n\widetilde{\gamma} \rfloor}\right\}$, and $H_{n}(\widetilde{\gamma})=\left\{Z_{\lfloor n\widetilde{\gamma} \rfloor+1},\dots,Z_{n}\right\}$. The estimated breakfraction is defined as $\widehat{\gamma}_{n} = \widehat{\tau}_{n}/{n}$.
By definition of $\widehat{\tau}_{n}$ and $\widehat{\gamma}_{n}$,
\begin{align} \label{eq14}
	\rho_{n}(\widehat{\gamma}_{n}) \geq \rho_{n}(\gamma) \quad \text{which implies} \quad \frac{\widehat{\tau}_{n}(n-\widehat{\tau}_{n})}{n^{2}}d_{n}(\widehat{\gamma}_{n}) \geq \frac{\lfloor n\gamma \rfloor(n-\lfloor n\gamma \rfloor)}{n^{2}}d_{n}(\gamma).
\end{align}
Now, by Lemma 3.2, $\frac{\lfloor n\gamma \rfloor(n-\lfloor n\gamma \rfloor)}{n^{2}}d_{n}(\gamma) \xrightarrow[]{} \gamma(1-\gamma)d(F_{1},F_{2})$ $[\mathbb{P}]~a.s$. Using this in equation (\ref{eq14}), we obtain
\begin{align} \label{eq16}
	\frac{\widehat{\tau}_{n}(n-\widehat{\tau}_{n})}{n^{2}}d_{n}(\widehat{\gamma}_{n}) \geq \gamma(1-\gamma)d(F_{1},F_{2}) - o(1).
\end{align}
Since the function $\widetilde{\gamma}(1 - \widetilde{\gamma})S(\widetilde{\gamma})$ attains its maximum at $\widetilde{\gamma} = \gamma$, we have
\begin{align*}
	0 &\leq \gamma(1-\gamma)S(\gamma)d(F_{1},F_{2}) - \widehat{\gamma}_{n}(1-\widehat{\gamma}_{n})S(\widehat{\gamma}_{n})d(F_{1},F_{2}) \\
	&\leq \frac{\widehat{\tau}_{n}(n-\widehat{\tau}_{n})}{n^{2}}d_{n}(\widehat{\gamma}_{n}) + o(1) - \frac{\widehat{\tau}_{n}(n-\widehat{\tau}_{n})}{n^{2}}S(\widehat{\gamma}_{n})d(F_{1},F_{2}) \: \: [\text{follows from (\ref{eq16}) and $S(\gamma) = 1$}].
\end{align*}
The RHS of the above inequality converges to $0$ $[\mathbb{P}]~a.s.$ since $d_{n}(\widehat{\gamma}_{n})-S(\widehat{\gamma}_{n})d(F_{1},F_{2}) \xrightarrow[]{} 0$ $[\mathbb{P}]~a.s.$ (by Lemma 3.2) and the scaling term $\widehat{\tau}_{n}(n-\widehat{\tau}_{n})/n^2$ is bounded above by $1$. Therefore, 
\begin{align} \label{eq17}
	\widehat{\gamma}_{n}(1-\widehat{\gamma}_{n})S(\widehat{\gamma}_{n})d(F_{1},F_{2}) \xrightarrow[]{} \gamma(1-\gamma)S(\gamma)d(F_{1},F_{2})\quad [\mathbb{P}] ~ a.s.
\end{align}
Moreover,
\begin{align*}
	0 &\leq \gamma(1-\gamma)S(\gamma)d(F_{1},F_{2}) - \widehat{\gamma}_{n}(1-\widehat{\gamma}_{n})S(\widehat{\gamma}_{n})d(F_{1},F_{2}) \\
	&\leq \gamma(1-\gamma)d(F_{1},F_{2}) - \frac{\widehat{\tau}_{n}(n-\widehat{\tau}_{n})}{n^{2}}d_{n}(\widehat{\gamma}_{n}) + o(1).
\end{align*}
The last inequality follows from the facts that $d_{n}(\widehat{\gamma}_{n}) - S(\widehat{\gamma}_{n})d(F_{1},F_{2}) \xrightarrow[]{} 0 \quad [\mathbb{P}] ~ a.s.$ and $S(\gamma) = 1$.
This implies,
\begin{align} \label{eq18}
	\frac{\widehat{\tau}_{n}(n-\widehat{\tau}_{n})}{n^{2}}d_{n}(\widehat{\gamma}_{n}) \leq \gamma(1-\gamma)d(F_{1},F_{2}) + o(1).
\end{align}
Hence, combining (\ref{eq16}) and (\ref{eq18}),
\begin{align*}
	T_{n}=\frac{\widehat{\tau}_{n}(n-\widehat{\tau}_{n})}{n^{2}}d_{n}(\widehat{\gamma}_{n}) \xrightarrow[]{} \gamma(1-\gamma)d(F_{1},F_{2}) \quad [\mathbb{P}] ~ a.s.
\end{align*}
This completes the proof of Theorem 3.3.

\subsection*{ Proof of Theorem 3.4}
\textit{(a)} Under $H_{1\gamma}$ for some prefixed $\gamma \in (0,1)$, define the following sets
\begin{align*}
	&P_{1,n} = \{(i,j):Z_i, Z_j \sim F_1\},\: P_{2,n} = \{(i,j):Z_i \sim F_1, Z_j \sim F_2\} \\&\text{ and } P_{3,n} = \{(i,j):Z_i, Z_j \sim F_2\}.
\end{align*}
Now, consider the following sets
\begin{align} \label{sets with P probability 1}
	&A_1 := \{\omega \in \Omega: |\frac{1}{|S_{1,n}(\omega)|}\sum\limits_{(i,j) \in S_{1,n}(\omega)}k(Z_i(\omega),Z_j(\omega))-\mu_{k}^{X}| \xrightarrow[]{} 0 \: \text{for any} \: S_{1,n}(\omega) \subseteq P_{1,n}   \text{with} \: |S_{1,n}(\omega)| \xrightarrow[]{} \infty\},\nonumber\\
	&A_2 := \{\omega \in \Omega: |\frac{1}{|S_{2,n}(\omega)|}\sum\limits_{(i,j) \in S_{2,n}(\omega)}k(Z_i(\omega),Z_j(\omega))-\mu_{k}^{XY}| \xrightarrow[]{} 0 \: \text{for any} \: S_{2,n}(\omega) \subseteq P_{2,n}  \text{with} \: |S_{2,n}(\omega)| \xrightarrow[]{} \infty\},\nonumber\\
	&A_3 := \{\omega \in \Omega: |\frac{1}{|S_{3,n}(\omega)|}\sum\limits_{(i,j) \in S_{3,n}(\omega)}k(Z_i(\omega),Z_j(\omega))-\mu_{k}^{Y}| \xrightarrow[]{} 0 \: \text{for any} \: S_{3,n}(\omega) \subseteq P_{3,n}  \text{with} \: |S_{3,n}(\omega)| \xrightarrow[]{} \infty\}. 
\end{align}
Note that $\mathbb{P}(A_1)=1, \mathbb{P}(A_2)=1$ and $\mathbb{P}(A_3)=1$ which implies $\mathbb{P}(A_1 \cap A_2 \cap A_3)=1$. Now, fix $\omega \in A_1 \cap A_2 \cap A_3$. Recall that for the $r$-th permutation of $\{Z_1, Z_2, \dots, Z_n\}$, we denote the permuted observations by 
\(\{ Z^{(r)}_{1}, Z^{(r)}_{2}, \ldots, Z^{(r)}_{n} \}\). For fixed $\omega \in \Omega$, the realization of $\{ Z^{(r)}_{1}, Z^{(r)}_{2}, \ldots, Z^{(r)}_{n} \}$ is $\{ Z^{(r)}_{1}(\omega), Z^{(r)}_{2}(\omega), \ldots, Z^{(r)}_{n}(\omega) \}$ and 
\begin{align} \label{quantities for fixed omega}
&G_{n}^{(r)}(\omega)(\widetilde{\gamma})=\{ Z^{(r)}_{1}(\omega), Z^{(r)}_{2}(\omega), \ldots, Z^{(r)}_{\lfloor n\widetilde{\gamma}\rfloor}(\omega) \},\nonumber\\ &H_{n}^{(r)}(\omega)(\widetilde{\gamma})=\{ Z^{(r)}_{\lfloor n\widetilde{\gamma}\rfloor+1}(\omega), Z^{(r)}_{\lfloor n\widetilde{\gamma}\rfloor+2}(\omega), \ldots, Z^{(r)}_{n}(\omega) \},\nonumber\\
&d_n^{(r)}(\omega)(\widetilde{\gamma}) = d\left(\widehat{P}_{G_{n}^{(r)}(\omega)(\widetilde{\gamma})},\widehat{P}_{H_{n}^{(r)}(\omega)(\widetilde{\gamma})}\right), \: \:\rho_n^{(r)}(\omega)(\widetilde{\gamma}) = \frac{\lfloor n\widetilde{\gamma}\rfloor(n-\lfloor n\widetilde{\gamma}\rfloor)}{n^2} d_n^{(r)}(\omega)(\widetilde{\gamma}), \nonumber\\
    &\widehat{\tau}_{n}^{(r)}(\omega)=\underset{\lceil n\delta_{n}  \rceil \leq t \leq \lfloor n(1-\delta_{n}  )\rfloor}{\arg\max} \rho_n^{(r)}(\omega)(t/n), \widehat{\gamma}_{n}^{(r)}(\omega)=\widehat{\tau}_{n}^{(r)}(\omega)/n. 
\end{align}
Define $W_n$ as the number of $Z_{i}^{(r)}(\omega)$ in the set $\{Z_{1}^{(r)}(\omega),Z_{2}^{(r)}(\omega),\dots,Z_{\lfloor na_n\rfloor}^{(r)}(\omega)\}$ coming from first population $F_1$, $\alpha_n = W_{n}/\lfloor na_n\rfloor$, $\beta_n = (\lfloor n\gamma\rfloor - W_n)/(n-\lfloor na_n\rfloor)$ where $\{a_{n}\}$ is any sequence of positive numbers with $n^{-\delta} < a_n < 1-n^{-\delta}$ for $n \in \mathbb{N}$ where $0<\delta<1$.
The random variable $W_n$ follows the Hypergeometric distribution with parameters $(n,\lfloor na_n\rfloor, \lfloor n\gamma \rfloor)$, i.e., for $\max{\{0,\lfloor na_n \rfloor + \lfloor n\gamma \rfloor - n\}} \leq w \leq \min{\{\lfloor na_{n}\rfloor, \lfloor n\gamma\rfloor\}}$,
\begin{align*}
	\mathbb{P}^*(W_n = w) = \frac{\binom{\lfloor n\gamma \rfloor}{w}\binom{n - \lfloor n\gamma \rfloor}{\lfloor na_{n} \rfloor - w}}{\binom{n}{\lfloor na_{n} \rfloor}}.
\end{align*}
Let us define the following sets
\begin{align*}
	&C_1 := \{i \in \{1,2,\dots,\lfloor na_n\rfloor\} : Z_{i}^{(r)}(\omega) \text{ is a realization of a random variable from $F_1$}\},\\
	&C_2 := \{i \in \{1,2,\dots,\lfloor na_n\rfloor\} : Z_{i}^{(r)}(\omega) \text{ is a realization of a random variable from $F_2$}\},\\
	&C_3 := \{i \in \{\lfloor na_n\rfloor+1,\dots,n\} : Z_{i}^{(r)}(\omega) \text{ is a realization of a random variable from $F_1$}\}, \\
	&C_4 := \{i \in \{\lfloor na_n\rfloor+1,\dots,n\}: Z_{i}^{(r)}(\omega) \text{ is a realization of a random variable from $F_2$}\},\\
	&C_5 := \{i \in \{1,2,\dots,J\}: Z_{i}^{(r)}(\omega) \text{ comes from first population}\} \text{ where } \\&J = \inf \{j \in \{1,2,\dots,n\}:\sum\limits_{i=1}^{j} \mathbb{I}(Z_{i}^{(r)}(\omega) \text{ is a realization of a random variable from $F_1$})=\lfloor \gamma na_n\rfloor\}\\
	&C_6 := \{i \in \{1,2,\dots,J\}: Z_{i}^{(r)}(\omega) \text{ is a realization of a random variable from $F_2$}\},\\
	&C_7 := \{i \in \{J+1,J+2,\dots,n\}: Z_{i}^{(r)}(\omega) \text{ is a realization of a random variable from $F_1$}\},\\
	&C_8 := \{i \in \{J+1,J+2,\dots,n\}: Z_{i}^{(r)}(\omega) \text{ is a realization of a random variable from $F_2$}\}.
\end{align*}
The sets $C_5, C_6, C_7$ and $C_8$ are random through the randomness of $J$. Recall that $(\Omega^{*}, \mathcal{F}^{*}, \mathbb{P}^{*})$ denote a probability space that governs the distribution induced by the permutation procedure, mentioned in Section 3 of the main article. For notational simplicity, when $\omega^* \in \Omega^*$ is fixed, we denote the sets by $C_i$ instead of $C_i(\omega^*)$ for $i=5,6,7,8$. 

Note that
\begin{align} \label{difference of two d measures}
	&d_{n}^{(r)}(\omega)(a_{n})-(\alpha_n - \beta_n)^{2}d(F_1,F_2)  \nonumber\\&= \alpha_{n}^{2} \left\{\frac{1}{W_{n}^{2}}\sum\limits_{i,i' \in C_1}k(Z_i^{(r)}(\omega),Z_{i'}^{(r)}(\omega))-\mu_{k}^{X} \right\} + \beta_{n}^{2} \left\{ \frac{1}{(\lfloor n\gamma \rfloor -  W_{n})^{2}}\sum\limits_{j,j' \in C_3}k(Z_j^{(r)}(\omega),Z_{j'}^{(r)}(\omega))-\mu_{k}^{X}\right\} \nonumber \\&- 2\alpha_{n}\beta_{n}\left\{\frac{1}{W_{n}(\lfloor n\gamma \rfloor - W_{n})}\sum\limits_{i \in C_1, j \in C_3}k(Z_i^{(r)}(\omega),Z_{j}^{(r)}(\omega))-\mu_{k}^{X}\right\} \nonumber\\&+ (1-\alpha_{n})^{2}\left\{\frac{1}{(\lfloor na_{n}\rfloor-W_{n})^{2}} \sum\limits_{i,i' \in C_2}k(Z_i^{(r)}(\omega),Z_{i'}^{(r)}(\omega))-\mu_{k}^{Y}\right\} \nonumber \\
	&+(1-\beta_{n})^{2}\left\{\frac{1}{(n - \lfloor n\gamma \rfloor - \lfloor na_{n} \rfloor +  W_{n})^{2}}\sum\limits_{j,j' \in C_4}k(Z_j^{(r)}(\omega),Z_{j'}^{(r)}(\omega)) - \mu_{k}^{Y}\right\} \nonumber \\&- 2(1-\alpha_{n})(1-\beta_{n})\left\{\frac{1}{(\lfloor na_{n} \rfloor - W_{n})(n -\lfloor n\gamma \rfloor - \lfloor na_{n} \rfloor + W_{n})}\sum\limits_{i \in C_2, j \in C_4}k(Z_i^{(r)}(\omega),Z_{j}^{(r)}(\omega)) \right. \nonumber\\&- \left. \mu_{k}^{Y}\right\}+2\alpha_{n}(1-\alpha_{n})\left\{\frac{1}{W_{n}(\lfloor na_{n} \rfloor - W_{n})}\sum\limits_{i \in C_1, i' \in C_2}k(Z_i^{(r)}(\omega),Z_{i'}^{(r)}(\omega))-\mu_{k}^{XY}\right\} \nonumber \\ &+ 2\beta_{n}(1-\beta_{n})\left\{\frac{1}{(\lfloor n\gamma \rfloor - W_{n})(n -\lfloor n\gamma \rfloor - \lfloor na_{n} \rfloor \nonumber + W_{n})}\sum\limits_{j \in C_3, j' \in C_4}k(Z_j^{(r)}(\omega),Z_{j'}^{(r)}(\omega))-\mu_{k}^{XY}\right\} \nonumber \\ & - 2\alpha_{n}(1-\beta_{n}) \left\{\frac{1}{W_{n}(n -\lfloor n\gamma \rfloor - \lfloor na_{n} \rfloor + W_{n})}\sum\limits_{i \in C_1, j \in C_4}k(Z_i^{(r)}(\omega),Z_{j}^{(r)}(\omega))-\mu_{k}^{XY}\right\} \nonumber \\ & - 2\beta_{n}(1-\alpha_{n})\left\{\frac{1}{(\lfloor n\gamma \rfloor - W_{n})(\lfloor na_{n} \rfloor - W_{n})}\sum\limits_{i \in C_2, j \in C_3}k(Z_i^{(r)}(\omega),Z_{j}^{(r)}(\omega))-\mu_{k}^{XY}\right\}. 
\end{align}
Now, for any $\epsilon > 0$,
\begin{align*} 
	\mathbb{P}^*(|W_n/\lfloor na_n\rfloor-\gamma| > \epsilon) &= \mathbb{P}^*(W_n > \lfloor na_n\rfloor(\lfloor n\gamma\rfloor/n+\gamma+\epsilon-\lfloor n\gamma\rfloor/n)) + \mathbb{P}^*(W_n < \lfloor na_n\rfloor(\lfloor n\gamma\rfloor/n+\gamma-\epsilon-\lfloor n\gamma\rfloor/n)) \\&\leq exp(-2\lfloor na_n\rfloor(\gamma+\epsilon-\lfloor n\gamma\rfloor/n)^2)+exp(-2\lfloor na_n\rfloor(\lfloor n\gamma\rfloor/n + \epsilon - \gamma)^2).
\end{align*}
The last inequality follows from \citet{hoeffding1963probability}.
Now, $\gamma-\lfloor n\gamma\rfloor/n \leq 1/n$ and there exists an integer $M_1 \in \mathbb{N}$ such that $1/n < \epsilon/2$ for all $n \geq M_1$. Then for $n \geq M_1$,
\begin{align*}
	\mathbb{P}^*(|W_n/\lfloor na_n\rfloor-\gamma| > \epsilon) &\leq exp(-2\lfloor na_n\rfloor \epsilon^2) + exp(-\lfloor na_n\rfloor \epsilon^2/2)\\
	&\leq exp(-2n^{1-\delta}\epsilon^2) + exp(-n^{1-\delta}\epsilon^2/2) \quad [\text{as } na_n \geq n^{1-\delta}].
\end{align*}
Hence, $ \sum\limits_{n=1}^{\infty}\mathbb{P}^*(|W_n/\lfloor na_n\rfloor-\gamma| > \epsilon) < \infty$ and therefore, $W_n/\lfloor na_n\rfloor \xrightarrow[]{} \gamma \quad [\mathbb{P}^*] ~ a.s.$ Similarly, it can be proved that $(\lfloor n\gamma\rfloor - W_n)/(n-\lfloor na_n\rfloor) \xrightarrow[]{} \gamma \quad [\mathbb{P}^*] ~ a.s.$, $(\lfloor na_n\rfloor-W_n)/\lfloor na_n\rfloor \xrightarrow[]{} 1-\gamma \quad [\mathbb{P}^*] ~ a.s.$ and $(n-\lfloor n\gamma\rfloor -\lfloor na_n\rfloor+W_n)/(n-\lfloor na_n\rfloor)\xrightarrow[]{} 1-\gamma \quad [\mathbb{P}^*] ~ a.s.$  \\\\
Now,
\begin{align} \label{upper_bound_first_term}
	\left|\frac{1}{W_{n}^{2}}\sum\limits_{i,i' \in C_1}k(Z_i^{(r)}(\omega),Z_{i'}^{(r)}(\omega)) - \mu_{k}^{X}\right| \nonumber\\
	& \leq \left|\frac{1}{W_{n}^{2}}\sum\limits_{i,i' \in C_1}k(Z_i^{(r)}(\omega),Z_{i'}^{(r)}(\omega))-\frac{1}{\lfloor \gamma na_{n}\rfloor^2}\sum\limits_{i,i' \in C_1}k(Z_i^{(r)}(\omega),Z_{i'}^{(r)}(\omega))\right| \nonumber\\& \quad + \left|\frac{1}{\lfloor \gamma na_{n}\rfloor^2}\sum\limits_{i,i' \in C_1}k(Z_i^{(r)}(\omega),Z_{i'}^{(r)}(\omega)) - \frac{1}{\lfloor \gamma na_{n}\rfloor^2}\sum\limits_{i,i' \in C_5}k(Z_i^{(r)}(\omega),Z_{i'}^{(r)}(\omega))\right| \nonumber\\& \quad +\left|\frac{1}{\lfloor \gamma n a_{n}\rfloor^2}\sum\limits_{i,i' \in C_5}k(Z_i^{(r)}(\omega),Z_{i'}^{(r)}(\omega)) - \mu_{k}^{X}\right|\nonumber\\
	& \leq \frac{|W_{n}^2-\lfloor \gamma n a_n\rfloor^2|}{W_{n}^{2}\lfloor \gamma n a_n\rfloor^2} W_{n}^{2} + \frac{(W_n - \lfloor \gamma n a_n\rfloor)^2 + 2\lfloor \gamma n a_n \rfloor |W_n - \lfloor \gamma n a_n\rfloor|}{\lfloor \gamma n a_n\rfloor^2} \nonumber \\& + \left|\frac{1}{\lfloor \gamma n a_{n}\rfloor^2}\sum\limits_{i,i' \in C_5}k(Z_i^{(r)}(\omega),Z_{i'}^{(r)}(\omega)) - \mu_{k}^{X}\right|.  
\end{align}
The last inequality of \eqref{upper_bound_first_term} follows from the fact that $|k(x,y)|\leq 1$ for all $x,y$.\\
Now, \begin{align} \label{first term}
	\frac{|W_n^2-\lfloor \gamma n a_n\rfloor^2|}{\lfloor \gamma n a_n\rfloor^2}=(W_n/\lfloor \gamma n a_n\rfloor+1)|W_n/\lfloor \gamma n a_n\rfloor-1|. 
\end{align}
Note that,
\begin{align*}
	\frac{W_n}{\lfloor \gamma na_n\rfloor}+1 \leq \frac{W_n}{\gamma na_n-1}+1=\frac{W_n}{\gamma na_n}+1+\frac{W_n}{\gamma na_n(\gamma na_n-1)} &\leq \frac{1}{\gamma}+1+\frac{1}{\gamma(\gamma na_n-1)} \leq \frac{1}{\gamma}+1+\frac{1}{\gamma(\gamma n^{1-\delta}-1)}.
\end{align*}
There exists an integer $M_2 \in \mathbb{N}$ such that $1/(\gamma n^{1-\delta}-1)<\gamma$ for $n \geq M_2$. Hence, for $n \geq M_2$, $W_{n}/\lfloor \gamma na_n\rfloor+1 \leq 2+1/\gamma$ $[\mathbb{P}^*] ~ a.s.$ \\
On the other hand,
\begin{align*}
	\left|\frac{W_n}{\lfloor \gamma n a_n\rfloor}-1 \right| \leq \frac{\lfloor na_n\rfloor}{\lfloor \gamma n a_n\rfloor} \left|\frac{W_{n}}{\lfloor na_n\rfloor} - \gamma \right|+ \left|\frac{\gamma\lfloor na_n\rfloor}{\lfloor \gamma n a_n\rfloor}-1 \right|.
\end{align*}
Now, $\lfloor na_n\rfloor/\lfloor \gamma na_n\rfloor \leq na_n/(\gamma na_n - 1)=1/\gamma + 1/\gamma(\gamma na_n-1) \leq 1/\gamma+1/\gamma(\gamma n^{1-\delta}-1)$. Hence, for $n \geq M_2$, $\lfloor na_n\rfloor/\lfloor \gamma na_n\rfloor \leq 1+1/\gamma$. 
As $W_n/\lfloor na_n \rfloor \xrightarrow[]{} \gamma \quad [\mathbb{P}^*] ~ a.s.$ and $|\gamma \lfloor na_n\rfloor/\lfloor \gamma na_n\rfloor-1| \leq 1/\lfloor \gamma na_n\rfloor \leq 1/(\gamma n^{1-\delta}-1)$, 
\begin{align} \label{convergence of w/ngamma}
	\left|\frac{W_n}{\lfloor \gamma na_n\rfloor}-1 \right| \xrightarrow[]{} 0 \quad [\mathbb{P}^*]~a.s.    
\end{align}
Since $W_n/\lfloor \gamma na_n\rfloor+1 \leq 2+1/\gamma$ with $[\mathbb{P}^*] ~ a.s.$ for $n \geq M_2$, \eqref{first term} and \eqref{convergence of w/ngamma} imply
\begin{align} \label{convergence of w2}
	\frac{|W_n^2-\lfloor \gamma n a_n\rfloor^2|}{\lfloor \gamma n a_n\rfloor^2} \xrightarrow[]{} 0 \quad [\mathbb{P}^*]~a.s.    
\end{align}
Also, it follows from \eqref{convergence of w/ngamma} that 
\begin{align}\label{convergence of other terms}
	\frac{(W_n - \lfloor \gamma n a_n\rfloor)^2}{\lfloor \gamma n a_n\rfloor^2} = \left|\frac{W_n}{\lfloor \gamma na_n\rfloor}-1\right|^2 \xrightarrow[]{} 0 \quad [\mathbb{P}^*]~a.s.,\quad         \frac{2\lfloor \gamma na_n\rfloor|W_n-\lfloor \gamma na_n\rfloor|}{\lfloor \gamma na_n\rfloor^2} \xrightarrow[]{} 0 \quad [\mathbb{P}^*]~a.s. 
\end{align}
Now, fix any $\omega^* \in \Omega^*$. Then 
\begin{align*}
	\frac{1}{\lfloor \gamma n a_{n}\rfloor^2}\sum\limits_{i,i' \in C_5}k(Z_i^{(r)}(\omega),Z_{i'}^{(r)}(\omega)) = \frac{1}{\lfloor \gamma n a_{n}\rfloor^2}\sum\limits_{(i,j) \in S}k(Z_{i}(\omega),Z_{j}(\omega))
\end{align*}
with $[\mathbb{P}^*]~ a.s.$ for some $S \subseteq P_{1, n}$ with $|S|=\lfloor \gamma na_n\rfloor$. Since, $\lfloor \gamma na_n\rfloor \xrightarrow[]{} \infty$ and  $\omega \in A_1$, for any fixed $\omega^* \in \Omega^*$,
\begin{align} \label{convergence of v statistic for fixed omega*}
	\frac{1}{\lfloor \gamma n a_{n}\rfloor^2}\sum\limits_{i,i' \in C_5}k(Z_i^{(r)}(\omega),Z_{i'}^{(r)}(\omega)) - \mu_{k}^{X} \xrightarrow[]{} 0.
\end{align}
Since \eqref{convergence of v statistic for fixed omega*} holds for any $\omega^* \in \Omega^*$, 
\begin{align} \label{convergence of v statistic}
	\frac{1}{\lfloor \gamma n a_{n}\rfloor^2}\sum\limits_{i,i' \in C_5}k(Z_i^{(r)}(\omega),Z_{i'}^{(r)}(\omega)) - \mu_{k}^{X} \xrightarrow[]{} 0 \quad [\mathbb{P}^*]~a.s. 
\end{align}
Hence, \eqref{convergence of w2}, \eqref{convergence of other terms} and \eqref{convergence of v statistic} together imply that all the three terms of RHS of \eqref{upper_bound_first_term} converge to $0$ $[\mathbb{P}^*]~a.s.$ Therefore,
\begin{align} \label{a.s. convergence of first term}
	\frac{1}{W_{n}^{2}}\sum\limits_{i,i' \in C_1}k(Z_i^{(r)}(\omega),Z_{i'}^{(r)}(\omega))-\mu_{k}^{X} \xrightarrow[]{} 0 \quad [\mathbb{P}^*]~a.s.
\end{align}
Similarly, it can be proved that
\begin{align} \label{a.s. convergence of second to tenth term}
	&\frac{1}{(\lfloor n\gamma \rfloor -  W_{n})^{2}}\sum\limits_{j,j' \in C_3}k(Z_j^{(r)}(\omega),Z_{j'}^{(r)}(\omega))-\mu_{k}^{X} \xrightarrow[]{} 0 \quad [\mathbb{P}^*] ~ a.s.
	\nonumber\\            &\frac{1}{W_{n}(\lfloor n\gamma \rfloor - W_{n})}\sum\limits_{i \in C_1, j \in C_3}k(Z_i^{(r)}(\omega),Z_{j}^{(r)}(\omega))-\mu_{k}^{X} \xrightarrow[]{} 0 \quad [\mathbb{P}^*] ~ a.s.
	\nonumber\\            &\frac{1}{(\lfloor na_{n}\rfloor-W_{n})^{2}} \sum\limits_{i,i' \in C_2}k(Z_i^{(r)}(\omega),Z_{i'}^{(r)}(\omega))-\mu_{k}^{Y} \xrightarrow[]{} 0 \quad [\mathbb{P}^*] ~ a.s.
	\nonumber\\            &\frac{1}{(n - \lfloor n\gamma \rfloor - \lfloor na_{n} \rfloor +  W_{n})^{2}}\sum\limits_{j,j' \in C_4}k(Z_j^{(r)}(\omega),Z_{j'}^{(r)}(\omega))-\mu_{k}^{Y} \xrightarrow[]{} 0 \quad [\mathbb{P}^*] ~ a.s.
	\nonumber\\            &\frac{1}{(\lfloor na_{n} \rfloor - W_{n})(n -\lfloor n\gamma \rfloor - \lfloor na_{n} \rfloor + W_{n})}\sum\limits_{i \in C_2, j \in C_4}k(Z_i^{(r)}(\omega),Z_{j}^{(r)}(\omega))-\mu_{k}^{Y} \xrightarrow[]{} 0 \quad [\mathbb{P}^*] ~ a.s.
	\nonumber\\            &\frac{1}{W_{n}(\lfloor na_{n} \rfloor - W_{n})}\sum\limits_{i \in C_1, i' \in C_2}k(Z_i^{(r)}(\omega),Z_{i'}^{(r)}(\omega))-\mu_{k}^{XY} \xrightarrow[]{} 0 \quad [\mathbb{P}^*] ~ a.s.
	\nonumber\\            &\frac{1}{(\lfloor n\gamma \rfloor - W_{n})(n -\lfloor n\gamma \rfloor - \lfloor na_{n} \rfloor \nonumber + W_{n})}\sum\limits_{j \in C_3, j' \in C_4}k(Z_j^{(r)}(\omega),Z_{j'}^{(r)}(\omega))-\mu_{k}^{XY} \xrightarrow[]{} 0 \quad [\mathbb{P}^*] ~ a.s. \nonumber\\
	&\frac{1}{W_{n}(n -\lfloor n\gamma \rfloor - \lfloor na_{n} \rfloor + W_{n})}\sum\limits_{i \in C_1, j \in C_4}k(Z_i^{(r)}(\omega),Z_{j}^{(r)}(\omega))-\mu_{k}^{XY} \xrightarrow[]{} 0 \quad [\mathbb{P}^*] ~ a.s. \nonumber\\
	& \frac{1}{(\lfloor n\gamma \rfloor - W_{n})(\lfloor na_{n} \rfloor - W_{n})}\sum\limits_{i \in C_2, j \in C_3}k(Z_i^{(r)}(\omega),Z_{j}^{(r)}(\omega))-\mu_{k}^{XY} \xrightarrow[]{} 0 \quad [\mathbb{P}^*] ~ a.s. 
\end{align}
Since, $0 \leq \alpha_{n},\beta_{n} \leq 1$ $[\mathbb{P}^*] ~ a.s.$, \eqref{difference of two d measures}, \eqref{a.s. convergence of first term} and \eqref{a.s. convergence of second to tenth term} together imply that 
\[
d_{n}^{(r)}(\omega)(a_n)-(\alpha_n-\beta_n)^{2}d(F_1,F_2) \xrightarrow[]{} 0 \quad [\mathbb{P}^*] ~ a.s.
\]
Since $\frac{\lfloor na_n\rfloor(n-\lfloor na_n\rfloor)}{n^2} \leq 1$, 
\begin{align} \label{convergence of the difference}
	\rho_{n}^{(r)}(\omega)(a_n)-\frac{\lfloor na_n\rfloor(n-\lfloor na_n\rfloor)}{n^2}(\alpha_n-\beta_n)^{2}d(F_1,F_2) \xrightarrow[]{} 0 \quad [\mathbb{P}^*] ~ a.s.
\end{align}
Now,
\begin{align*}
	\frac{\lfloor na_n\rfloor(n-\lfloor na_n\rfloor)}{n^2}(\alpha_n -\beta_n)^2 = \frac{\lfloor na_n\rfloor(n-\lfloor na_n\rfloor)}{n^2}\left\{\frac{nW_n}{\lfloor na_n\rfloor(n-\lfloor na_n\rfloor)}-\frac{\lfloor n\gamma\rfloor}{n-\lfloor na_n\rfloor}\right\}^2.
\end{align*}
\\\\
For any $\epsilon^*>0$,
\begin{align*}
	\mathbb{P}^*\left(\frac{\lfloor na_n\rfloor(n-\lfloor na_n\rfloor)}{n^2}(\alpha_n -\beta_n)^2 > \epsilon^* \right) &= \mathbb{P}^*\left(W_{n}>\lfloor na_n\rfloor\left(\frac{\lfloor na_n\rfloor \lfloor n\gamma\rfloor}{n} + \sqrt{\frac{(n-\lfloor na_n\rfloor)\epsilon^*}{\lfloor na_n \rfloor}}\right)\right) \\&\quad + \mathbb{P}^*\left(W_{n}<\lfloor na_n\rfloor\left(\frac{\lfloor na_n\rfloor \lfloor n\gamma\rfloor}{n} - \sqrt{\frac{(n-\lfloor na_n\rfloor)\epsilon^*}{\lfloor na_n \rfloor}}\right)\right)\\
	& \leq 2\: exp\left(-\lfloor na_n\rfloor\frac{(n-\lfloor na_n\rfloor)\epsilon^*}{\lfloor na_n \rfloor}\right) \\ &[\text{follows from \citet{hoeffding1963probability}}]\\
	&=2\:exp(-(n-\lfloor na_n\rfloor)\epsilon^*)\\
	& \leq 2\:exp(-n^{1-\delta}\epsilon^*) \quad [\text{as } a_n < 1-n^{-\delta}].
\end{align*}
Therefore, $\sum\limits_{n=1}^{\infty} \mathbb{P}^*\left(\frac{\lfloor na_n\rfloor(n-\lfloor na_n\rfloor)}{n^2}(\alpha_n -\beta_n)^2 > \epsilon^* \right) < \infty$, and hence, 
$$\frac{\lfloor na_n\rfloor(n-\lfloor na_n\rfloor)}{n^2}(\alpha_n -\beta_n)^2 \xrightarrow[]{} 0 \quad [\mathbb{P}^*] ~ a.s.$$ 
Since $d(F_1,F_2) \leq 4$,
\begin{align} \label{convergence of limit}
	\frac{\lfloor na_n\rfloor(n-\lfloor na_n\rfloor)}{n^2}(\alpha_n -\beta_n)^2 d(F_1,F_2) \xrightarrow[]{} 0 \quad [\mathbb{P}^*] ~ a.s.    
\end{align}
Equations \eqref{convergence of the difference} and \eqref{convergence of limit} together imply that $\rho_{n}^{(r)}(\omega)(a_n)  \xrightarrow[]{} 0 \quad [\mathbb{P}^*] ~ a.s.$ for any sequence $\{a_n\}$ such that $n^{-\delta} < a_n < 1-n^{-\delta}$ for $n \in \mathbb{N}$ where $\rho_{n}^{(r)}(\omega)(a_n)$ is defined in \eqref{quantities for fixed omega}. 
Now, if $a_n \leq n^{-\delta}$ or $a_n \geq 1-n^{-\delta}$, $\frac{\lfloor na_n\rfloor(n-\lfloor na_n\rfloor)}{n^2} \leq n^{-\delta}$. Therefore, $\rho_{n}^{(r)}(\omega)(a_n) \leq 4n^{-\delta} \xrightarrow[]{} 0$. Hence $\rho_{n}^{(r)}(\omega)(a_n) \xrightarrow[]{} 0 \quad [\mathbb{P}^*] ~ a.s.$ for any sequence $\{a_n\}$ such that $0 \leq a_n \leq 1$ for all $n$.
Now for fixed $\omega \in \Omega$, $\mathbb{P}^{*}(\widehat{\gamma}_{n}^{(r)}(\omega) < 1)=1$.\\ Now, choose $\omega^*$ from the set  $\{\omega^* \in \Omega^* : \rho_{n}^{(r)}(\omega,\omega^*)(a_n) \xrightarrow[]{} 0 \text{ for any sequence $\{a_n\}$ such that $0 \leq a_n \leq 1$ for all $n \geq 1$}\}$ (this set has $\mathbb{P}^*$ measure 1) where \\ $\rho_{n}^{(r)}(\omega, \omega^*)(a_n)$ denotes the realized value of $\rho_{n}^{(r)}(\omega)(a_n)$ for fixed $\omega^* \in \Omega^*$. Then \\$\rho_{n}^{(r)}(\omega,\omega^*)(\widehat{\gamma}_{n}^{(r)}(\omega,\omega^*)) \xrightarrow[]{} 0$ where $\widehat{\gamma}_{n}^{(r)}(\omega,\omega^*)$ denotes the realization of $\widehat{\gamma}_{n}^{(r)}(\omega)$ for fixed $\omega^* \in \Omega^*$. Since, $\rho_{n}^{(r)}(\omega,\omega^*)(\widehat{\gamma}_{n}^{(r)}(\omega,\omega^*)) \xrightarrow[]{} 0$ for any $\omega^* \in \Omega^*$ chosen from a set with $\mathbb{P}^*$ measure 1, 
\begin{align} \label{P* convergence of rho measure}
	\rho_{n}^{(r)}(\omega)(\widehat{\gamma}_{n}^{(r)}(\omega)) \xrightarrow[]{} 0 \quad [\mathbb{P}^*] ~ a.s.
\end{align}
Since \eqref{P* convergence of rho measure} holds for any $\omega \in A_1 \cap A_2 \cap A_3$ and the set $\mathbb{P}(A_1 \cap A_2 \cap A_3)=1$,
\begin{align} \label{P tilde convergence of rho measure}
	T_{n}^{(r)}=\rho_{n}^{(r)}(\widehat{\gamma}_{n}^{(r)}) \xrightarrow[]{} 0 \quad [\widetilde{\mathbb{P}}] ~ a.s.
\end{align}
\textit{(b)} By part (a) of Theorem 3.4,
\begin{align} \label{upper_bound_T_perm}
\underset{1 \leq r \leq R}{\max}T_{n}^{(r)} \xrightarrow[]{} 0 \: \: [\widetilde{\mathbb{P}}]~a.s.
\end{align}
As $T_{n}$ converges to the positive constant $\gamma(1-\gamma)d(F_1,F_2)$ $[\mathbb{P}]~ a.s.$ under $H_{1\gamma}$ (Theorem 3.3), \eqref{upper_bound_T_perm} implies that the permutation p-value $p_n=\sum\limits_{r=1}^{R}\mathbb{I}(T_{n}^{(r)} > T_{n})/R \xrightarrow[]{} 0\: \: [\widetilde{\mathbb{P}}]~a.s.$ under $H_{1\gamma}$ as $\mathbb{I}(T_{n}^{(r)} > T_{n}) \xrightarrow[]{} 0\,\,  [\widetilde{\mathbb{P}}]~a.s.$  for every $r=1,2,\dots,R$. Since this is true for every $\gamma \in (0,1)$, $p_n \xrightarrow[]{} 0$ under $H_{1}$. This completes the proof of Theorem 3.4.

\subsection*{ Proof of Theorem 3.5} 
Under single changepoint setup, for some prefixed $\gamma \in (0,1)$, simple algebraic calculation gives that 
$$
\left| \gamma(1-\gamma)S(\gamma) - \widetilde{\gamma}(1-\widetilde{\gamma})S(\widetilde{\gamma}) \right|=\begin{cases}
	\frac{(1-\gamma)(\gamma-\widetilde{\gamma})}{1-\widetilde{\gamma}} & \text{if $\widetilde{\gamma} \leq \gamma$},\\
	\frac{\gamma(\widetilde{\gamma}-\gamma)}{\widetilde{\gamma}} & \text{if $\widetilde{\gamma} > \gamma$},
\end{cases}
$$
where $S(\widetilde{\gamma})$ is defined in \eqref{S_gamma}. Note that $S(\gamma)=1$. 
This implies
\begin{align*}
\left| \gamma(1-\gamma)d(F_{1},F_{2}) - \widetilde{\gamma}(1-\widetilde{\gamma})S(\widetilde{\gamma})d(F_{1},F_{2}) \right| \geq \left| 
\gamma - \widetilde{\gamma} \right| \min(\gamma, 1 - \gamma)d(F_{1},F_{2}).
\end{align*}
Hence, for every $\epsilon > 0$
\begin{align}\label{epsmin1}
	\left| \gamma(1-\gamma)d(F_{1},F_{2}) - \widetilde{\gamma}(1-\widetilde{\gamma})S(\widetilde{\gamma})d(F_{1},F_{2}) \right| > \epsilon\min(\gamma, 1 - \gamma)d(F_{1},F_{2}) \: \: \text{for all} \: \: \widetilde{\gamma}\: \text{such that} \:\left|\widetilde{\gamma} - \gamma \right| > \epsilon. 
\end{align}
By (\ref{eq17}), there exists $N^{*} \in \mathbb{N}$ such that
\begin{align}\label{epsmin2}
	\left| \gamma(1-\gamma)d(F_{1},F_{2}) - \widehat{\gamma}_{n}(1-\widehat{\gamma}_{n})S(\widehat{\gamma}_{n})d(F_{1},F_{2}) \right| < \epsilon\min(\gamma, 1 - \gamma)d(F_{1},F_{2}) ~
	[\mathbb{P}] ~ a.s. \: \text{whenever } n > N^{*}.
\end{align}
If $ \left|\widehat{\gamma}_{n} - \gamma \right| > \epsilon$ $[\mathbb{P}] ~ a.s.$ for some $n > N^{*}$, then \eqref{epsmin1} holds for $\widetilde{\gamma} = \widehat{\gamma}_n$ which contradicts \eqref{epsmin2}. Therefore,
$ \left|\widehat{\gamma}_{n} - \gamma \right| < \epsilon$ $[\mathbb{P}] ~ a.s.$ for all $n > N^{*}$.
In other words, $\widehat{\gamma}_{n} \xrightarrow[]{} \gamma$ $[\mathbb{P}] ~ a.s.$ This completes the proof of Theorem 3.5.
\subsection*{ Proof of Lemma 4.1}
Let $\widetilde{\gamma} \in [\gamma^{(i-1)},\gamma^{(i)}]$ and $\widetilde{s}$ = $\lfloor n\widetilde{\gamma} \rfloor$. Then
\begin{align*}
	G_{n}(\widetilde{\gamma}) = \left\{Z_1,Z_2,\dots,Z_{\widetilde{s}}\right\} \: \: \text{and} \: \: H_{n}(\widetilde{\gamma}) = \left\{Z_{\widetilde{s}+1},Z_{\widetilde{s}+2},\dots,Z_{n}\right\}. 
\end{align*}
Although there are actually $i-1$ many changepoints present in $G_{n}(\widetilde{\gamma})$, this set can be considered to contain one changepoint as $G_{n}(\widetilde{\gamma})$ can be considered as a possible realization of mixture distributions of $Q_{1}^{i}$ and $F_{i}$ where
\begin{align*}
Q_{1}^{i} := \sum\limits_{j=1}^{i-1} \frac{\gamma^{(j)}-\gamma^{(j-1)}}{\gamma^{(i-1)}} F_{j}.    
\end{align*}
 Applying equation (\ref{eq10}) for $G_{n}(\widetilde{\gamma})$, we get
\begin{align} \label{eq20}
	\underset{\widetilde{\gamma} \in [\gamma^{(i-1)}, \gamma^{(i)}]}{\sup} \left|\frac{1}{\widetilde{s}^{2}}\sum\limits_{j=1}^{\widetilde{s}}\sum\limits_{j'=1}^{\widetilde{s}}k(Z_{j},Z_{j'})  - \left\{\left(\frac{\gamma^{(i-1)}}{\widetilde{\gamma}}\right)^{2}\mu_{k}^{X} + \left(\frac{\widetilde{\gamma}-\gamma^{(i-1)}}{\widetilde{\gamma}}\right)^{2}\mu_{k}^{U} +  \frac{2\gamma^{(i-1)}(\widetilde{\gamma}-\gamma^{(i-1)})}{\widetilde{\gamma}^{2}}\mu_{k}^{XU}\right\} \right| \xrightarrow[]{} \: 0 \quad [\mathbb{P}] ~ a.s., 
\end{align}
where $\mu_{k}^{X}=\mathbb{E}[k(X,X')], \mu_{k}^{U}=\mathbb{E}[k(U,U')]$ and $\mu_{k}^{XU}=\mathbb{E}[k(X,U)]$ with $X,X' \stackrel{i.i.d}{\sim} Q_{1}^{i}, U,U' \stackrel{i.i.d}{\sim} F_{i}$ and $X$ and $U$ are independent.
Similarly for $H_{n}(\widetilde{\gamma})$,
\begin{flalign} \label{eq21}
	\underset{\widetilde{\gamma} \in [\gamma^{(i-1)}, \gamma^{(i)}]}{\sup} \left|\frac{1}{(n-\widetilde{s})^{2}}\sum\limits_{j=\widetilde{s}+1}^{n}\sum\limits_{j'=\widetilde{s}+1}^{n}k(Z_{j},Z_{j'}) - \left\{\left(\frac{\gamma^{(i)}-\widetilde{\gamma}}{1-\widetilde{\gamma}}\right)^{2}\mu_{k}^{U} + \left(\frac{1-\gamma^{(i)}}{1-\widetilde{\gamma}}\right)^{2}\mu_{k}^{Y}  +  \frac{2(\gamma^{(i)}-\widetilde{\gamma})(1-\gamma^{(i)})}{1-\widetilde{\gamma}^{2}}\mu_{k}^{YU}\right\} \right| \xrightarrow[]{} \: 0 \quad [\mathbb{P}] ~ a.s., 
\end{flalign}
where $\mu_{k}^{Y}=\mathbb{E}[k(Y,Y')]$ and $\mu_{k}^{YU}=\mathbb{E}[k(Y,U)]$ with $Y,Y' \stackrel{i.i.d}{\sim} Q_{2}^{i}$, $X,Y$ and $U$ are independent and
\begin{align*}
    Q_{2}^{i} := \sum\limits_{j=i+1}^{K_0+1} \frac{\gamma^{(j)}-\gamma^{(j-1)}}{1-\gamma^{(i)}} F_{j}.
\end{align*}
For the cross-product term, from the strong law of large numbers for $V$-statistics, it follows that
\begin{flalign} \label{eq22}
	\underset{\widetilde{\gamma} \in [\gamma^{(i-1)}, \gamma^{(i)}]}{\sup} &\left|\frac{1}{\widetilde{s}(n-\widetilde{s})}\sum\limits_{j=1}^{\widetilde{s}}\sum\limits_{j'=\widetilde{s}+1}^{n}k(Z_{j},Z_{j'}) - \left\{\frac{\gamma^{(i-1)}(\gamma^{(i)}-\widetilde{\gamma})}{\widetilde{\gamma}(1-\widetilde{\gamma})}\mu_{k}^{XU} + \frac{(\widetilde{\gamma}-\gamma^{(i-1)})(1-\gamma^{(i)})}{\widetilde{\gamma}(1-\widetilde{\gamma})}\mu_{k}^{U} \right. \right. \nonumber\\ &\left. \left. +  \frac{\gamma^{(i-1)}(1-\gamma^{(i)})}{\widetilde{\gamma}(1-\widetilde{\gamma})}\mu_{k}^{XY} + \frac{(\widetilde{\gamma}-\gamma^{(i-1)})(1-\gamma^{(i)})}{\widetilde{\gamma}(1-\widetilde{\gamma})}\mu_{k}^{YU}\right\} \right| \xrightarrow[]{} \: 0 \quad [\mathbb{P}] ~ a.s. 
\end{flalign}
Note that
\begin{align*}
&\left(\frac{\gamma^{(i-1)}}{\widetilde{\gamma}}\right)^{2}\mu_{k}^{X} + \left(\frac{\widetilde{\gamma}-\gamma^{(i-1)}}{\widetilde{\gamma}}\right)^{2}\mu_{k}^{U} +  \frac{2\gamma^{(i-1)}(\widetilde{\gamma}-\gamma^{(i-1)})}{\widetilde{\gamma}^{2}}\mu_{k}^{XU} +  \left(\frac{\gamma^{(i)}-\widetilde{\gamma}}{1-\widetilde{\gamma}}\right)^{2}\mu_{k}^{U} + \left(\frac{1-\gamma^{(i)}}{1-\widetilde{\gamma}}\right)^{2}\mu_{k}^{Y} \\ &+  \frac{2(\gamma^{(i)}-\widetilde{\gamma})(1-\gamma^{(i)})}{1-\widetilde{\gamma}^{2}}\mu_{k}^{YU} -2\left\{\frac{\gamma^{(i-1)}(\gamma^{(i)}-\widetilde{\gamma})}{\widetilde{\gamma}(1-\widetilde{\gamma})}\mu_{k}^{XU} + \frac{(\widetilde{\gamma}-\gamma^{(i-1)})(1-\gamma^{(i)})}{\widetilde{\gamma}(1-\widetilde{\gamma})}\mu_{k}^{U} \right.\\&\left. +  \frac{\gamma^{(i-1)}(1-\gamma^{(i)})}{\widetilde{\gamma}(1-\widetilde{\gamma})}\mu_{k}^{XY} + \frac{(\widetilde{\gamma}-\gamma^{(i-1)})(1-\gamma^{(i)})}{\widetilde{\gamma}(1-\widetilde{\gamma})}\mu_{k}^{YU}\right\}\\
&=d\left(\frac{\gamma^{(i-1)}}{\widetilde{\gamma}}Q_{1}^{i} + \frac{\widetilde{\gamma}-\gamma^{(i-1)}}{\widetilde{\gamma}}F_{i}, \frac{\gamma^{(i)}-\widetilde{\gamma}}{1-\widetilde{\gamma}}F_{i} + \frac{1-\gamma^{(i)}}{1-\widetilde{\gamma}}Q_{2}^{i}\right)
\end{align*}
and 
\begin{align*}
    d_{n}(\widetilde{\gamma}) = \frac{1}{\widetilde{s}^{2}}\sum\limits_{j=1}^{\widetilde{s}}\sum\limits_{j'=1}^{\widetilde{s}}k(Z_{j},Z_{j'}) + \frac{1}{(n-\widetilde{s})^{2}}\sum\limits_{j=\widetilde{s}+1}^{n}\sum\limits_{j'=\widetilde{s}+1}^{n}k(Z_{j},Z_{j'})-\frac{2}{\widetilde{s}(n-\widetilde{s})}\sum\limits_{j=1}^{\widetilde{s}}\sum\limits_{j'=\widetilde{s}+1}^{n}k(Z_{j},Z_{j'})
\end{align*}
Hence, by triangle inequality,
\begin{align*}
    &\underset{\widetilde{\gamma} \in [\gamma^{(i-1)}, \gamma^{(i)}]}{\sup} \left| d_{n}(\widetilde{\gamma}) - d\left(\frac{\gamma^{(i-1)}}{\widetilde{\gamma}}Q_{1}^{i} + \frac{\widetilde{\gamma}-\gamma^{(i-1)}}{\widetilde{\gamma}}F_{i}, \frac{\gamma^{(i)}-\widetilde{\gamma}}{1-\widetilde{\gamma}}F_{i} + \frac{1-\gamma^{(i)}}{1-\widetilde{\gamma}}Q_{2}^{i}\right) \right|\\
    & \leq \underset{\widetilde{\gamma} \in [\gamma^{(i-1)}, \gamma^{(i)}]}{\sup} \left|\frac{1}{\widetilde{s}^{2}}\sum\limits_{j=1}^{\widetilde{s}}\sum\limits_{j'=1}^{\widetilde{s}}k(Z_{j},Z_{j'})   -  \left\{\left(\frac{\gamma^{(i-1)}}{\widetilde{\gamma}}\right)^{2}\mu_{k}^{X} + \left(\frac{\widetilde{\gamma}-\gamma^{(i-1)}}{\widetilde{\gamma}}\right)^{2}\mu_{k}^{U} +  \frac{2\gamma^{(i-1)}(\widetilde{\gamma}-\gamma^{(i-1)})}{\widetilde{\gamma}^{2}}\mu_{k}^{XU}\right\} \right| \\&+\underset{\widetilde{\gamma} \in [\gamma^{(i-1)}, \gamma^{(i)}]}{\sup} \left|\frac{1}{(n-\widetilde{s})^{2}}\sum\limits_{j=\widetilde{s}+1}^{n}\sum\limits_{j'=\widetilde{s}+1}^{n}k(Z_{j},Z_{j'}) - \left\{\left(\frac{\gamma^{(i)}-\widetilde{\gamma}}{1-\widetilde{\gamma}}\right)^{2}\mu_{k}^{U} + \left(\frac{1-\gamma^{(i)}}{1-\widetilde{\gamma}}\right)^{2}\mu_{k}^{Y} \right.\right.  \\ &\left.\left. +  \frac{2(\gamma^{(i)}-\widetilde{\gamma})(1-\gamma^{(i)})}{1-\widetilde{\gamma}^{2}}\mu_{k}^{YU}\right\} \right| + \underset{\widetilde{\gamma} \in [\gamma^{(i-1)}, \gamma^{(i)}]}{\sup} \left|\frac{1}{\widetilde{s}(n-\widetilde{s})}\sum\limits_{j=1}^{\widetilde{s}}\sum\limits_{j'=\widetilde{s}+1}^{n}k(Z_{j},Z_{j'}) \right.\\&- \left. \left\{\frac{\gamma^{(i-1)}(\gamma^{(i)}-\widetilde{\gamma})}{\widetilde{\gamma}(1-\widetilde{\gamma})}\mu_{k}^{XU} + \frac{(\widetilde{\gamma}-\gamma^{(i-1)})(1-\gamma^{(i)})}{\widetilde{\gamma}(1-\widetilde{\gamma})}\mu_{k}^{U}  +  \frac{\gamma^{(i-1)}(1-\gamma^{(i)})}{\widetilde{\gamma}(1-\widetilde{\gamma})}\mu_{k}^{XY} +  \frac{(\widetilde{\gamma}-\gamma^{(i-1)})(1-\gamma^{(i)})}{\widetilde{\gamma}(1-\widetilde{\gamma})}\mu_{k}^{YU}\right\} \right|
\end{align*}
Equations (\ref{eq20}), (\ref{eq21}) and (\ref{eq22}) together imply that 
\begin{align}
	\label{eq30}
    \underset{\widetilde{\gamma} \in [\gamma^{(i-1)}, \gamma^{(i)}]}{\sup} \left| d_{n}(\widetilde{\gamma}) - d\left(\frac{\gamma^{(i-1)}}{\widetilde{\gamma}}Q_{1}^{i} + \frac{\widetilde{\gamma}-\gamma^{(i-1)}}{\widetilde{\gamma}}F_{i}, \frac{\gamma^{(i)}-\widetilde{\gamma}}{1-\widetilde{\gamma}}F_{i} + \frac{1-\gamma^{(i)}}{1-\widetilde{\gamma}}Q_{2}^{i}\right) \right| \xrightarrow[]{} \: 0 \quad [\mathbb{P}] ~ a.s. 
\end{align}
If $\widetilde{\gamma}<\gamma^{(1)}$, the set $G_{n}(\widetilde{\gamma})=\{Z_{1},Z_{2},\dots,Z_{\widetilde{s}}\}$ can be considered to be a set of observations from $F_1$ only while $H_{n}(\widetilde{\gamma})=\{Z_{\widetilde{s}+1},Z_{\widetilde{s}+2},\dots,Z_{n}\}$ can be considered as a set of observations from the mixture distribution of $F_1$ and $Q_{2}^{1}$ where 
\begin{align*}
    Q_{2}^{1} := \sum\limits_{j=2}^{K_0+1} \frac{\gamma^{(j)}-\gamma^{(j-1)}}{1-\gamma^{(1)}} F_{j}.
\end{align*}
By Lemma 3.2,
\begin{align}
    \label{eq31}\underset{\widetilde{\gamma} \in [\delta_n, \gamma^{(1)}]}{\sup} \left| d_{n}  (\widetilde{\gamma}) - d\left(F_{1},\frac{\gamma^{(1)}-\widetilde{\gamma}}{1-\widetilde{\gamma}}F_{1}+\frac{1-\gamma^{(1)}}{1-\widetilde{\gamma}}Q_{2}^{1} \right)\right| \xrightarrow[]{}  0 \: \: [\mathbb{P}]~a.s.
\end{align}
Similarly, it can also be proved that
\begin{align}
    \label{eq32}\underset{\widetilde{\gamma} \in [\gamma^{(K_0)},1-\delta_{n}]}{\sup} \left| d_{n}(\widetilde{\gamma}) -  d\left(\frac{\gamma^{(K_0)}}{\widetilde{\gamma}}Q_{1}^{K_0+1}+\frac{\widetilde{\gamma}-\gamma^{(K_0)}}{\widetilde{\gamma}}F_{K_0+1},F_{K_0+1} \right)\right| \xrightarrow[]{}  0 \: \: [\mathbb{P}]~a.s.,
\end{align}
where
\begin{align*}
    Q_{1}^{K_0+1} :=\sum\limits_{j=1}^{K_0} \frac{\gamma^{(j)}-\gamma^{(j-1)}}{\gamma^{(K_0)}} F_{j}
\end{align*}
Combining \eqref{eq30}, \eqref{eq31} and \eqref{eq32}, it follows that
\begin{align*}
      	\underset{\widetilde{\gamma} \in [\delta_n, 1-\delta_n]}{\sup} \left| d_{n}(\widetilde{\gamma}) - d_{**}(\widetilde{\gamma})\right|  \xrightarrow[]{} \: 0 \: \: \text{ $[\mathbb{P}]$ a.s.}.
        \end{align*}
        where $d_{**}(\widetilde{\gamma})$ is defined in Lemma 4.1 of the main paper.
        This completes the proof of Lemma 4.1.

\subsection*{ Proof of Lemma 4.2}
Let us define $S_{i}(\widetilde{\gamma})=\widetilde{\gamma}(1-\widetilde{\gamma})d_{**}(\widetilde{\gamma})\mathbb{I}(\widetilde{\gamma} \in [\gamma^{(i-1)},\gamma^{(i)}])$ which represents the population version of $\rho$ distance when the observations are split around $\lfloor n\widetilde{\gamma}\rfloor$. Note that 
\begin{equation}
\label{S_i for mcp}
S_{i}(\widetilde{\gamma}) = \newline
\begin{cases}
	\widetilde{\gamma}(1 - \widetilde{\gamma})d\left(\frac{\gamma^{(i-1)}}{\widetilde{\gamma}}Q_{1}^{i} + \frac{\widetilde{\gamma}-\gamma^{(i-1)}}{\widetilde{\gamma}}F_{i}, \frac{\gamma^{(i)}-\widetilde{\gamma}}{1-\widetilde{\gamma}}F_{i} + \frac{1-\gamma^{(i)}}{1-\widetilde{\gamma}}Q_{2}^{i}\right)\mathbb{I}(\widetilde{\gamma} \in [\gamma^{(i-1)},\gamma^{(i)}]) \quad \\ \hspace{9.8cm} \text{if} \quad 2 \leq i \leq K_0,\\
	\widetilde{\gamma}(1 - \widetilde{\gamma})d\left(F_{1},\frac{\gamma^{(1)}-\widetilde{\gamma}}{1-\widetilde{\gamma}}F_{1}+\sum\limits_{j=2}^{K_0+1} \frac{\gamma^{(j)}-\gamma^{(j-1)}}{1-\widetilde{\gamma}} F_{j}\right)\mathbb{I}(\widetilde{\gamma} \in (0,\gamma^{(1)}]) \quad \text{if} \quad i=1,\\
	\widetilde{\gamma}(1 - \widetilde{\gamma})d\left(\sum\limits_{j=1}^{K_0} \frac{\gamma^{(j)}-\gamma^{(j-1)}}{\widetilde{\gamma}}F_{j}+\frac{\widetilde{\gamma}-\gamma^{(K_0)}}{\widetilde{\gamma}}F_{K_0+1},F_{K_0+1}\right)\mathbb{I}(\widetilde{\gamma} \in [\gamma^{(K_0)},1)) \,\, \text{ if } \,\, i=K_0+1.
\end{cases}
\end{equation}
\\
Carrying out some algebraic calculations, we obtain for $i \in \{2,3,\dots,K_0\}$,
\begin{align*}
	S_{i}(\widetilde{\gamma}) =& \left\{\gamma^{(i-1)}\gamma^{(i)} - \gamma^{(i-1)^{2}} - \gamma^{(i-1)} + \frac{\gamma^{(i-1)^{2}}}{\widetilde{\gamma}}\right\}d(Q_{1}^{i},F_{i}) + \gamma^{(i-1)}(1 - \gamma^{(i)})d(Q_{1}^{i},Q_{2}^{i})\\&+ \left\{\frac{\widetilde{\gamma}}{1-\widetilde{\gamma}}(1 - \gamma^{(i)})^{2} - \gamma^{(i-1)}(1 - \gamma^{(i)})\right\}d(F_{i},Q_{2}^{i}),
\end{align*}
when $\widetilde{\gamma} \in [\gamma^{(i-1)},\gamma^{(i)}]$ for some $i \in \{2,3,\dots,K_{0}\}$.
By taking the first and the second order derivatives of $S_{i}(\widetilde{\gamma})$, we obtain
\begin{align} \label{eq23}
	S_{i}'(\widetilde{\gamma}) = \frac{(1-\gamma^{(i)})^{2}}{(1-\widetilde{\gamma})^{2}}d(F_{i},Q_{2}^{i}) - \frac{\gamma^{(i-1)^{2}}}{\widetilde{\gamma}^{2}}d(Q_{1}^{i},F_{i})
\end{align}
\begin{center}
	and
\end{center}
\begin{align} \label{eq24}
	S_{i}''(\widetilde{\gamma}) = 2\left\{\frac{(1-\gamma^{(i)})^{2}}{(1-\widetilde{\gamma})^{3}}d(F_{i},Q_{2}^{i}) + \frac{\gamma^{(i-1)^{2}}}{\widetilde{\gamma}^{3}}d(Q_{1}^{i},F_{i})\right\}  > 0.  
\end{align}
Hence, for $i \in \{2,3,\dots,K_{0}\}$, $S_{i}(\widetilde{\gamma})$ is convex in the interval $[\gamma^{(i-1)},\gamma^{(i)}]$ and therefore, $S_{i}(\widetilde{\gamma})$ is maximized either at $\widetilde{\gamma}=\gamma^{(i-1)}$ or at $\widetilde{\gamma}=\gamma^{(i)}$.
\par Note that $S_{1}(\widetilde{\gamma})=\{\widetilde{\gamma}(1-\gamma^{(1)})^{2}/(1-\widetilde{\gamma})\}d(F_1,Q_{2}^{1})\mathbb{I}(\widetilde{\gamma} \in (0,\gamma^{(1)}])$ is an increasing function of $\widetilde{\gamma}$. Hence, $S_{1}(\widetilde{\gamma})$ is maximized at $\widetilde{\gamma}=\gamma^{(1)}$. On the other hand, $S_{K_0+1}(\widetilde{\gamma})=\{\gamma^{(K_0)^{2}}(1-\widetilde{\gamma})/\widetilde{\gamma}\}d(Q_{1}^{K_0+1},F_{K_0+1})\mathbb{I}(\widetilde{\gamma} \in [\gamma^{(K_0)},1))$ is a decreasing function of $\widetilde{\gamma}$. Hence, $S_{K_0+1}(\widetilde{\gamma})$ is maximized at $\widetilde{\gamma}=\gamma^{(K_0)}$. This completes the proof of Lemma 4.2.

\subsection*{ Proof of Theorem 4.3}
Let $M(\widetilde{
	\gamma
})=\sum\limits_{i=1}^{K_0+1}S_{i}(\widetilde{\gamma})-\sum\limits_{i=1}^{K_0}S_{i+1}(\widetilde{\gamma})\mathbb{I}(\widetilde{\gamma}=\gamma^{(i)})$, where $S_{i}(\widetilde{\gamma})$ is defined in \eqref{S_i for mcp}. 
Note that $M(\widetilde{\gamma})=S_{i}(\widetilde{\gamma})$ if $\widetilde{\gamma} \in [\gamma^{(i-1)},\gamma^{(i)}]$. By Lemma 4.2, $\underset{\widetilde{\gamma} \in (0,1)}{\arg\max} M(\widetilde{\gamma}) \subseteq \{\gamma^{(1)},\gamma^{(2)},\dots,\gamma^{(K_0)}\}$.
\\
Let $\underset{\widetilde{\gamma} \in (0,1)}{\arg\max} M(\widetilde{\gamma})=\mathcal{C}$. Then for any $j \in \{1,2,\dots,K_0\}$,
\begin{align*}
	S_{j}(\gamma^{(j)})&=\gamma^{(j)}(1-\gamma^{(j)})d\left(\frac{\gamma^{(j-1)}}{\gamma^{(j)}}Q_{1}^{j} + \frac{\gamma^{(j)}-\gamma^{(j-1)}}{\gamma^{(j)}}F_{j}, \frac{1-\gamma^{(j)}}{1-\gamma^{(j)}}Q_{2}^{j}\right)\\
	&=\gamma^{(j)}(1-\gamma^{(j)})d\left(\sum\limits_{i=1}^{j}\frac{\gamma^{(i)}-\gamma^{(i-1)}}{\gamma^{(j)}}F_{i}, \sum\limits_{i=j+1}^{K_0+1}\frac{\gamma^{(i)}-\gamma^{(i-1)}}{1-\gamma^{(j)}}F_{i}\right).
\end{align*}
Now,
\begin{align*}
	S_{j+1}(\gamma^{(j)})&=\gamma^{(j)}(1-\gamma^{(j)})d\left(\frac{\gamma^{(j)}}{\gamma^{(j)}}Q_{1}^{j+1}, \frac{\gamma^{(j+1)}-\gamma^{(j)}}{1-\gamma^{(j)}}F_{j+1}+\frac{1-\gamma^{(j+1)}}{1-\gamma^{(j)}}Q_{2}^{j+1}\right)\\
	&= \gamma^{(j)}(1-\gamma^{(j)})d\left(\sum\limits_{i=1}^{j}\frac{\gamma^{(i)}-\gamma^{(i-1)}}{\gamma^{(j)}}F_{i}, \sum\limits_{i=j+1}^{K_0+1}\frac{\gamma^{(i)}-\gamma^{(i-1)}}{1-\gamma^{(j)}}F_{i}\right)\\
	&= S_{j}(\gamma^{(j)}).
\end{align*}
Therefore, for any $j \in \{1,2,\dots,K_0\}$, $S_{j}(\gamma^{(j)})=S_{j+1}(\gamma^{(j)})$.         Since $\frac{\widehat{\tau}_{n}(n-\widehat{\tau}_{n})}{n^2}-\widehat{\gamma}_{n}(1-\widehat{\gamma}_{n}) = 0$, $\rho_{n}(\widehat{\gamma}_{n})-M(\widehat{\gamma}_{n}) \xrightarrow[]{} 0 \quad [\mathbb{P}] ~ a.s.$ (Follows from Lemma 4.1). 
Therefore, for any $\epsilon^{*}>0$, there exists $N^{*} \in \mathbb{N}$ such that whenever $n \geq N^{*}$,
\begin{align*}
	M(\widehat{\gamma}_{n})+\frac{\epsilon^{*}}{2} > \rho_{n}(\widehat{\gamma}_{n}) \geq \underset{\widetilde{\gamma} \in \mathcal{C}}{\max} \rho_{n}(\widetilde{\gamma}) > M_{\max} - \frac{\epsilon^{*}}{2} \quad \text{$[\mathbb{P}]$ a.s.},
\end{align*}
where $M_{\max}=\underset{\widetilde{\gamma} \in (0,1)}{\max}M(\widetilde{\gamma})$.
Hence, $|M_{\max}-M(\widehat{\gamma}_{n})|<\epsilon^{*}$ whenever $n \geq N^{*}$. Since $\epsilon^{*}$ is an arbitrary positive number,
\begin{align} \label{eq36}
	M(\widehat{\gamma}_{n})-M_{\max} \xrightarrow[]{} 0 \quad [\mathbb{P}] ~ a.s.
\end{align}
Therefore, $T_n=\rho_{n}(\widehat{\gamma}_{n}) \xrightarrow[]{} M_{\max}>0$ (as $\rho_{n}(\widehat{\gamma}_{n})-M(\widehat{\gamma}_{n}) \xrightarrow[]{} 0 \quad [\mathbb{P}] ~ a.s.$).\\\\
For $\gamma^{(j)} \in \mathcal{C}$, let us define
\begin{align*}
	\gamma^{j}_{\min}=\min\{\gamma \in [\gamma^{(j-1)},\gamma^{(j)}] : M(\widetilde{\gamma}) \text{ is increasing in the interval } [\gamma,\gamma^{(j)}]\}
\end{align*}
\begin{center}
	and
\end{center}
\begin{align*}
	\gamma^{j+1}_{\min}=\max\{\gamma \in [\gamma^{(j)},\gamma^{(j+1)}] : M(\widetilde{\gamma}) \text{ is decreasing in the interval } [\gamma^{(j)},\gamma]\}.
\end{align*}
These $\gamma^{j}_{\min}$ and $\gamma^{j+1}_{\min}$ always exist as $M(\widetilde{\gamma})$ is convex in the intervals $[\gamma^{(j-1)},\gamma^{(j)}]$ and $[\gamma^{(j)},\gamma^{(j+1)}]$ and $M(\gamma^{(j)}) \geq M(\widetilde{\gamma})$ for all $\widetilde{\gamma}$.\\
If $\mathcal{C}=\{\gamma^{(1)},\gamma^{(2)},\dots,\gamma^{(K_0)}\}$, then $\cup_{j:\gamma^{(j)} \in \mathcal{C}}[\gamma^{j}_{\min},\gamma^{j+1}_{\min}]=[0,1]$ and hence\\ $\widehat{\gamma}_{n} \in \cup_{j:\gamma^{(j)} \in \mathcal{C}}[\gamma^{j}_{\min},\gamma^{j+1}_{\min}]$.\\ 
Now, assume that $\mathcal{C} \subset \{\gamma^{(1)},\gamma^{(2)},\dots,\gamma^{(K_0)}\}$. 
Let $\gamma^{*} = \underset{\widetilde{\gamma} \in \{\gamma^{(1)},\gamma^{(2)},\dots,\gamma^{(K_0)}\}\backslash \mathcal{C}}{\arg\max}M(\widetilde{\gamma})$. In other words, $\gamma^{*}$ be the element (or set of elements) of the set of true breakfractions at which the function $M(\widetilde{\gamma})$ attains second maximum value. If $\widehat{\gamma}_{n} \not\in \cup_{j:\gamma^{(j)} \in \mathcal{C}}[\gamma^{j}_{\min},\gamma^{j+1}_{\min}]$, then $M(\gamma^{*}) \geq M(\widehat{\gamma}_{n})$ and for every $j$ such that $\gamma^{(j)} \in \mathcal{C}$,
\begin{align*}
	M_{\max}-M(\widehat{\gamma}_{n}) = M(\gamma^{(j)})-M(\widehat{\gamma}_{n}) \geq M(\gamma^{(j)})-M(\gamma^{*}) > 0,
\end{align*}
which contradicts equation \eqref{eq36}. Hence, there exists 
$\widetilde{N} \in \mathbb{N}$ such that $\widehat{\gamma}_{n} \in \cup_{j:\gamma^{(j)} \in \mathcal{C}}[\gamma^{j}_{\min},\gamma^{j+1}_{\min}]$ whenever $n \geq \widetilde{N}$. 
\\
For each $j_1,j_2 \in \{1,2,\dots,K_0\}$ such that $\gamma^{(j_1)}, \gamma^{(j_2)} \in \mathcal{C}$, the intersection of the intervals $[\gamma^{j_1}_{\min},\gamma^{j_1+1}_{\min}]$ and $[\gamma^{j_2}_{\min},\gamma^{j_2+1}_{\min}]$ is at most singleton set. 
Let $\gamma^{(j_n)} \in \mathcal{C}$ such that $\widehat{\gamma}_{n} \in [\gamma^{j_n}_{\min},\gamma^{j_n+1}_{\min}]$ for $n \geq \widetilde{N}$. If $\widehat{\gamma}_{n} \in [\gamma^{j_n}_{\min},\gamma^{(j_n)}]$ then,  
\begin{align} \label{eq27}
 \left|M_{\max}-M(\widehat{\gamma}_{n})\right| 
	&=\left| S_{j_n}(\gamma^{(j_n)}) - S_{j_n}(\widehat{\gamma}_{n}) \right| \nonumber\\&= \left| \left( 
	\frac{\gamma^{(j_n-1)^{2}}}{\gamma^{(j_n)}}  - \frac{\gamma^{(j_n-1)^{2}}}{\widehat{\gamma}_{n}}\right)d(Q_{1}^{j_n},F_{j_n}) + \left\{1-\gamma^{(j_n)} - \frac{(1-\gamma^{(j_n)})^{2}}{1-\widehat{\gamma}_{n}}\right\}d(F_{j_n},Q_{2}^{j_n}) \right| \nonumber\\
	&= \left| \frac{(1-\gamma^{(j_n)})(\gamma^{(j_n)}-\widehat{\gamma}_{n})}{1-\widehat{\gamma}_{n}}d(F_{j_n},Q_{2}^{j_n}) - \frac{\gamma^{(j_n-1)^{2}}(\gamma^{(j_n)}-\widehat{\gamma}_{n})}{\widehat{\gamma}_{n}\gamma^{(j_n)}}d(Q_{1}^{j_n},F_{j_n}) \right| \nonumber\\
	&= \left|\gamma^{(j_n)} - \widehat{\gamma}_{n} \right| \left| \frac{1-\gamma^{(j_n)}}{1-\widehat{\gamma}_{n}}d(F_{j_n},Q_{2}^{j_n}) - \frac{\gamma^{(j_n-1)^{2}}}{\widehat{\gamma}_{n}\gamma^{(j_n)}}d(Q_{1}^{j_n},F_{j_n}) \right|. 
\end{align}
Now,
\begin{align*}
	\frac{1-\gamma^{(j_n)}}{1-\widehat{\gamma}_{n}}d(F_{j_n},Q_{2}^{j_n}) - \frac{\gamma^{(j_n-1)^{2}}}{\widehat{\gamma}_{n}\gamma^{(j_n)}}d(Q_{1}^{j_n},F_{j_n}) &\geq \frac{1-\gamma^{(j_n)}}{1-\gamma^{j_n}_{\min}}d(F_{j_n},Q_{2}^{j_n}) - \frac{\gamma^{(j_n-1)^{2}}}{\gamma^{j_n}_{\min}\gamma^{(j_n)}}d(Q_{1}^{j_n},F_{j_n})   \\
	&=\frac{S_{j_n}(\gamma^{(j_n)})-S_{j_n}(\gamma^{j_n}_{\min})}{\gamma^{(j_n)}-\gamma^{j_n}_{\min}}\\
	& \geq \underset{j:\gamma^{(j)} \in \mathcal{C}}{\min} \{M(\gamma^{(j)})-M(\gamma^{j}_{\min})\} = A \quad \text{(say)}.
\end{align*}
Note that the function $M(\widetilde{\gamma})$ is convex in the interval $[\gamma^{(j)-1},\gamma^{(j)}]$ and it is strictly increasing in the interval $[\gamma_{\min}^{j},\gamma^{(j)}]$ for any $\gamma^{(j)} \in \mathcal{C}$. Therefore, $A>0$. Then $\left|M_{\max}-M(\widehat{\gamma}_{n})\right| \geq A\left|\gamma^{(j_n)} - \widehat{\gamma}_{n} \right|$ where $A>0$. Thus, whenever $\left|\gamma^{(j_n)} - \widehat{\gamma}_{n} \right|>\epsilon$ then $\left|M_{\max}-M(\widehat{\gamma}_{n})\right| > A \epsilon$.\\
Similarly, if $\widehat{\gamma}_{n} \in [\gamma^{(j_n)}, \gamma^{j_n+1}_{\min}]$, then,
\begin{align*}
|M_{\max}-M(\widehat{\gamma}_{n})|
	&=|S_{j_n+1}(\gamma^{(j_n)})-S_{j_n+1}(\widehat{\gamma}_{n})| \\&= |\widehat{\gamma}_{n}-\gamma^{(j_n)}|\left\{\frac{\gamma^{(j_n)}}{\widehat{\gamma}_{n}}d(Q_{1}^{j_n+1},F_{j_n+1}) - \frac{(1-\gamma^{(j_n+1)})^2}{(1-\gamma^{(j_n)})(1-\widehat{\gamma}_{n})}d(F_{j_n+1},Q_{2}^{j_n+1})\right\}
	\\ & \geq |\widehat{\gamma}_{n}-\gamma^{(j_n)}|\left\{\frac{\gamma^{(j_n)}}{\gamma_{\min}^{j_n+1}}d(Q_{1}^{j_n+1},F_{j_n+1}) - \frac{(1-\gamma^{(j_n+1)})^2}{(1-\gamma^{(j_n)})(1-\gamma_{\min}^{j_n+1})}d(F_{j_n+1},Q_{2}^{j_n+1})\right\}
	\\ &= |\widehat{\gamma}_{n}-\gamma^{(j_n)}| \frac{|S_{j_n+1}(\gamma^{(j_n)})-S_{j_n+1}(\gamma_{\min}^{j_n+1})|}{|\gamma_{\min}^{j_n+1}-\widehat{\gamma}_{n}|}
	\\
	& \geq |\widehat{\gamma}_{n}-\gamma^{(j_n)}| \underset{j:\gamma^{(j)} \in \mathcal{C}}{\min} \{M(\gamma^{(j)})-M(\gamma^{j}_{\min})\} = A|\widehat{\gamma}_{n}-\gamma^{(j_n)}|.
\end{align*}
 Hence, there exists a positive constant $B$ such that $\left|M_{\max}-M(\widehat{\gamma}_{n})\right| > A \epsilon$ whenever $\left|\gamma^{(j_n)} - \widehat{\gamma}_{n} \right| > \epsilon$.\par Thus, whenever $\left|\gamma^{(j_n)} - \widehat{\gamma}_{n} \right|>\epsilon$,  $\left|M_{\max}-M(\widehat{\gamma}_{n})\right| > A \epsilon$ which contradicts \eqref{eq36}. Hence, for every $\epsilon > 0$ there exists $N \in \mathbb{N}$ such that
\begin{align} \label{eq37}
	\left|\gamma^{(j_n)} - \widehat{\gamma}_{n} \right|<\epsilon \: \text{ $[\mathbb{P}] ~ a.s.$ whenever $n \geq N$.}
\end{align}
As $\gamma^{(j_n)} \in \mathcal{C}$ for all $n$, it follows from \eqref{eq37} that
\begin{align*}
	\underset{j:\gamma^{(j)} \in \mathcal{C}}{\min}\left|\gamma^{(j)} - \widehat{\gamma}_{n} \right|<\epsilon \: \text{ $[\mathbb{P}] ~ a.s.$ whenever $n \geq N$}.
\end{align*}
As $\mathcal{C} \subseteq \{\gamma^{(1)},\gamma^{(2)},\dots,\gamma^{(K_0)}\}$ and $\mathcal{C}$ is non-empty,
\begin{align*}
    \underset{1 \leq i \leq K_0}{\min} |\widehat{\gamma}_{n} - \gamma^{(i)}| \xrightarrow[]{} 0 \quad \text{$[\mathbb{P}]$ a.s.}
\end{align*}
This completes the proof of Theorem 4.3.
\subsection*{ Proof of Theorem 5.2}
Consider the multiple changepoints setup described at the beginning of Section 4 of the main paper. Let us define the following sets
\begin{align*}
	P_{r,n} = \{(i,j):Z_i, Z_j \sim F_r\}\: \text{and } P_{r,s,n} = \{(i,j):Z_i \sim F_r, Z_j \sim F_s\},
\end{align*}
for $r,s=1,2,\dots,K_0+1$ and $r \neq s$. Let $\mu_{k}^{r}=\mathbb{E}[k(X,X')]$ and $\mu_{k}^{r,s}=\mathbb{E}[k(X,Y)]$ where $X,X' \stackrel{i.i.d}{\sim} F_{r}$, $Y \sim F_{s}$ and $X, Y$ are independent. Now for any $S_{r,n} \subseteq P_{r,n}$ with $|S_{r,n}| \xrightarrow[]{} \infty$,
\begin{align*}
	\left|\frac{1}{|S_{r,n}|}\sum\limits_{(i,j) \in S_{r,n}}k(Z_i,Z_j)-\mu_{k}^{r}\right| \xrightarrow[]{} 0 \: \: \text{$[\mathbb{P}]$ a.s.},
\end{align*}
for $r=1,2,\dots,K_0+1$. Similarly, 
\begin{align*}
	\left|\frac{1}{|S_{r,s,n}|}\sum\limits_{(i,j) \in S_{r,s,n}}k(Z_i,Z_j)-\mu_{k}^{r,s}\right| \xrightarrow[]{} 0 \:\: \text{$[\mathbb{P}]$ a.s.},
\end{align*}
for any $S_{r,s,n} \subseteq P_{r,s,n}$ with $|S_{r,s,n}| \xrightarrow[]{} \infty$ for $r,s=1,2,\dots,K_0+1 (r \neq s)$.\\
Now, consider the following sets
\begin{align} \label{sets with P probability 1 asymptotic}
	&A_r := \{\omega \in \Omega: |\frac{1}{|S_{r,n}(\omega)|}\sum\limits_{(i,j) \in S_{r,n}(\omega)}k(Z_i(\omega),Z_j(\omega))-\mu_{k}^{r}| \xrightarrow[]{} 0 \: \text{for any } S_{r,n}(\omega) \subseteq P_{r,n}  \text{ with } |S_{r,n}(\omega)| \xrightarrow[]{} \infty\}.\nonumber\\
	&A_{r,s} := \{\omega \in \Omega: |\frac{1}{|S_{r,s,n}(\omega)|}\sum\limits_{(i,j) \in S_{r,s,n}(\omega)}k(Z_i(\omega),Z_j(\omega))-\mu_{k}^{r,s}| \xrightarrow[]{} 0 \: \text{for any }  S_{r,s,n}(\omega) \subseteq P_{r,s,n}(\omega)    \text{ with } |S_{r,s,n}(\omega)| \xrightarrow[]{} \infty\}. 
\end{align}
Fix $\omega \in \cap_{r=1}^{K_0+1} \cap_{s=1}^{K_0+1}\{A_{r} \cap A_{r,s}\}$. Let $C(\omega)$ be a subset of $\{Z_{1}(\omega),Z_{2}(\omega),\dots,Z_{n}(\omega)\}$ with cardinality $\widetilde{n}$ of the form $\{Z_{n^*+1}(\omega),Z_{n^*+2}(\omega),\dots,Z_{n^*+\widetilde{n}}(\omega)\}$ for some $n^* \in \{0,1,\dots,n-\widetilde{n}\}$ such that $\widetilde{n}/n \xrightarrow[]{} \zeta$ for some $\zeta \in (0,1]$. Consider the partition of $C(\omega)$ into two sets 
	$G_{\widetilde{n}}(\omega)(\widetilde{\gamma}) = \{Z_{n^*+1}(\omega),Z_{n^*+2}(\omega),\dots,Z_{n^*+\lfloor \widetilde{n}\widetilde{\gamma}\rfloor}(\omega)\}$ and \\$H_{\widetilde{n}}(\omega)(\widetilde{\gamma})=\{Z_{n^*+\lfloor \widetilde{n}\widetilde{\gamma}\rfloor+1}(\omega),Z_{n^*+\lfloor \widetilde{n}\widetilde{\gamma}\rfloor+2}(\omega),\dots,Z_{n^*+\widetilde{n}}(\omega)\}$.\\
Define,
\begin{align*}
	&d_{\widetilde{n}}(\omega)(\widetilde{\gamma})=d \left(\widehat{P}_{G_{\widetilde{n}}(\omega)(\widetilde{\gamma})},\widehat{P}_{H_{\widetilde{n}}(\omega)(\widetilde{\gamma})} \right), \; \; \rho_{\widetilde{n}}(\omega)(\widetilde{\gamma})=\frac{\lfloor \widetilde{n}\widetilde{\gamma}\rfloor(\widetilde{n}-\lfloor \widetilde{n}\widetilde{\gamma}\rfloor)}{\widetilde{n}^2}d_{\widetilde{n}}(\omega)(\widetilde{\gamma}),\\
	&  T_{\widetilde{n}}(\omega)=\underset{\lceil \widetilde{n}\delta_{\widetilde{n}}\rceil \leq t \leq \lfloor \widetilde{n}(1-\delta_{\widetilde{n}})\rfloor}{\max} \rho_{\widetilde{n}}(\omega)(t/\widetilde{n}),\; \widehat{\tau}_{\widetilde{n}}(\omega) = \underset{\lceil \widetilde{n}\delta_{\widetilde{n}}\rceil \leq t \leq \lfloor \widetilde{n}(1-\delta_{\widetilde{n}})\rfloor}{\arg \max} \rho_{\widetilde{n}}(\omega)(t/\widetilde{n}) \; \text{ and } \\& \widehat{\gamma}_{\widetilde{n}}(\omega)=\frac{\widehat{\tau}_{\widetilde{n}}(\omega)}{\widetilde{n}},
\end{align*}
where $\delta_{\widetilde{n}}$ be a sequence of positive numbers such that $\delta_{\widetilde{n}} \xrightarrow[]{} 0$ and $\widetilde{n}\delta_{\widetilde{n}} \xrightarrow[]{} \infty$.\\
Before proving part \textit{(a)} of Theorem 5.2, let us consider the following lemmas which will play crucial role in proving the theorem.
\newtheorem{supplemma}{Lemma}
\renewcommand{\thesupplemma}{S\arabic{supplemma}}
\begin{supplemma} \label{Tn under asymptotic null}
	Borrowing the notations of the proof of Theorem 5.2, if $$Z_{n^*+ a_n+1}(\omega),Z_{n^*+ a_n+2}(\omega),\dots,Z_{n^*+ b_n}(\omega)$$ are realizations of random variables from any $F_i$ for $1,\dots,K_{0}+1$ and fixed $\omega \in \cap_{r=1}^{K_0+1} \cap_{s=1}^{K_0+1}\{A_{r} \cap A_{r,s}\}$, where $a_n/\widetilde{n} \xrightarrow[]{} 0$ and $b_n/\widetilde{n} \xrightarrow[]{} 1$, then $T_{\widetilde{n}}(\omega) \xrightarrow[]{} 0$.
\end{supplemma}
\begin{proof}
	Define, $g_n = \max{\{a_n/\widetilde{n},\widetilde{n}^{-\lambda}\}}, h_n = \max{\{1-b_n/\widetilde{n},\widetilde{n}^{-\lambda}\}}$ for some $\lambda \in (0,1)$.\\
	For $\widetilde{\gamma} \leq \sqrt{g_n}$,
	\begin{align*}
		\frac{\lfloor \widetilde{n}\widetilde{\gamma}\rfloor(\widetilde{n}-\lfloor \widetilde{n}\widetilde{\gamma}\rfloor)}{\widetilde{n}^2} \leq \sqrt{g_n} \xrightarrow[]{} 0.
	\end{align*}
	For $\widetilde{\gamma} \geq 1-\sqrt{h_n}$,
	\begin{align*}
		\frac{\lfloor \widetilde{n}\widetilde{\gamma}\rfloor(\widetilde{n}-\lfloor \widetilde{n}\widetilde{\gamma}\rfloor)}{\widetilde{n}^2} \leq \frac{\widetilde{n}-\lfloor \widetilde{n}\widetilde{\gamma}\rfloor}{\widetilde{n}} \leq \frac{\widetilde{n}- \widetilde{n}\widetilde{\gamma}+1}{\widetilde{n}} \leq \sqrt{h_n}+\frac{1}{\widetilde{n}} \xrightarrow[]{} 0.
	\end{align*}
	As, $d_{\widetilde{n}}(\omega)(\widetilde{\gamma})$ is bounded,
	\begin{align} \label{Tn in boundaries under asymptotic null}
		\underset{\widetilde{\gamma} \in (0,\sqrt{g_n}]\cup[1-\sqrt{h_n},1)}{\sup}\rho_{\widetilde{n}}(\omega)(\widetilde{\gamma}) \xrightarrow[]{} 0.
	\end{align}
	Let $\mu_{k}^{Z}=\mathbb{E}[k(Z,Z')]$, $Z, Z' \stackrel{i.i.d}{\sim} F_{i}$. Now, take $\widetilde{\gamma} \in (\sqrt{g_n},1-\sqrt{h_n})$ and denote $\lfloor \widetilde{n}\widetilde{\gamma}\rfloor$ by $\widetilde{r}$. Then
	\begin{align} \label{d under asymptotic null}
		d_{\widetilde{n}}(\omega)(\widetilde{\gamma}) &= \frac{1}{\widetilde{r}^2}\sum\limits_{i=1}^{\widetilde{r}}\sum\limits_{i'=1}^{\widetilde{r}}k(Z_{n^*+i}(\omega),Z_{n^*+i'}(\omega)) + \frac{1}{(\widetilde{n} - \widetilde{r})^2}\sum\limits_{j=\widetilde{r}+1}^{\widetilde{n}}\sum\limits_{j'=\widetilde{r}+1}^{\widetilde{n}}k(Z_{n^*+j}(\omega),Z_{n^*+j'}(\omega)) \nonumber\\&\quad- \frac{2}{\widetilde{r}(\widetilde{n} - \widetilde{r})}\sum\limits_{i=1}^{\widetilde{r}}\sum\limits_{j=\widetilde{r}+1}^{\widetilde{n}}k(Z_{n^*+i}(\omega),Z_{n^*+j}(\omega))\nonumber\\
		&=\frac{1}{\widetilde{r}^2}\sum\limits_{i=1}^{a_n}\sum\limits_{i'=1}^{a_n}k(Z_{n^*+i}(\omega),Z_{n^*+i'}(\omega))+\left\{\frac{1}{\widetilde{r}^2}\sum\limits_{i=a_n+1}^{\widetilde{r}}\sum\limits_{i'=a_n+1}^{\widetilde{r}}k(Z_{n^*+i}(\omega),Z_{n^*+i'}(\omega))-\mu_{k}^{Z}\right\} \nonumber\\&\quad+\frac{2}{\widetilde{r}^2}\sum\limits_{i=1}^{a_n}\sum\limits_{i'=a_n+1}^{\widetilde{r}}k(Z_{n^*+i}(\omega),Z_{n^*+i'}(\omega))+\left\{\frac{1}{(\widetilde{n} - \widetilde{r})^2}\sum\limits_{j=\widetilde{r}+1}^{ b_n}\sum\limits_{j'=\widetilde{r}+1}^{ b_n}k(Z_{n^*+j}(\omega),Z_{n^*+j'}(\omega))-\mu_{k}^{Z}\right\} \nonumber\\&\quad+\frac{1}{(\widetilde{n} - \widetilde{r})^2}\sum\limits_{j= b_n+1}^{\widetilde{n}}\sum\limits_{j'= b_n+1}^{\widetilde{n}}k(Z_{n^*+j}(\omega),Z_{n^*+j'}(\omega))+\frac{2}{(\widetilde{n} - \widetilde{r})^2}\sum\limits_{j=\widetilde{r}+1}^{ b_n}\sum\limits_{j'= b_n+1}^{\widetilde{n}}k(Z_{n^*+j}(\omega),Z_{n^*+j'}(\omega)) \nonumber\\&\quad-\frac{2}{\widetilde{r}(\widetilde{n} - \widetilde{r})}\sum\limits_{i=1}^{a_n}\sum\limits_{j=\widetilde{r}+1}^{ b_n}k(Z_{n^*+i}(\omega),Z_{n^*+j}(\omega))-\frac{2}{\widetilde{r}(\widetilde{n} - \widetilde{r})}\sum\limits_{i=1}^{a_n}\sum\limits_{j= b_n+1}^{\widetilde{n}}k(Z_{n^*+i}(\omega),Z_{n^*+j}(\omega)) \nonumber\\
		&\quad-\left\{\frac{2}{\widetilde{r}(\widetilde{n} - \widetilde{r})}\sum\limits_{i=a_n+1}^{\widetilde{r}}\sum\limits_{j=\widetilde{r}+1}^{ b_n}k(Z_{n^*+i}(\omega),Z_{n^*+j}(\omega))-2\mu_{k}^{Z}\right\} \nonumber\\&\quad -\frac{2}{\widetilde{r}(\widetilde{n} - \widetilde{r})}\sum\limits_{i=a_n+1}^{\widetilde{r}}\sum\limits_{j= b_n+1}^{\widetilde{n}}k(Z_{n^*+i}(\omega),Z_{n^*+j}(\omega)), \nonumber\\
        &= I_{1} + I_{2} + 2I_{3} + I_{4} + I_{5} + 2I_{6} - 2I_{7} - 2I_{8} - 2I_{9} - 2I_{10}. 
	\end{align}
 
	Now, 
	\begin{align} \label{convergence of I1}
		I_{1}=\frac{1}{\widetilde{r}^2}\sum\limits_{i=1}^{a_n}\sum\limits_{i'=1}^{a_n}k(Z_{n^*+i}(\omega),Z_{n^*+i'}(\omega)) \leq \frac{a_n^2}{\widetilde{r}^2} \leq \frac{a_n}{\widetilde{n}} \xrightarrow[]{} 0, \ \text{as $k$ is bounded.}
        \end{align}
        Next, note that,
        \begin{align*}
        &\widetilde{r}-a_n = \widetilde{r}\left(1-\frac{a_n}{\widetilde{r}}\right) \geq \widetilde{r}\left(1-\sqrt{\frac{a_n}{\widetilde{n}}}\right) \geq \lfloor \widetilde{n}^{1-\lambda/2}\rfloor\left(1-\sqrt{\frac{a_n}{\widetilde{n}}}\right)  \xrightarrow[]{} \infty, \ \frac{\widetilde{r}-a_{n}}{\widetilde{r}} \xrightarrow[]{} 1,
        \end{align*}
      since $\widetilde{n}^{1-\lambda/2} \xrightarrow[]{} \infty$ and $a_{n}/\widetilde{n} \xrightarrow[]{} 0$.  Hence,
        \begin{align} 
        \label{Convergence of I2}       
		 I_{2} = \frac{1}{\widetilde{r}^2}\sum\limits_{i=a_n+1}^{\widetilde{r}}\sum\limits_{i'=a_n+1}^{\widetilde{r}}k(Z_{n^*+i}(\omega),Z_{n^*+i'}(\omega)) - \mu_{k}^{Z}\xrightarrow[]{} 0.
        \end{align}
\noindent
        Further,
        \begin{align} \label{Convergence of I3}
	 I_{3}=\frac{1}{\widetilde{r}^2}\sum\limits_{i=1}^{a_n}\sum\limits_{i'=a_n+1}^{\widetilde{r}}k(Z_{n^*+i}(\omega),Z_{n^*+i'}(\omega)) \leq \frac{a_n(\widetilde{r}-a_n)}{\widetilde{r}^2} \leq \frac{a_n}{\widetilde{r}} \leq \sqrt{\frac{a_n}{n}} \xrightarrow[]{} 0. 
        \end{align}
        Observe that
        \begin{align*}
        &\frac{b_n-\widetilde{r}}{\widetilde{n}-\widetilde{r}} = 1-\frac{\widetilde{n}-b_n}{\widetilde{n}-\widetilde{r}} \geq 1-\frac{\widetilde{n}-b_n}{\widetilde{n}\sqrt{h_n}}=1-\frac{1-b_n/\widetilde{n}}{\sqrt{h_n}} \geq 1-\frac{1-b_n/\widetilde{n}}{\sqrt{1-b_n/\widetilde{n}}} = 1-\sqrt{1-b_n/\widetilde{n}}, \\
        & b_{n}-\widetilde{r} = (\widetilde{n}-\widetilde{r}) \left(1-\sqrt{1-b_n/\widetilde{n}} \right) \geq \lfloor \widetilde{n}\sqrt{h_{n}}\rfloor  \left(1-\sqrt{1-b_n/\widetilde{n}} \right) \geq \lfloor \widetilde{n}^{1-\lambda/2} \rfloor\left(1-\sqrt{1-b_n/\widetilde{n}} \right) \xrightarrow[]{} \infty, \nonumber\\
        & \frac{b_{n}-\widetilde{r}}{\widetilde{n}-\widetilde{r}} \xrightarrow[]{} 1, 
        \end{align*}
   since $\widetilde{n}^{1-\lambda/2} \xrightarrow[]{} \infty$ and $b_{n}/\widetilde{n} \xrightarrow[]{} 1$.  Hence,
        \begin{align} \label{Convergence of I4}
 I_{4} = \frac{1}{(\widetilde{n} - \widetilde{r})^2}\sum\limits_{j=\widetilde{r}+1}^{ b_n}\sum\limits_{j'=\widetilde{r}+1}^{ b_n}k(Z_{n^*+j}(\omega),Z_{n^*+j'}(\omega)) - \mu_{k}^{Z} \xrightarrow[]{} 0.
        \end{align}
        Next, note that
        \begin{align} \label{Convergence of I5}
         I_{5} = \frac{1}{(\widetilde{n} - \widetilde{r})^2}\sum\limits_{j= b_n+1}^{\widetilde{n}}\sum\limits_{j'= b_n+1}^{\widetilde{n}}k(Z_{n^*+j}(\omega),Z_{n^*+j'}(\omega)) \leq \frac{(\widetilde{n}-b_n)^2}{(\widetilde{n} - \widetilde{r})^2} \xrightarrow[]{} 0, 
         \end{align}
         \begin{align} \label{Convergence of I6}
		I_{6} = \frac{1}{(\widetilde{n} - \widetilde{r})^2}\sum\limits_{j=\widetilde{r}+1}^{ b_n}\sum\limits_{j'= b_n+1}^{\widetilde{n}}k(Z_{n^*+j}(\omega),Z_{n^*+j'}(\omega)) \leq \frac{(b_n-\widetilde{r})(\widetilde{n}-b_n)}{(\widetilde{n} - \widetilde{r})^2} \leq \frac{\widetilde{n}-b_n}{\widetilde{n} - \widetilde{r}} \xrightarrow[]{} 0, 
        \end{align}
        \begin{align} \label{Convergence of I7}
		I_{7} = \frac{1}{\widetilde{r}(\widetilde{n} - \widetilde{r})}\sum\limits_{i=1}^{a_n}\sum\limits_{j=\widetilde{r}+1}^{ b_n}k(Z_{n^*+i}(\omega),Z_{n^*+j}(\omega)) \leq \frac{a_n}{\widetilde{r}}\frac{b_n-\widetilde{r}}{\widetilde{n}-\widetilde{r}} \leq \frac{a_n}{\widetilde{r}} \xrightarrow[]{} 0, \text{ and }
        \end{align}
        \begin{align} \label{convergence of I8}
        & I_{8} = \frac{1}{\widetilde{r}(\widetilde{n} - \widetilde{r})}\sum\limits_{i=1}^{a_n}\sum\limits_{j= b_n+1}^{\widetilde{n}}k(Z_{n^*+i}(\omega),Z_{n^*+j}(\omega)) \leq \frac{a_n(\widetilde{n}-b_n)}{\widetilde{r}(\widetilde{n}-\widetilde{r})} \leq \frac{a_n}{\widetilde{r}} \xrightarrow[]{} 0. 
        \end{align}
Furthermore,
        \begin{align*}        	\frac{(\widetilde{r}-a_n)(b_n - \widetilde{r})}{\widetilde{r}(\widetilde{n}-\widetilde{r})} \xrightarrow[]{} 1, \quad \frac{1}{(\widetilde{r}-a_n)(b_n - \widetilde{r})}\sum\limits_{i=a_n+1}^{\widetilde{r}}\sum\limits_{j=\widetilde{r}+1}^{ b_n}k(Z_{n^*+i}(\omega),Z_{n^*+j}(\omega)) \xrightarrow[]{} \mu_{k}^{Z}. 
        \end{align*}
        \noindent
        Thus,
	\begin{align} \label{Convergence of I9}    I_{9} = \frac{1}{\widetilde{r}(\widetilde{n} - \widetilde{r})}\sum\limits_{i=a_n+1}^{\widetilde{r}}\sum\limits_{j=\widetilde{r}+1}^{ b_n}k(Z_{n^*+i}(\omega),Z_{n^*+j}(\omega))-\mu_{k}^{Z} \xrightarrow[]{} 0,
    \end{align}
    Also,
    \begin{align} \label{Convergence of I10}
		I_{10} = \frac{1}{\widetilde{r}(\widetilde{n} - \widetilde{r})}\sum\limits_{i=a_n+1}^{\widetilde{r}}\sum\limits_{j= b_n+1}^{\widetilde{n}}k(Z_{n^*+i}(\omega),Z_{n^*+j}(\omega)) \leq \frac{(\widetilde{r}-a_n)(\widetilde{n}-b_n)}{\widetilde{r}(\widetilde{n} - \widetilde{r})} \leq \frac{\widetilde{n}-b_n}{\widetilde{n} - \widetilde{r}} \xrightarrow[]{} 0.
	\end{align}
	All these convergences are uniform over $\widetilde{\gamma} \in (\sqrt{g_n},1-\sqrt{h_n})$ (by the similar argument used in the proof of Theorem 3.1). Hence,  $d_{\widetilde{n}}(\omega)(\widetilde{\gamma}) \xrightarrow[]{} 0$ uniformly over $\widetilde{\gamma} \in (\sqrt{g_n},1-\sqrt{h_n})$ (follows from equation \eqref{d under asymptotic null} to \eqref{Convergence of I10}). Hence,
	\begin{align} \label{Tn in middle under asymptotic null}
		\underset{\widetilde{\gamma} \in (\sqrt{g_n},1-\sqrt{h_n})}{\sup} \rho_{\widetilde{n}}(\omega)(\widetilde{\gamma}) &= \underset{\widetilde{\gamma} \in (\sqrt{g_n},1-\sqrt{h_n})}{\sup} \frac{\lfloor \widetilde{n}\widetilde{\gamma}\rfloor(\widetilde{n}-\lfloor \widetilde{n}\widetilde{\gamma}\rfloor)}{\widetilde{n}^2} d_{\widetilde{n}}(\omega)(\widetilde{\gamma})\\
        & \leq \underset{\widetilde{\gamma} \in (\sqrt{g_n},1-\sqrt{h_n})}{\sup} \frac{\lfloor \widetilde{n}\widetilde{\gamma}\rfloor(\widetilde{n}-\lfloor \widetilde{n}\widetilde{\gamma}\rfloor)}{\widetilde{n}^2} \underset{\widetilde{\gamma} \in (\sqrt{g_n},1-\sqrt{h_n})}{\sup} d_{\widetilde{n}}(\omega)(\widetilde{\gamma}) \nonumber\\ 
        &\xrightarrow[]{} 0. \nonumber
	\end{align}
	Therefore,
	\begin{align*}
		&\underset{\widetilde{\gamma} \in (\delta_{\widetilde{n}},1-\delta_{\widetilde{n}})}{\sup} \rho_{\widetilde{n}}(\omega)(\widetilde{\gamma}) \leq \underset{\widetilde{\gamma} \in (0,\sqrt{g_n}] \cup [1-\sqrt{h_n},1)}{\sup} \rho_{\widetilde{n}}(\omega)(\widetilde{\gamma})+\underset{\widetilde{\gamma} \in (\sqrt{g_n},1-\sqrt{h_n})}{\sup} \rho_{\widetilde{n}}(\omega)(\widetilde{\gamma}) \xrightarrow[]{} 0 \\ &(\text{follows from \eqref{Tn in boundaries under asymptotic null} and \eqref{Tn in middle under asymptotic null}}).
	\end{align*}
	Hence,
	\begin{align*}
		T_{\widetilde{n}}(\omega)=\rho_{\widetilde{n}}(\omega)(\widehat{\gamma}_{\widetilde{n}}) \xrightarrow[]{} 0.
	\end{align*}
    This completes the proof of Lemma \ref{Tn under asymptotic null}.
\end{proof}

\begin{supplemma} \label{Tn under asymptotic AMOC}
Borrowing the notations of the proof of Theorem 5.2, suppose that
\begin{align*}
& Z_{n^*+a_n+1}(\omega),Z_{n^*+a_n+2}(\omega),\dots,Z_{n^*+\tau_n}(\omega) \text{ are realizations from any $F_i$ and }\\
& Z_{n^*+\tau_n+1}(\omega),  Z_{n^*+\tau_n+2}(\omega),  \dots,Z_{n^*+ b_n}(\omega) \text{ are realizations from
    $F_{i+1}$,} 
\end{align*}    
    for fixed $\omega \in \cap_{r=1}^{K_0+1} \cap_{s=1}^{K_0+1}\{A_{r} \cap A_{r,s}\}$ with $F_i \neq F_{i+1}$ where $a_n/\widetilde{n} \xrightarrow[]{} 0, b_n/\widetilde{n} \xrightarrow[]{} 1$ and $n^*+\tau_n=\lfloor n\gamma\rfloor$ for some $\gamma \in \{\gamma^{(1)},\gamma^{(2)},\dots,\gamma^{(K_0)}\}$. If $\tau_{n}/\widetilde{n} \xrightarrow[]{} \eta \in (0,1)$, then
	\begin{align*} \widehat{\gamma}_{\widetilde{n}}(\omega) \xrightarrow[]{} \eta   \text{ 
			and } \: T_{\widetilde{n}}(\omega) \xrightarrow[]{} \eta(1-\eta)d(F_i, F_{i+1}).
	\end{align*}
\end{supplemma}
\begin{proof}
	Following similar argument used in the proof of the previous lemma, 
	\begin{align} \label{Tn in boundaries under asymptotic AMOC}
		\underset{\widetilde{\gamma} \in (0,\sqrt{g_n}]\cup[1-\sqrt{h_n},1)}{\sup}\rho_{\widetilde{n}}(\omega)(\widetilde{\gamma}) \xrightarrow[]{} 0.
	\end{align}
	Now, take $\widetilde{\gamma} \in (\sqrt{g_n},1-\sqrt{h_n})$. Denote $\lfloor \widetilde{n}\widetilde{\gamma}\rfloor$ by $\widetilde{r}$. Then,
	\begin{align*}
		d_{\widetilde{n}}(\omega)(\widetilde{\gamma}) &= \frac{1}{\widetilde{r}^2}\sum\limits_{i=1}^{\widetilde{r}}\sum\limits_{i'=1}^{\widetilde{r}}k(Z_{n^*+i}(\omega),Z_{n^*+i'}(\omega)) + \frac{1}{(\widetilde{n}- \widetilde{r})^2}\sum\limits_{j=\widetilde{r}+1}^{\widetilde{n}}\sum\limits_{j'=\widetilde{r}+1}^{\widetilde{n}}k(Z_{n^*+j}(\omega),Z_{n^*+j'}(\omega)) \\&- \frac{2}{\widetilde{r}(\widetilde{n} - \widetilde{r})}\sum\limits_{i=1}^{\widetilde{r}}\sum\limits_{j=\widetilde{r}+1}^{\widetilde{n}}k(Z_{n^*+i}(\omega),Z_{n^*+j}(\omega)).
	\end{align*}
	Let 
	\begin{align*}
		\widehat{\tau}_{\widetilde{n}}^*(\omega) = \underset{\lceil \widetilde{n}\sqrt{g_n}\rceil \leq t \leq \lfloor \widetilde{n}(1-\sqrt{h_n})\rfloor}{\arg \max} \rho_{\widetilde{n}}(\omega)(t/\widetilde{n}) \: \text{ and } \: \widehat{\gamma}_{\widetilde{n}}^*(\omega)=\widehat{\tau}_{\widetilde{n}}^*(\omega)/\widetilde{n}.
	\end{align*}
	Applying the law of large numbers for V-statistic, it can be proved that
	\begin{align} \label{uniform convergence of d in asymptotic AMOC}
		\underset{\widetilde{\gamma} \in (\sqrt{g_n},1-\sqrt{h_n})}{\sup}|d_{\widetilde{n}}(\omega)(\widetilde{\gamma})-S_{*}(\widetilde{\gamma})d(F_i, F_{i+1})| \xrightarrow[]{} 0,
	\end{align}
	where $S_{*}(\widetilde{\gamma})=(\frac{1-\eta}{1-\widetilde{\gamma}})^{2}\mathbb{I}(\widetilde{\gamma} \leq \eta)+(\frac{\eta}{\widetilde{\gamma}})^{2}\mathbb{I}(\widetilde{\gamma} > \eta)$ (This can be proved by similar line of argument used in the proof of Lemma 3.2).
	Note that $\widetilde{\gamma}(1-\widetilde{\gamma})S_{*}(\widetilde{\gamma})$ is maximum when $\widetilde{\gamma}=\eta$ and the maximum value is $\eta(1-\eta)$ (as $S_{*}(\eta)=1$).
	Now, by definition of $\widehat{\tau}_{\widetilde{n}}^*(\omega)$ and $\widehat{\gamma}_{\widetilde{n}}^*(\omega)$,
	\begin{align} \label{empirical measure at estimated and true breakfraction under asymptotic AMOC}
		\rho_{\widetilde{n}}(\omega)(\widehat{\gamma}_{\widetilde{n}}^*(\omega)) \geq \rho_{\widetilde{n}}(\omega)(\eta)\geq \eta(1-\eta)d(F_i, F_{i+1})-o(1).
	\end{align}
	The last inequality follows from \eqref{uniform convergence of d in asymptotic AMOC} and $\tau_n/\widetilde{n} \xrightarrow[]{} \eta$.\\
	Also,
	\begin{align} \label{limit at estimated and true breakfraction under asymptotic AMOC}
		0 &\leq \eta(1-\eta)S_{*}(\eta)d(F_i, F_{i+1})-\frac{\widehat{\tau}_{\widetilde{n}}^*(\omega)(\widetilde{n}-\widehat{\tau}_{\widetilde{n}}^*(\omega))}{\widetilde{n}^2}S_{*}(\widehat{\gamma}_{\widetilde{n}}^{*}(\omega))d(F_i, F_{i+1}) \nonumber\\&\leq \eta(1-\eta)d(F_i, F_{i+1})-\rho_{\widetilde{n}}(\omega)(\widehat{\gamma}_{\widetilde{n}}^*(\omega)) + o(1). 
	\end{align}
	The last inequality follows from \eqref{uniform convergence of d in asymptotic AMOC} with $\widetilde{\gamma}=\widehat{\gamma}_{\widetilde{n}}^*(\omega)$. Hence, $\rho_{\widetilde{n}}(\omega)(\widehat{\gamma}_{\widetilde{n}}^*(\omega)) \leq \eta(1-\eta)d(F_i, F_{i+1})+o(1)$. Therefore, $\rho_{\widetilde{n}}(\omega)(\widehat{\gamma}_{\widetilde{n}}^*(\omega)) \xrightarrow[]{} \eta(1-\eta)d(F_i, F_{i+1})$ (follows from \eqref{empirical measure at estimated and true breakfraction under asymptotic AMOC} and \eqref{limit at estimated and true breakfraction under asymptotic AMOC}).\\
	Also,
	\begin{align*}
		\underset{\lceil \widetilde{n}\delta_{\widetilde{n}}\rceil \leq t \leq \lfloor \widetilde{n}(1-\delta_{\widetilde{n}})\rfloor}{\max} \rho_{\widetilde{n}}(\omega)(t/\widetilde{n})=\underset{\lceil \widetilde{n}\sqrt{g_n}\rceil \leq t \leq \lfloor \widetilde{n}(1-\sqrt{h_n})\rfloor}{\max} \rho_{\widetilde{n}}(\omega)(t/\widetilde{n}) \: \: \text{eventually (due to \eqref{Tn in boundaries under asymptotic AMOC})},
	\end{align*}
	and therefore,
	$\widehat{\tau}_{\widetilde{n}}(\omega)=\widehat{\tau}_{\widetilde{n}}^*(\omega)$ eventually. Hence, $T_{\widetilde{n}}(\omega)=\rho_{\widetilde{n}}(\omega)(\widehat{\gamma}_{\widetilde{n}}(\omega)) \xrightarrow[]{} \eta(1-\eta)d(F_i, F_{i+1})$.
	\par
	Carrying out some algebraic calculations, we obtain 
	$$
	\left| \eta(1-\eta) - \widetilde{\gamma}(1-\widetilde{\gamma})S_{*}(\widetilde{\gamma}) \right|=\begin{cases}
		\frac{(\eta-\widetilde{\gamma})(1-\eta)}{1-\widetilde{\gamma}} & \text{if $\widetilde{\gamma} \leq \eta$},\\
		\frac{\eta(\widetilde{\gamma}-\eta)}{\widetilde{\gamma}} & \text{if $\widetilde{\gamma} > \eta$}.
	\end{cases}
	$$
	So, $| \eta(1-\eta) - \widetilde{\gamma}(1-\widetilde{\gamma})S_{*}(\widetilde{\gamma})| \geq |\eta-\widetilde{\gamma}|\min{\{\eta,1-\eta\}}$. In other words, if $|\eta-\widetilde{\gamma}| > \epsilon$ then, $| \eta(1-\eta)d(F_i, F_{i+1}) - \widetilde{\gamma}(1-\widetilde{\gamma})S_{*}(\widetilde{\gamma})d(F_i, F_{i+1})| > \epsilon \min{\{\eta,1-\eta\}}d(F_i, F_{i+1})$, where $\epsilon>0$ is any positive number.  \\
	Since, $| \eta(1-\eta)d(F_i, F_{i+1}) - \widehat{\gamma}_{\widetilde{n}}^*(\omega)(1-\widehat{\gamma}_{\widetilde{n}}^*(\omega))S_{*}(\widehat{\gamma}_{\widetilde{n}}^*(\omega))d(F_i, F_{i+1})| \xrightarrow[]{} 0$ (as the R.H.S of \eqref{limit at estimated and true breakfraction under asymptotic AMOC} converges to $0$ and $S_{*}(\eta)=1$), we have $|\widehat{\gamma}_{\widetilde{n}}^*(\omega)-\eta| \xrightarrow[]{} 0$. 
	As $\widehat{\tau}_{\widetilde{n}}^*(\omega)=\widehat{\tau}_{\widetilde{n}}(\omega)$ eventually, $\widehat{\gamma}_{\widetilde{n}}(\omega) \xrightarrow[]{} \eta$.\\
    This completes the proof of Lemma \ref{Tn under asymptotic AMOC}.
\end{proof}
\begin{supplemma} \label{Tn under asymptotic MCP}
	Borrowing the notations of the proof of Theorem 5.2, suppose that 
    \begin{align*}
    & Z_{n^*+a_n+1}(\omega),Z_{n^*+a_n+2}(\omega),\dots,Z_{n^*+\tau_1}(\omega) \text{ are realizations  from } F_{i_1},\\
    & Z_{n^*+\tau_1+1}(\omega),Z_{n^*+\tau_1+2}(\omega),\dots, Z_{n^*+\tau_2}(\omega) \text{ are realizations from } F_{i_1+1}, \\
    & \hspace{6cm}\vdots \\
     & Z_{n^*+\tau_M+1}(\omega),Z_{n^*+\tau_M+2}(\omega),\dots,Z_{n^*+b_n}(\omega) \text{ are realizations from } F_{i_M+1},
     \end{align*}
     for fixed $\omega \in \cap_{r=1}^{K_0+1} \cap_{s=1}^{K_0+1}\{A_{r} \cap A_{r,s}\}$ where $a_n/\widetilde{n} \xrightarrow[]{} 0, b_n/\widetilde{n} \xrightarrow[]{} 1,F_{i_j} \neq F_{i_j+1}$ and $n^*+\tau_j=\lfloor n\gamma^{(i_j)}\rfloor$ for $j=1,2,\dots,M$. If $\tau_{j}/\widetilde{n} \xrightarrow[]{} \eta^{(j)} \in (0,1)$ for $j=1,2,\dots,M$ such that $\eta^{(1)}<\eta^{(2)}<\dots<\eta^{(M)}$, $T_{\widetilde{n}}(\omega)$ converges to a positive quantity and
	\begin{align*}
		\underset{1 \leq i \leq M}{\min} |\widehat{\gamma}_{\widetilde{n}}(\omega) - \eta^{(i)}| \xrightarrow[]{} 0.
	\end{align*}
\end{supplemma}
\begin{proof}
	Using similar argument used in the proof of previous lemma, it can be proved that
	\begin{align*}
		\underset{\lceil \widetilde{n}\delta_{\widetilde{n}}\rceil \leq t \leq \lfloor \widetilde{n}(1-\delta_{\widetilde{n}})\rfloor}{\max} \rho_{\widetilde{n}}(\omega)(t/\widetilde{n})=\underset{\lceil \widetilde{n}\sqrt{g_n}\rceil \leq t \leq \lfloor \widetilde{n}(1-\sqrt{h_n})\rfloor}{\max} \rho_{\widetilde{n}}(\omega)(t/\widetilde{n}) \: \: \text{eventually},
	\end{align*}
	and therefore,
	\begin{align*}
		\widehat{\tau}_{\widetilde{n}}(\omega)=\underset{\lceil \widetilde{n}\sqrt{g_n}\rceil \leq t \leq \lfloor \widetilde{n}(1-\sqrt{h_n})\rfloor}{\arg \max} \rho_{\widetilde{n}}(\omega)(t/\widetilde{n}) \: \: \text{eventually}.
	\end{align*}
	The remaining part can be proved using similar argument used in the proof of Theorem 4.3.
\end{proof}

We are now ready to prove Theorem 5.2.\\
\textit{(a)} We need to show that for $K < K_0$, each of the estimated breakfractions converges to one of the true breakfractions $[\mathbb{P}]$ a.s. The sets of observations produced after completion of the $j$-th stage of Algorithm DESC-S are denoted by $C_{j}^{(1)},C_{j}^{(2)},\dots,C_{j}^{(j+1)}$ and the estimated breakfractions are denoted by $\widehat{\gamma}_{n,j}^{(1)},\widehat{\gamma}_{n,j}^{(2)},\dots,\widehat{\gamma}_{n,j}^{(j)}$ with $\widehat{\gamma}_{n,j}^{(1)}<\widehat{\gamma}_{n,j}^{(2)}<\dots<\widehat{\gamma}_{n,j}^{(j)}$.
\par Assume that $K_0 \geq 2$. Fix $\omega \in \cap_{r=1}^{K_0+1} \cap_{s=1}^{K_0+1}\{A_{r} \cap A_{r,s}\}$. For $K=1$, if $C_{1}^{(1)}(\omega)$ and $C_{1}^{(2)}(\omega)$ be the two sets of observations obtained after completion of the first stage of Algorithm DESC-S applied on the set $\{Z_{1}(\omega),Z_{2}(\omega),\dots,Z_{n}(\omega)\}$, $|C_{1}^{(1)}(\omega)|/n \xrightarrow[]{} \gamma$ for some $\gamma \in \{\gamma^{(1)},\gamma^{(2)},\dots,\gamma^{(K_0)}\}$ (follows from Lemma \ref{Tn under asymptotic MCP} with $n^*=0,a_n=0,b_n=\widetilde{n}=n$ and $\tau_i = \lfloor n\gamma^{(i)}\rfloor$ for $i=1,2,\dots,K_0$). We first show that $\underset{1 \leq i \leq 2}{\max}\:\underset{1 \leq i' \leq K_0}{\min} |\widehat{\gamma}_{n,2}^{(i)}(\omega)-\gamma^{(i')}| \xrightarrow[]{} 0$.  If $\gamma=\gamma^{(1)}$, $\rho(D_{1}^{(1)}(\omega),D_{1}^{(2)}(\omega)) \xrightarrow[]{} 0$ (follows from Lemma \ref{Tn under asymptotic null} applied on $C_{1}^{(1)}(\omega)$ with $n^*=a_n=0$, $b_n=|C_{1}^{(1)}(\omega)|$ if $\lfloor n\gamma^{(1)}\rfloor \geq |C_{1}^{(1)}(\omega)|$ and $b_n=\lfloor n\gamma^{(1)}\rfloor$ if $\lfloor n\gamma^{(1)}\rfloor < |C_{1}^{(1)}(\omega)|$) where $D_{1}^{(1)}(\omega)$ and $D_{1}^{(2)}(\omega)$ be two sets obtained by splitting $C_{1}^{(1)}(\omega)$. On the other hand, if $\gamma > \gamma^{(1)}$, $\rho(D_{1}^{(1)}(\omega),D_{1}^{(2)}(\omega))$ converges to a positive quantity (follows from Lemma \ref{Tn under asymptotic AMOC} and \ref{Tn under asymptotic MCP} applied on $C_{1}^{(1)}(\omega)$ with $n^*=a_n=0$, $b_n=|C_{1}^{(1)}(\omega)|$ if $\lfloor n\gamma\rfloor \geq |C_{1}^{(1)}(\omega)|$ and $b_n=\lfloor n\gamma\rfloor$ if $\lfloor n\gamma\rfloor < |C_{1}^{(1)}(\omega)|$) and for some $\gamma^* \in \{\gamma^{(1)},\gamma^{(2)},\dots,\gamma^{(K_0)}\}$ with $\gamma^* < \gamma$,
\begin{align*}
	\frac{|D_{1}^{(1)}(\omega)|}{|C_{1}^{(1)}(\omega)|} - \frac{\lfloor n\gamma^*\rfloor}{|C_{1}^{(1)}(\omega)|} \xrightarrow[]{} 0\quad \text{which implies} \quad \frac{|D_{1}^{(1)}(\omega)|}{|C_{1}^{(1)}(\omega)|} \xrightarrow[]{} \frac{\gamma^*}{\gamma}.
\end{align*}
 Hence,
\begin{align*}
	\frac{|D_{1}^{(1)}(\omega)|}{n}=\frac{|D_{1}^{(1)}(\omega)|}{|C_{1}^{(1)}(\omega)|}\frac{|C_{1}^{(1)}(\omega)|}{n} \xrightarrow[]{} \frac{\gamma^*}{\gamma}\gamma = \gamma^*.
\end{align*}
Therefore, if $\gamma>\gamma^{(1)}$, $|D_{1}^{(1)}(\omega)|/n \xrightarrow[]{} \gamma^*$ for some $\gamma^* \in \{\gamma^{(1)},\gamma^{(2)},\dots,\gamma^{(K_0)}\}$ with $\gamma^* < \gamma$. Similarly, if $\gamma=\gamma^{(K_0)}$, $\rho(D_{1}^{(3)}(\omega),D_{1}^{(4)}(\omega)) \xrightarrow[]{} 0$. On the other hand, if $\gamma < \gamma^{(K_0)}$, $\rho(D_{1}^{(3)}(\omega),D_{1}^{(4)}(\omega))$ converges to a positive constant and $\{|C_{1}^{(1)}(\omega)|+|D_{1}^{(3)}(\omega)|\}/n \xrightarrow[]{} \gamma'$ for some $\gamma' \in \{\gamma^{(1)},\gamma^{(2)},\dots,\gamma^{(K_0)}\}$ with $\gamma' > \gamma$. So, in the merging step, $D_{1}^{(1)}(\omega)$ and $D_{1}^{(2)}(\omega)$ will be merged and $\{|C_{1}^{(1)}(\omega)|+|D_{1}^{(3)}(\omega)|\}/n \xrightarrow[]{} \gamma'(>\gamma^{(1)})$ if $\gamma=\gamma^{(1)}$. On the other hand, $D_{1}^{(3)}(\omega)$ and $D_{1}^{(4)}(\omega)$ will be merged and $|D_{1}^{(1)}(\omega)|/n \xrightarrow[]{} \gamma^*(< \gamma^{(K_0)})$ if $\gamma=\gamma^{(K_0)}$. In the remaining cases, both $|D_{1}^{(1)}(\omega)|/n$ and $\{|C_{1}^{(1)}(\omega)|+|D_{1}^{(3)}(\omega)|\}/n$ converge to two different true breakfractions (follows from Lemma \ref{Tn under asymptotic AMOC} and \ref{Tn under asymptotic MCP}). After completion of the second stage of Algorithm DESC-S, one estimated breakfraction is $|C_{1}^{(1)}(\omega)|/n$. The other breakfraction is either $|D_{1}^{(1)}(\omega)|/n$ or $\{|C_{1}^{(1)}(\omega)|+|D_{1}^{(3)}(\omega)|\}/n$ and this estimated breakfraction also converges to one of the true breakfraction (which is different from $\gamma$). Therefore, $(C_{2}^{(1)}(\omega),C_{2}^{(2)}(\omega))=(D_{1}^{(1)}(\omega),D_{1}^{(2)}(\omega))$ or $(C_{1}^{(1)}(\omega),D_{1}^{(3)}(\omega))$ and $(\widehat{\gamma}_{n,2}^{(1)}(\omega),\widehat{\gamma}_{n,2}^{(2)}(\omega))=(|C_{2}^{(1)}(\omega)|/n,\{|C_{2}^{(1)}(\omega)|+|C_{2}^{(2)}(\omega)|\}/n)$. Hence, 
\begin{align*}
	\underset{1 \leq i \leq 2}{\max} \: \underset{1 \leq i' \leq K_0}{\min} |\widehat{\gamma}_{n,2}^{(i)}(\omega) - \gamma^{(i')}| \xrightarrow[]{} 0.
\end{align*}
Now, assume that for $j \leq K_0-2$,
\begin{align*}
	\underset{1 \leq i \leq j}{\max}\: \underset{1 \leq i' \leq K_0}{\min} |\widehat{\gamma}_{n,j}^{(i)}(\omega) - \gamma^{(i')}| \xrightarrow[]{} 0.
\end{align*}
The set of estimated $j$ changepoints splits the set of observations into $j+1$ groups $C_{j}^{(1)}(\omega), C_{j}^{(2)}(\omega), \dots$, $C_{j}^{(j+1)}(\omega)$ such that $\sum\limits_{i=1}^{l}|C_{j}^{(i)}(\omega)|/n \xrightarrow[]{} \gamma_{j}^{(l)}$ for $l=1,2,\dots,j$, where $\gamma_{j}^{(1)}<\gamma_{j}^{(2)}<\dots<\gamma_{j}^{(j)}$ be elements of the set of true breakfractions $\{\gamma^{(1)},\gamma^{(2)},\dots,\gamma^{(K_0)}\}$.
\\ Now, let us consider two types of sets of observations.
A set $C$ is referred to as ``pure'' if it either does not include any true changepoint or contains true changepoints such that the ratio $\tau/|C|$ converges either to $0$ or to $1$, where, $\tau$ denotes the position of the changepoint within $C$, and this condition must be satisfied for all changepoints present in the set. We refer to $C$ as ``impure'' if there exists at least one true changepoint for which the ratio $\tau/|C|$ converges to a positive value that lies between $0$ and $1$, excluding the endpoints. For $l=1,2,\dots,j$, if $\lfloor n\gamma_{j}^{(l-1)}\rfloor \geq \sum\limits_{i=1}^{l-1}|C_{j}^{(i)}(\omega)|$ then the ratio of relative location of $\lfloor n\gamma_{j}^{(l-1)}\rfloor$ in the set $C_{j}^{(l)}(\omega)$ and $|C_{j}^{(l)}(\omega|$ is
\begin{align*}
	\frac{\lfloor n\gamma_{j}^{(l-1)}\rfloor - \sum\limits_{i=1}^{l-1}|C_{j}^{(i)}(\omega)|}{|C_{j}^{(l)}(\omega)|} \xrightarrow[]{} \frac{\gamma_{j}^{(l-1)}-\gamma_{j}^{(l-1)}}{\gamma_{j}^{(l)}-\gamma_{j}^{(l-1)}}=0.
\end{align*}
On the other hand, if $\lfloor n\gamma_{j}^{(l)} \rfloor \leq \sum\limits_{i=1}^{l}|C_{j}^{(l)}(\omega)|$ then the ratio of relative location of $\lfloor n\gamma_{j}^{(l)}\rfloor$ in the set $C_{j}^{(l)}(\omega)$ and $|C_{j}^{(l)}(\omega|$ is
\begin{align*}
	\frac{ \lfloor n\gamma_{j}^{(l)}\rfloor-\sum\limits_{i=1}^{l-1}|C_{j}^{(i)}(\omega)|}{|C_{j}^{(l)}(\omega)|} \xrightarrow[]{} \frac{\gamma_{j}^{(l)}-\gamma_{j}^{(l-1)}}{\gamma_{j}^{(l)}-\gamma_{j}^{(l-1)}}=1.
\end{align*}
If $\sum\limits_{i=1}^{l-1}|C_{j}^{(i)}(\omega)| < \lfloor n\gamma^{(l^*)}\rfloor < \sum\limits_{i=1}^{l}|C_{j}^{(i)}(\omega)|$ such that $\gamma^{(l^*)} \in \{\gamma^{(1)},\gamma^{(2)},\dots,\gamma^{(K_0)}\}$ and $\gamma_{j}^{(l-1)} < \gamma^{(l^*)} < \gamma_{j}^{(l)}$, then the ratio of relative location of $\lfloor n\gamma_{j}^{(l^*)}\rfloor$ in the set $C_{j}^{(l)}(\omega)$ and $|C_{j}^{(l)}(\omega|$ is
\begin{align*}
	\frac{\lfloor n\gamma^{(l^*)}\rfloor-\sum\limits_{i=1}^{l-1} |C_{j}^{(i)}(\omega)|}{|C_{j}^{(l)}(\omega)|} \xrightarrow[]{} \frac{\gamma^{(l^*)}-\gamma_{j}^{(l-1)}}{\gamma_{j}^{(l)}-\gamma_{j}^{(l-1)}} \in (0,1).
\end{align*}
Hence, if $C_{j}^{(l)}(\omega)$ does not contain any true changepoint except $\lfloor n\gamma_{j}^{(l-1)}\rfloor$ and $\lfloor n\gamma_{j}^{(l)}\rfloor$ then $C_{j}^{(l)}(\omega)$ is a pure set.
Now, for $l=1,2,\dots,j+1$, $C_{j}^{(l)}(\omega)$ is split around the estimated changepoint to produce $D_{j}^{(2l-1)}(\omega)$ and $D_{j}^{(2l)}(\omega)$. If $C_{j}^{(l)}(\omega)$ is a ``pure'' set, $\rho(D_{j}^{(2l-1)}(\omega),D_{j}^{(2l)}(\omega)) \xrightarrow[]{} 0$ (follows from Lemma \ref{Tn under asymptotic null}). On the other hand, if $C_{j}^{(l)}(\omega)$ is ``impure'', $\rho(D_{j}^{(2l-1)}(\omega),D_{j}^{(2l)}(\omega))$ converges to a positive quantity (follows from Lemma \ref{Tn under asymptotic AMOC} and \ref{Tn under asymptotic MCP}). As the maximum number of ``pure'' sets among $C_{1}^{(j)}(\omega), C_{2}^{(j)}(\omega), \dots, C_{j+1}^{(j)}(\omega)$ is $j$, and $j$ many mergings of consecutive pairs of sets are to be done in the $(j+1)$-th stage, $D_{j}^{(2l-1)}(\omega)$ and $D_{j}^{2l}(\omega)$ will be merged if $C_{j}^{(l)}(\omega)$ is ``pure''. So, after the completion of the $(j+1)$-th stage, the output sets will be
\begin{align*}
C_{j}^{(1)}(\omega),C_{j}^{(2)}(\omega),\dots,C_{j}^{(l'-1)}(\omega),D_{j}^{(2l'-1)}(\omega),D_{j}^{(2l')}(\omega),C_{j}^{(l'+1)}(\omega),\dots,C_{j}^{(j+1)}(\omega)    
\end{align*}
 where $C_{j}^{(l')}(\omega)$ is an ``impure'' set for some $l' \in \{1,2,\dots,j+1\}$. Hence, the number of sets of observations will be $j+2$, i.e., the number of estimated breakfractions is $j+1$ and the $i$-th breakfraction is 
\begin{eqnarray}
	\widehat{\gamma}_{n,j+1}^{(i)}(\omega) = \left \{
	\begin{array}{l@{\quad} l }
		\frac{\sum\limits_{r=1}^{i}|C_{j}^{(r)}(\omega)|}{n} ; & \text{for $i = 1, 2, \dots, l'-1$},\\\\
		\frac{\sum\limits_{r=1}^{l'-1}|C_{j}^{(r)}(\omega)|+|D_{j}^{(2l'-1)}(\omega)|}{n} ; & \text{for $i = l'$},\\\\
		\frac{\sum\limits_{r=1}^{i-1}|C_{j}^{(r)}(\omega)|}{n} ; & \text{for $i = l'+1, l'+2, \dots, j+1$.}
	\end{array}
	\right . 
\end{eqnarray}
Since $C_{j}^{(l')}(\omega)$ is impure, $|D_{j}^{(2l'-1)}(\omega)|/|C_{j}^{(l')}(\omega)|$ converges to $(\gamma^*-\gamma_{j}^{(l'-1)})/(\gamma_{j}^{(l')}-\gamma_{j}^{(l'-1)})$ (follows from Lemma \ref{Tn under asymptotic AMOC} and \ref{Tn under asymptotic MCP}) for some $\gamma^* \in \{\gamma^{(1)},\gamma^{(2)},\dots,\gamma^{(K_0)}\}$ such that $\gamma_{j}^{(l'-1)} < \gamma^* < \gamma_{j}^{(l')}$. Hence, $(\sum\limits_{i=1}^{l'-1}|C_{j}^{(i)}(\omega)|+|D_{j}^{(2l'-1)}(\omega)|)/n$ converges to $\gamma^*$.
Therefore, after completion of the $(j+1)$-th stage, each of the estimated breakfractions converges to a true breakfraction and no two different estimated breakfractions converge to same true breakfraction. Hence, 
\begin{align*}
	\underset{1 \leq i \leq j+1}{\max} \:\underset{1 \leq i' \leq K_0}{\min} |\widehat{\gamma}_{n,j+1}^{(i)}(\omega) - \gamma^{(i')}| \xrightarrow[]{} 0.
\end{align*}
By the argument of mathematical induction, for any $K<K_0$
\begin{align} \label{POP underestimation for fixed omega}
	\underset{1 \leq i \leq K}{\max} \:\underset{1 \leq i' \leq K_0}{\min} |\widehat{\gamma}_{n,K}^{(i)}(\omega) - \gamma^{(i')}| \xrightarrow[]{} 0.
\end{align}
Since, \eqref{POP underestimation for fixed omega} holds for any $\omega \in \cap_{r=1}^{K_0+1} \cap_{s=1}^{K_0+1}(A_{r} \cap A_{r,s})$,
\begin{align*}
	\underset{1 \leq i \leq K}{\max} \:\underset{1 \leq i' \leq K_0}{\min} |\widehat{\gamma}_{n,K}^{(i)} - \gamma^{(i')}| \xrightarrow[]{} 0 \quad [\mathbb{P}] ~ a.s.
\end{align*}
This completes the proof of part \textit{(a)} of Theorem 5.2.\\\\
\textit{(b)} We will show that if $K = K_0$, $\widehat{\gamma}_{n,K}^{(i)} \xrightarrow[]{} \gamma^{(i)}$ $[\mathbb{P}] ~ a.s.$ for all $i=1,2,\dots,K_{0}$.
\par By part \textit{(a)} of Theorem 5.2, after completion of the $(K_0-1)$-th stage of Algorithm DESC-S, the cardinality of the set of estimated breakfractions is $K_0-1$ and each of the estimated breakfractions converges to one of the true breakfractions $[\mathbb{P}]~a.s.$ Also, no two different elements of the set of estimated breakfractions converge to the same true breakfraction. Fix $\omega \in \cap_{r=1}^{K_0+1} \cap_{s=1}^{K_0+1}(A_{r} \cap A_{r,s})$. Let $C_{K_0-1}^{(1)}(\omega),C_{K_0-1}^{(2)}(\omega),\dots,C_{K_0-1}^{(K_0)}(\omega)$ be the output sets obtained after the $(K_0-1)$-th stage. Also, $\widehat{\gamma}_{n,K_0-1}^{(i)}(\omega)=\sum\limits_{r=1}^{i}|C_{K_0-1}^{(r)}(\omega)|/n \xrightarrow[]{} \gamma_{K_0-1}^{(i)}$ (follows from part \textit{(a)}) where
\begin{eqnarray}
	\gamma_{K_0-1}^{(i)} = \left \{
	\begin{array}{l@{\quad} l }
		\gamma^{(i)} ; & \text{for $ i= 1, 2, \dots, l-1$},\\\\
		\gamma^{(i+1)} ; & \text{for $i = l,l+1,\dots,K_0-1$,}
	\end{array}
	\right.  
\end{eqnarray}
for some $l \in \{1,2,\dots,K_0-1\}$. So, among the sets $C_{K_0-1}^{(1)}(\omega),C_{K_0-1}^{(2)}(\omega),\dots,C_{K_0-1}^{(K_0)}(\omega)$, the only ``impure'' set is $C_{K_0-1}^{(l)}(\omega)$. Now, for $i=1,2,\dots,K_0$, the set $C_{K_0-1}^{(i)}(\omega)$ is split around the estimated changepoint to produce $D_{K_0-1}^{(2i-1)}(\omega)$ and $D_{K_0-1}^{(2i)}(\omega)$. For $i \neq l$, $\rho(D_{K_0-1}^{(2i-1)}(\omega),D_{K_0-1}^{(2i)}(\omega)) \xrightarrow[]{} 0$ as $C_{K_0-1}^{(i)}(\omega)$ is ``pure'' (follows from Lemma \ref{Tn under asymptotic null}). Also, $\rho(D_{K_0-1}^{2l-1}(\omega),D_{K_0-1}^{(2l)}(\omega))$ converges to a positive quantity and $|D_{K_0-1}^{(2l-1)}(\omega)|/|C_{K_0-1}^{(l)}(\omega)| \xrightarrow[]{} (\gamma^{(l)}-\gamma^{(l-1)})/(\gamma^{(l+1)}-\gamma^{(l-1)})$ (follows from Lemma \ref{Tn under asymptotic AMOC}). Hence, in the merging step, $D_{K_0-1}^{(2i-1)}(\omega)$ and $D_{K_0-1}^{(2i)}(\omega)$ will be merged to reform $C_{K_0-1}^{(i)}(\omega)$ again for $i \neq l$ (as $K_0-1$ pairs of consecutive sets are to be merged in the $K_0$-th stage). Therefore, after completion of the $K_0$-th stage, the output sets will be $C_{K_0-1}^{(1)}(\omega),\dots,C_{K_0-1}^{(l-1)}(\omega),D_{K_0-1}^{(2l-1)}(\omega)$, $D_{K_0-1}^{(2l)}(\omega),C_{K_0-1}^{(l+1)}(\omega),\dots,C_{K_0-1}^{(K_0)}(\omega)$ and the number of estimated breakfraction is $K_0$ where the $i$-th breakfraction is
\begin{eqnarray}
	\widehat{\gamma}_{n,K_0}^{(i)}(\omega) = \left \{
	\begin{array}{l@{\quad} l }
		\frac{\sum\limits_{r=1}^{i}|C_{K_0-1}^{(r)}(\omega)|}{n} ; & \text{for $i = 1, 2, \dots, l-1$},\\\\
		\frac{\sum\limits_{r=1}^{l-1}|C_{K_0-1}^{(r)}(\omega)|+|D_{K_0-1}^{(2l-1)}(\omega)|}{n} ; & \text{for $i = l$},\\\\
		\frac{\sum\limits_{r=1}^{i-1}|C_{j}^{(r)}(\omega)|}{n} ; & \text{for $i = l+1, l+2, \dots, K_0$.}
	\end{array}
	\right . 
\end{eqnarray}
Note that, $\widehat{\gamma}_{n,K_0}^{(i)}(\omega) \xrightarrow[]{} \gamma^{(i)}$ for $i=1,2,\dots,l-1,l+1,\dots,K_0$. Also,
\begin{align*}
	\widehat{\gamma}_{n,K_0}^{(l)}(\omega)=\frac{\sum\limits_{r=1}^{l-1}|C_{K_0-1}^{(r)}(\omega)|+|D_{K_0-1}^{(2l-1)}(\omega)|}{n} \xrightarrow[]{} \gamma^{(l-1)} + \frac{\gamma^{(l)}-\gamma^{(l-1)}}{\gamma^{(l+1)}-\gamma^{(l-1)}}(\gamma^{(l+1)}-\gamma^{(l-1)})=\gamma^{(l)}.
\end{align*}
Hence,
$\widehat{\gamma}_{n,K_0}^{(i)}(\omega) \xrightarrow[]{} \gamma^{(i)}$ for all $i=1,2,\dots,K_0$. As this holds for any $\omega \in B \cap \{\cap_{r=1}^{K_0+1} \cap_{s=1}^{K_0+1}(A_{r} \cap A_{r,s})\}$,
\begin{align} \label{Convergence in supervised case}
\widehat{\gamma}_{n,K_0}^{(i)} \xrightarrow[]{} \gamma^{(i)} \quad [\mathbb{P}] ~ a.s.,
\end{align}
for all $i=1,2,\dots,K_0$.\\
Now,
\begin{align*}
    d_H(\widehat{\boldsymbol{\gamma}}_{[K_0]},\boldsymbol{\gamma}_{[K_0]}) = \max\{\underset{1 \leq i \leq K_0}{\max} |\widehat{\gamma}_{n,K_0}^{(i)}-\boldsymbol{\gamma}_{[K_0]}|,\underset{1 \leq i \leq K_0}{\max}|\widehat{\boldsymbol{\gamma}}_{[K_0]}-\gamma^{(i)}|\},
\end{align*}
where $|x-Y|=\underset{y \in Y}{\inf} |x-y|=|Y-x|$ and $\widehat{\boldsymbol{\gamma}}_{[K_0]},\boldsymbol{\gamma}_{[K_0]}$ are defined in Theorem 5.2. 
\par
Note that for $i,j=1,2,\dots,K_{0}$,
\begin{align*} 
 |\gamma^{(i)} - \gamma^{(j)}| \leq  |\widehat{\gamma}_{n,K_0}^{(i)} - \gamma^{(i)}| + |\widehat{\gamma}_{n,K_0}^{(i)} - \gamma^{(j)}|,  
\end{align*}
which implies for $i \neq j$,
\begin{align} \label{abs diff bet estimated and different true breakfraction}
    \underset{n \to \infty}{\liminf}|\widehat{\gamma}_{n,K_0}^{(i)} - \gamma^{(j)}| \geq |\gamma^{(i)} - \gamma^{(j)}| > 0 \quad [\mathbb{P}] ~ a.s.
\end{align}
Hence, for each $i=1,2,\dots,K_0$, $|\widehat{\gamma}_{n,K_0}^{(i)} - \gamma^{(i')}|$ is minimum when $i'=i$ for sufficiently large $n$ (follows from \eqref{Convergence in supervised case} and \eqref{abs diff bet estimated and different true breakfraction}). Therefore, for sufficiently large $n$,
\begin{align*}
    \underset{1 \leq i \leq K_0}{\max} |\widehat{\gamma}_{n,K_0}^{(i)}-\boldsymbol{\gamma}_{[K_0]}| = \underset{1 \leq i \leq K_0}{\max} \: \underset{1 \leq i' \leq K_0}{\min} |\widehat{\gamma}_{n,K_0}^{(i)} - \gamma^{(i')}| = \underset{1 \leq i \leq K_0}{\max} |\widehat{\gamma}_{n,K_0}^{(i)} - \gamma^{(i)}|. 
\end{align*}
Similarly, for every $i'=1,2,\dots,K_{0}$, $|\widehat{\gamma}_{n,K_0}^{(i)} - \gamma^{(i')}|$ is minimum when $i=i'$ for sufficiently large $n$ (follows from \eqref{Convergence in supervised case} and \eqref{abs diff bet estimated and different true breakfraction}). Therefore, for sufficiently large $n$,
\begin{align*}
    \underset{1 \leq i' \leq K_0}{\max}|\widehat{\boldsymbol{\gamma}}_{[K_0]}-\gamma^{(i')}| = \underset{1 \leq i' \leq K_0}{\max} \:\underset{1 \leq i \leq K_0}{\min} |\widehat{\gamma}_{n,K_0}^{(i)} - \gamma^{(i')}| = \underset{1 \leq i' \leq K_0}{\max} |\widehat{\gamma}_{n,K_0}^{(i')} - \gamma^{(i')}|.
\end{align*}
 Hence, 
\begin{align*}
    d_H(\widehat{\boldsymbol{\gamma}}_{[K_0]},\boldsymbol{\gamma}_{[K_0]}) = \underset{1 \leq i \leq K_0}{\max}  |\widehat{\gamma}_{n,K_0}^{(i)} - \gamma^{(i)}|  \: \text{eventually,}
\end{align*}
and $d_H(\widehat{\boldsymbol{\gamma}}_{[K_0]},\boldsymbol{\gamma}_{[K_0]}) \xrightarrow[]{} 0$ $[\mathbb{P}] ~ a.s.$ (since $\widehat{\gamma}_{n,K_0}^{(i)} \xrightarrow[]{} \gamma^{(i)} \quad [\mathbb{P}] ~ a.s.$ for all $i=1,2,\dots,K_0$).\\
This completes the proof of part \textit{(b)} of Theorem 5.2.
\\\\ 
\textit{(c)} According to Algorithm DESC-S, the set of estimated breakfractions in the $j$-th stage will be a subset of estimated breakfractions in the $(j+1)$-th stage for $j=1,2,\dots$. Hence, for $K=K_0+1,K_0+2,\dots$, the set $\widehat{\boldsymbol{\gamma}}_{[K_0]}$ will be a subset of the estimated breakfractions in the $K$-th stage. Hence, for every $i'=1,2,\dots,K_0$,
\begin{align*}
    \underset{1 \leq i \leq K}{\min} |\widehat{\gamma}_{n,K}^{(i)} - \gamma^{(i')}| \leq \underset{1 \leq i \leq K_0}{\min} |\widehat{\gamma}_{n,K_0}^{(i)} - \gamma^{(i')}|
\end{align*}
Therefore, 
\begin{align*}
\underset{1 \leq i' \leq K_0}{\max} \:\underset{1 \leq i \leq K}{\min} |\widehat{\gamma}_{n,K}^{(i)} - \gamma^{(i')}| \leq \underset{1 \leq i' \leq K_0}{\max} \:\underset{1 \leq i \leq K_0}{\min} |\widehat{\gamma}_{n,K_0}^{(i)} - \gamma^{(i')}| \leq  d_H(\widehat{\boldsymbol{\gamma}}_{[K_0]},\boldsymbol{\gamma}_{[K_0]}).  
\end{align*}
Since the R.H.S of the above inequality converges to 0 $[\mathbb{P}] ~ a.s.$ (follows from part \textit{(b)} of Theorem 5.2),
\begin{align*}
    \underset{1 \leq i' \leq K_0}{\max} \:\underset{1 \leq i \leq K}{\min} |\widehat{\gamma}_{n,K}^{(i)} - \gamma^{(i')}| \xrightarrow[]{} 0 \quad \text{$[\mathbb{P}]~a.s.$} 
\end{align*}
This completes the proof of part \textit{(c)} of Theorem 5.2.
\section*{ Simulation Results}
        The performance of Algorithm DESC-S for models 1 to 12 (mentioned in the `Simulated Data' section of the main paper) are summarized in Table \ref{DESC-S in AMOC} and \ref{DESC-S in mcp}. These two tables illustrate that the changepoint locations identified by Algorithm DESC-S are in close agreement with the actual changepoint locations, or are found within one neighborhood of them, in almost all cases where the specified and actual number of changepoints match. In situations of underspecification, the estimated changepoints mainly represent a subset of the true changepoints, while in cases of overspecification, the estimated changepoints generally form a superset of the true changepoints. These findings are corroborated by Theorem 5.2.
\begin{table}
\centering
	\caption{Performance of Algorithm DESC-S in single changepoint setup ($K_0=1$): The numbers in each cell are (i) $\mathbb{P}(\widehat{L} \simeq L) \times 100\%$ when $K=1$ and (ii) $\mathbb{P}(\widehat{L} \supsetsim L) \times 100\%$ when $K=2$.}
	\label{DESC-S in AMOC}
	\scalebox{0.9}{
		\begin{tabular} {@{}cccc@{}}
			\hline
			\textbf{Model} &$n_1$ & $K=1$ &  $K=2$\\
			\hline
			1& 45& 100& 100\\
			
			($n=300$)& 150& 100& 100 \\
			
			& 240& 100& 100\\
			[6pt]
			1& 90& 100& 100\\
			
			($n=600$)& 300& 100& 100\\
			
			& 480& 100& 100\\
			[6pt]
			2& 45& 100& 100\\
			
			($n=300$)& 150& 100& 100\\
			
			& 240& 100& 100\\
			[6pt]
			2& 90& 100& 100\\
			
			($n=600$)& 300& 100& 100\\
			
			& 480& 100& 100\\
			[6pt]
			3& 45& 100& 100\\
			
			($n=300$)& 150& 100& 100\\
			
			& 240& 100& 100\\
			[6pt]
			3& 90& 100& 100\\
			
			($n=600$)& 300& 100& 100\\
			
			& 480& 100& 100\\
			[6pt]
			4& 45& 94& 94\\
			
			($n=300$)& 150& 91& 91\\
			
			& 240& 93& 93\\
			[6pt]
			4& 90& 90& 90\\
			
			($n=600$)& 300& 90& 90\\
			
			& 480& 94& 94\\
			[6pt]
			5& 45& 100& 100\\
			
			($n=300$)& 150& 100& 100\\
			
			& 240& 100& 100\\
			[6pt]
			5& 90& 100& 100\\
			
			($n=600$)& 300& 100& 100\\
			
			& 480& 100& 100\\
			[6pt]
			
			6& 45& 100& 100\\
			
			($n=300$)& 150& 100& 100\\
			
			& 240& 100& 100\\
			[6pt]
			6& 90& 100& 100\\
			
			($n=600$)& 300& 100& 100\\
			
			& 480& 100& 100\\
			[6pt]
			7& 45& 90& 90\\
			
			($n=300$)& 150& 100& 100\\
			
			& 240& 100& 100\\
			[6pt]
			7& 90& 98& 98\\
			
			($n=600$)& 300& 100& 100\\
			
			& 480& 97& 97\\
			\hline
		\end{tabular}
	}

\end{table}

\begin{table}
\centering
	\caption{Performance of Algorithm DESC-S in two changepoints setup ($K_0=2$): The three numbers in each cell are (i) $\mathbb{P}(\widehat{L} \subsetsim L) \times 100\%$ when $K=1$, (ii) $\mathbb{P}(\widehat{L} \simeq L) \times 100\% $ when $K=2$ and (iii) $\mathbb{P}(\widehat{L} \supsetsim L) \times 100\%$ when $K=3$.}
	\label{DESC-S in mcp}
    \scalebox{1.2}{
		\begin{tabular} {@{}ccccc@{}}
			\hline
			\textbf{Models} &$(n_1,n_2)$ &$K=1$ &$K=2$ & $K=3$ \\
			\hline
			8& 45,\:75& 100& 100& 100\\
			
			($n=300$)& 100,\:100& 100& 100& 100\\
			
			& 180,\:45& 100& 100& 100\\
			[6pt]
			8& 90,\:150& 100& 100& 100\\
			
			($n=600$)& 200,\:200& 100& 100& 100\\
			
			& 360,\:90& 100& 100& 100\\
			[6pt]
			9& 45,\:75& 90& 87& 87\\
			
			($n=300$)& 100,\:100& 90& 84& 84\\
			
			& 180,\:45& 86& 82& 82\\
			[6pt]
			& 90,\:150& 89& 85 & 85\\
			
			($n=600$)& 200,\:200& 81& 80& 80\\
			
			& 360,\:90& 87& 78& 78\\
			[6pt]
			10& 45,\:75& 100& 100& 100\\
			
			($n=300$)& 100,\:100& 100& 100& 100 \\
			
			& 180,\:45& 100& 100& 100  \\
			[6pt]
			10& 90,\:150& 100& 100& 100\\
			
			($n=600$)& 200,\:200& 100& 100& 100\\
			
			& 360,\:90& 100& 100& 100\\
			[6pt]
			11& 45,\:75& 98& 93& 93\\
			
			($n=300$)& 100,\:100& 93& 89& 89\\
			
			& 180,\:45& 81& 78& 78\\
			[6pt]
			11& 90,\:150& 92& 89& 89\\
			
			($n=600$)& 200,\:200& 96& 95& 95\\
			
			& 360,\:90& 82& 81& 81\\
			[6pt]
			12& 45,\:75& 100& 99& 99\\
			
			($n=300$)& 100,\:100& 97& 97& 97\\
			
			& 180,\:45& 81& 81& 81\\
			[6pt]
			12& 90,\:150& 100& 100& 100\\
			
			($n=600$)& 200,\:200& 99& 99& 99\\
			
			& 360,\:90& 88& 88& 88\\
			\hline
		\end{tabular}
        }

\end{table}

\section*{ Real Data Analysis}
\begin{figure}[t!]
	\centering
	\includegraphics[scale=0.4]{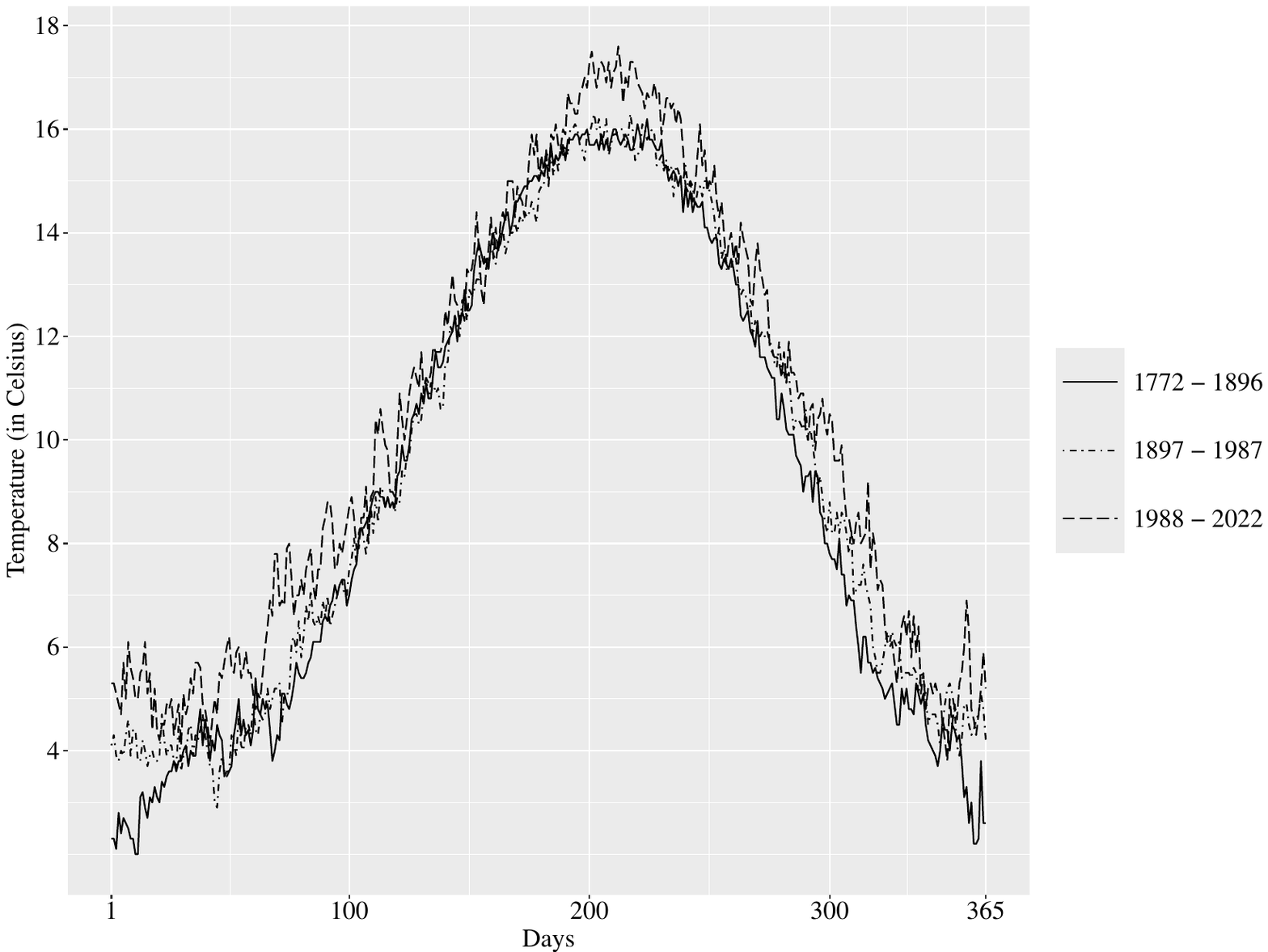}
	\caption{Median temperature curves for the segments $1772-1896$, $1897-1987$ and $1988-2022$ in the CET data obtained by Algorithm DESC-U.}
	\label{fig:DESC-U_full}
\end{figure}

  Figure \ref{fig:DESC-U_full} shows the median temperature curves for the three segments estimated by Algorithm DESC-U when applied on Central England Temperature (CET) dataset (mentioned in the `Real Data Analysis' section of the main paper) which clearly illustrates the warming trend.
\noindent

 
\end{document}